\documentclass[aps,pre,preprint,groupedaddress]{revtex4-2}

\usepackage{amsmath}
\usepackage{amssymb}
\usepackage{CJKutf8}
\usepackage{booktabs}
\usepackage{caption}

\usepackage{graphicx}% Include figure files
\usepackage{dcolumn}% Align table columns on decimal point
\usepackage{bm}% bold math
\usepackage[plain]{fancyref}
\usepackage{subfig}
\usepackage{chngcntr}
\usepackage{xcolor}
\usepackage{algorithm}
\usepackage[noend]{algpseudocode}
\makeatletter
\def\BState{\State\hskip-\ALG@thistlm}
\makeatother

\usepackage{hyperref}% add hypertext capabilities
\usepackage[mathlines]{lineno}% Enable numbering of text and display math

\newcommand*{\fancyrefapplabelprefix}{appsec}

\frefformat{main}{\fancyrefapplabelprefix}{%
Appendix \fancyrefdefaultspacing#2%
}%
\Frefformat{main}{\fancyrefapplabelprefix}{%
Appendix \fancyrefdefaultspacing#1%
}%

\frefformat{plain}{\fancyrefapplabelprefix}{Appendix #1}
\Frefformat{plain}{\fancyrefapplabelprefix}{Appendix #1}

\begin{document}

\preprint{APS/???-???}
\title{ Fitting In and Breaking Up: A Nonlinear Version of Coevolving Voter Models}% Force line breaks with \\
%\thanks{A footnote to the article title}%

\author{Yacoub H. Kureh}
%  \email{ykureh@math.ucla.edu}
\author{Mason A. Porter}
%  \email{mason@math.ucla.edu}
\affiliation{%
Department of Mathematics, University of California Los Angeles, Los Angeles, California 90095, USA
}%
\date{\today}

\begin{abstract}

We investigate a nonlinear version of coevolving voter models, in which node states and network structure update as a coupled stochastic dynamical process. Most prior work on coevolving voter models has focused on linear update rules with fixed and homogeneous rewiring and adopting probabilities. By contrast, in our nonlinear version, the probability that a node rewires or adopts is a function of how well it ``fits in'' within its neighborhood. To explore this idea, we incorporate a parameter $\sigma$ that represents the fraction of neighbors of an updating node that share its opinion state. In an update, with probability $\sigma^q$ (for some nonlinearity parameter $q$), the updating node rewires; with complementary probability $1-\sigma^q$, the updating node adopts a new opinion state. We study this mechanism using three rewiring schemes: after an updating node deletes a discordant edge, it then either (1) ``rewires-to-random" by choosing a new neighbor in a random process; (2) ``rewires-to-same" by choosing a new neighbor in a random process from nodes that share its state; or (3) ``rewires-to-none" by not rewiring at all (akin to ``unfriending'' on social media). We compare our nonlinear coevolving model to several existing linear models, and we find in our model that initial network topology plays a larger role in the dynamics and the choice of rewiring mechanism plays a smaller role. A particularly interesting feature of our model is that, under certain conditions, the opinion state that is held initially by a minority of the nodes can effectively spread to almost every node in a network if the minority nodes view themselves as the majority. In light of this observation, we relate our results to recent work on the majority illusion in social networks.

\end{abstract}

\pacs{Valid PACS appear here}% PACS, the Physics and Astronomy
% Classification Scheme.
%\keywords{Suggested keywords}%Use showkeys class option if keyword
%display desired
\maketitle
% \tableofcontents
%%%%%%%

%%%%%%%%%%%%%%%%%%%%%%%%
%%% Introduction
%%%%%%%%%%%%%%%%%%%%%%%%

\section{Introduction} \label{sec:CVMBackground}

The spread of opinions and the competition between different opinions is a vital aspect of societal discourse, and analysis of such phenomena has become increasingly prominent amidst the ubiquity of social media and intensifying political polarization~\cite{sunstein2018republic}. Such topics have been studied using a variety of lenses from numerous disciplines~\cite{del2016echo,john2015don,skoric2018predicts,geschke2019triple}, including longstanding efforts to develop mathematical frameworks for modeling opinion dynamics by employing ideas from subjects such as statistical physics and nonlinear dynamics~\cite{castellano2009statistical}. There has also been much cross-fertilization with research in the modeling of disease spreading~\cite{kiss2017mathematics}. For example, one can examine the ``virality'' of memes or seemingly contagious behaviors~\cite{lehmann2018complex}. 

In developing mathematical frameworks for studying opinion dynamics, accounting for social network structure can significantly improve both the accuracy of mathematical models and the understanding of spreading processes~\cite{opinion-review,lehmann2018complex,pastor-satorras2015}. 
For example, research on the severe acute respiratory syndrome (SARS) outbreaks by L. A. Meyers et al. illustrated the utility of accounting for social networks for assessing public-health strategies \cite{meyers2005network}. This stems from a network's influence on the properties of dynamical processes that take place on them~\cite{porter2016dynamical}. Additionally, networks themselves are typically not time-independent, as they often evolve in response to a dynamical process and in turn influence that process~\cite{sayama2013modeling,demirel2014moment}. For instance, in the spread of diseases, networks of interactions can change as a result of quarantines or different daily habits when somebody is ill. Similarly, in social media, individuals can choose to ``follow'' or ``unfollow'' other individuals (or other types of accounts) in response to posted content. The interplay of dynamics \emph{on} networks and dynamics \emph{of} networks \footnote{See~\cite{holme2012temporal,holme2015,holme2019} for reviews of research on time-dependent networks.} is a rapidly growing area of study in many disciplines \cite{porter2016dynamical,lehmann2018complex}. Although much of the prior work on such ``adaptive'' (also known as ``coevolving'') network models has focused on studying the complex behavior of simple models in abstract settings, there have also been efforts at incorporating further realism into the models~\cite{redner2018reality} and at applying such models to study empirical data in situations --- such as vote shares in historical United States elections~\cite{Fernandez-Gracia2014Voter} and the swarming dynamics of locusts~\cite{huepe2011adaptive} --- that can involve notions of ``opinions'' and consensus. 

A popular family of adaptive network models are \emph{coevolving voter models} (CVMs), in which node opinions (the node states) coevolve with network structure~\cite{porter2016dynamical}. (These models are also sometimes called ``adaptive voter models''.) Coevolving voter models combine the classical framework of voter models~\cite{clifford1973model, holley1975ergodic, cox1989coalescing}, in which individuals update their opinions based on their neighbors' opinions, with an evolving network structure (in which individuals change their relationships with other individuals in response to their opinions~\cite{ebel2002evolutionary, bornholdt2000topological, gross2006epidemic, gross2007adaptive}). A coevolving voter model consists of a network of individuals, two or more opinion states, and a rule (e.g., in the form of a stochastic process) for updating both the network and the states of its nodes or edges. We restrict our attention to a binary set of opinions, but one can also study models with more than two opinions~\cite{Holme2006Nonequilibrium,shi2013multiopinion} or with continuous opinions (e.g., using a bounded-confidence mechanism~\cite{brede2019}). There is also an interesting coevolving opinion model that includes states both on the nodes and on the edges \cite{saeedian2019absorbing}.

One of the motivations for studying coevolving voter models is their fascinating dynamics, and scholars have analyzed them using approaches from subjects like dynamical systems, statistical physics, partial differential equations, and probability theory~\cite{Holme2006Nonequilibrium,durrett2012graph,demirel2014moment,Silk2014Exploring,chodrow2018local}. A particularly interesting aspect of the dynamics of CVMs is the apparent phase transition that can occur in ``linear'' CVMs. In this context, ``linear'' refers to the linearity of the rewiring probability function $f_{\text{r}}(x)$ and adoption probability function $f_{\text{a}}(x)$: A vertex with a fraction $x$ of disagreeing neighbors has a probability $f_{\text{r}}(x) = \alpha x$ (for some parameter $\alpha$) to rewire and a probability $f_{\text{r}}(x) = (1-\alpha) x$ to adopt. The parameter $\alpha$ is sometimes called the ``rewiring rate'', and $(1-\alpha)$ is sometimes called the ``adoption rate''. 
{In some variants of linear CVMs (such as those in \cite{Holme2006Nonequilibrium,durrett2012graph}), there appears to be a phase transition as one increases the adoption rate (and thus decreases the rewiring rate) from a regime of ``rapid disintegration'' to a regime of ``prolonged persistence of the dynamics''. In the former regime, a network separates into components so quickly that the densities of the opinion states are unable to change significantly. 
In the latter regime, the system progresses slowly towards a steady state in which almost every node has the same opinion state.} Basu and Sly recently presented a mathematically rigorous proof of the existence of phase transitions for two variants of linear CVMs on dense Erd\H{o}s--R\'enyi (ER) networks (using the $G(N,p)$ model with $p = 1/2$)~\cite{basu2017evolving}. However, it has not yet been proven that a phase transition occurs for sparse networks or other classes of dense networks. In \fref{appsec:Linear_CVM_Simulations}, we review some of the existing computational results for linear CVMs on sparse ER networks and present computational results for these models on sparse networks that we construct from a stochastic block model (SBM). We demonstrate that, although these linear CVMs modify the structure of their associated network, their steady-state behavior appears to be insensitive to the examined initial network structures in a sense that we make precise in \fref{appsec:Linear_CVM_Simulations}. For over a decade, linear CVMs have been a challenging, popular, and fruitful topic to study. From a practical perspective, however, nonlinearity appears to be a critical ingredient for coevolving voter models. For example, Couzin et al.~\cite{couzin2011uninformed} successfully predicted the existence of novel collective behaviors of schooling fish using nonlinear adaptive-network models.

A key contribution of the present paper is the introduction of a nonlinear coevolving voter model. In our setting, individuals seek to achieve social harmony (i.e., having the same opinion state as all of their neighbors) by rewiring and adopting at rates that depend on the states of their neighbors in a nonlinear way. Each node is in one of two states, and neighboring nodes ``agree'' if they are in the same state and ``disagree'' if they are in different states. We refer to edges between agreeing neighbors as \emph{concordant} and edges between disagreeing neighbors as \emph{discordant}. When updating, a node that is not in a local consensus (i.e., it disagrees with at least one neighbor) performs one of two actions with respect to a disagreeing neighbor: (1) it adopts the opinion of the disagreeing neighbor, causing other neighbors who had been in agreement to now disagree; or (2) it abandons the edge that connects it to the disagreeing neighbor and possibly forms a new connection. Similar to the CVM in~\cite{Holme2006Nonequilibrium}, a node makes a random choice between options (1) and (2). However, unlike in their model, the probability to choose a given option is not homogeneous; instead, it depends nonlinearly on the states of the node's neighbors. In our model, nodes conduct a local survey of all of their neighbors. Those who agree with a large fraction of their neighbors (i.e., those who ``fit in'') are more likely to choose to remove the edge and possibly rewire, rather than adopting their disagreeing neighbor's opinion (which could place them in a local minority). Conversely, nodes that are in a local minority among their neighbors are more likely to adopt a neighbor's opinion (which could place them in a local majority), rather than remove the edge and possibly rewire.

A node's local survey provides it with a sample view of a population. The sample is susceptible to bias, as the nodes survey only their neighbors in a network. Under certain conditions, the local surveys can accurately estimate global statistics, such as an opinion's popularity, which is equal to the fraction of nodes that hold that opinion. However, it is possible to configure systems such that the sampling bias leads nodes to construe globally popular opinions as locally unpopular, and vice versa. Under certain conditions, we find in our nonlinear CVM that when a node's local surveys are so distorted that it perceives the minority opinion to instead be the majority opinion, almost every node eventually adopts the opinion that was initially unpopular. Consequently, such distorted sampling in local surveys, which we relate to recent work by Lerman et al.~\cite{lerman2016majority} on what they called the ``majority illusion'', has important implications for the dynamics of our nonlinear CVM.

\begin{figure}
  \includegraphics[width=\linewidth]{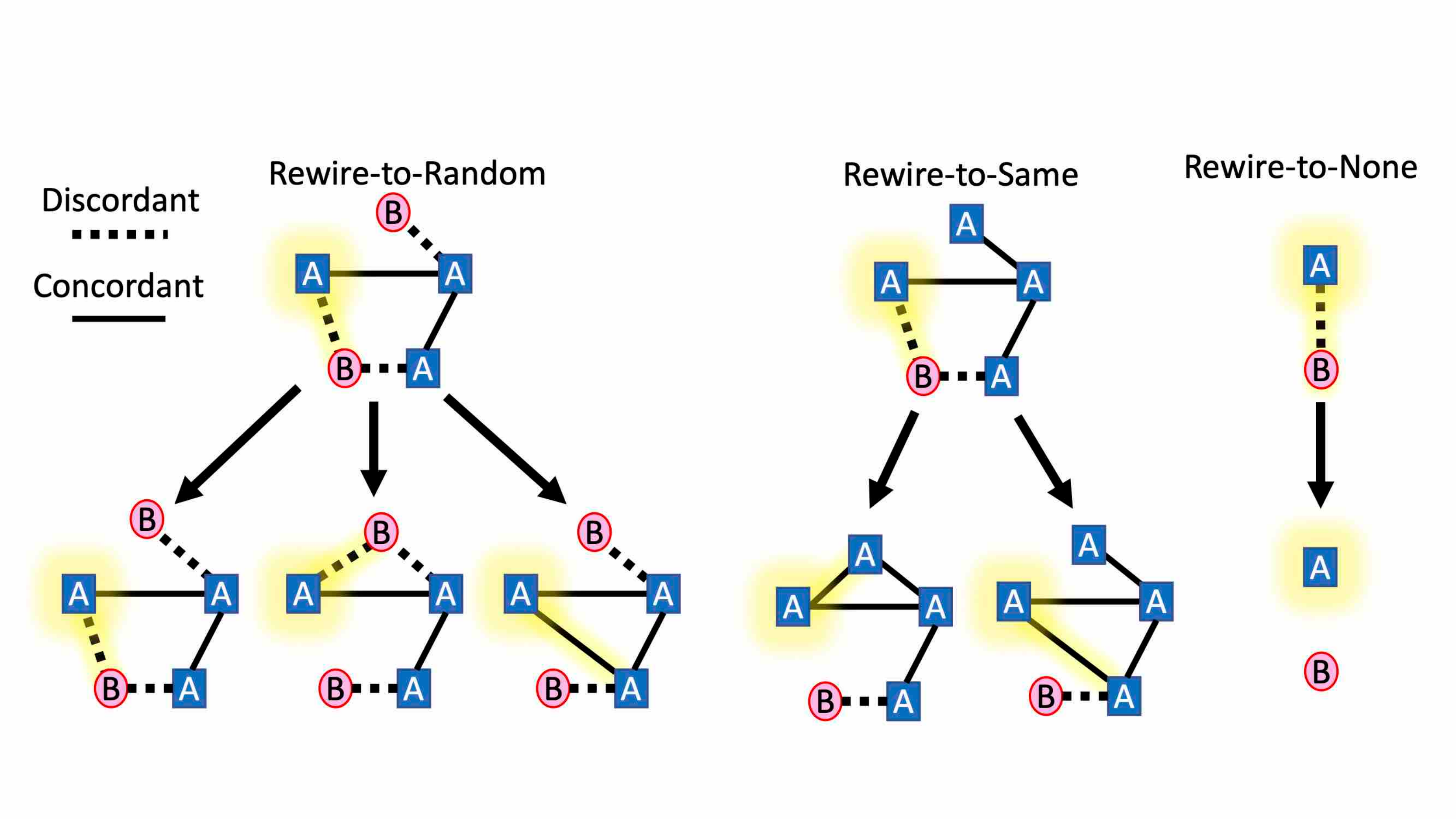}
  \caption{Illustration of the three rewiring schemes that we examine in our nonlinear CVM. 
  The highlighted edges in the top panel are the discordant edges that we will rewire, and the highlighted nodes are the focal nodes.  The bottom panel shows all possible rewiring outcomes for each scheme. (Left) Rewire-to-random scheme, in which the focal node either rewires to some node to which it is not adjacent or restores the original edge; see \fref{sec:RTR} for details. (Center) Rewire-to-same scheme, in which the focal node rewires only to nodes that share its opinion state; see \fref{sec:RTS} for details. (Right) Rewire-to-none scheme, in which we delete the discordant edge; see \fref{sec:RTN} for details.}\label{fig:rewiring_schemes}
\end{figure}

In our paper, we examine three different rewiring schemes, which we illustrate in \fref{fig:rewiring_schemes}. In \fref[plain]{sec:RTR}, we explore a ``rewire-to-random" (RTR) scheme, in which nodes sever connections (i.e., discordant edges) with disagreeing neighbors and replace them with new connections to nodes in a way that is agnostic to opinion states. In \fref[plain]{sec:RTS}, we explore a ``rewire-to-same" (RTS) scheme, in which nodes sever connections with disagreeing neighbors and replace them with new connections to nodes who share their opinion state. Finally, in \fref[plain]{sec:RTN}, we explore a ``rewire-to-none" (RTN) scheme, in which nodes sever connections with disagreeing neighbors without forming any new connections to other nodes. This third type of rewiring models behavior on social media in which individuals ``unfriend" (or ``unfollow'') someone after a disagreement~\cite{bode2016pruning, yang2017politics, zhu2017shield}. A fascinating feature of linear CVMs is that the choice of rewiring scheme has a dramatic impact on their steady-state properties~\cite{durrett2012graph}. However, for our nonlinear CVM, the choice between the above rewiring schemes does not seem to have major qualitative effects on their steady-state behavior. 

In \fref{sec:Conclusion}, we summarize our results and discuss possibilities for future work. We find that our nonlinear CVM has several features, such as a strong dependency on network structure, that distinguish it from previously-studied linear CVMs.
We give additional details, computations, and analysis in appendices. We provide details for our mean-field calculations in \fref{appsec:MFA_Sigma}. In \fref{appsec:RTS_RTN_Algos}, we give complete algorithms for our nonlinear CVM with rewire-to-same and rewire-to-none schemes. In \fref{appsec:Linear_CVM_Simulations}, we examine a linear CVM with the RTR, RTS, and RTN rewiring schemes to facilitate comparisons with our nonlinear CVM. We simulate our nonlinear CVM for a wider range of parameter values in \fref{appsec:RTR_big_q_appendix}.

%%%%%%%%%%%%%%%%%%%%%%%%
%%% Rewire-to-Random
%%%%%%%%%%%%%%%%%%%%%%%%

\section{Rewire-to-Random} \label{sec:RTR}

%%%%%%%%%%%%%%%%%%%%%%%% Rewire-to-Random Model Introduction

\subsection{Model}

We now formally introduce our nonlinear coevolving voter model (CVM), starting with the version with a ``rewire-to-random" scheme.  In all versions of our nonlinear CVM, we consider a space of binary opinion states $O=\{A,B\}$ and networks with finitely many nodes (i.e., vertices) and edges. The space of states for the system itself, which we will call ``system states'' to distinguish from the opinion states, is given by the space of triples $(V,E,S)$, where $V$ is the set of nodes (which represent individuals), $N := |V|$ denotes the number of nodes, $E$ is the set of edges (which represent undirected and unweighted ties between individuals), and $S: V\rightarrow O$ is a function that records the opinion state (or simply the ``state'') of each node. 

The systems that we study evolve between system states in a stochastic, memoryless fashion, so we can construe each such system as a first-order Markov chain. The set $V$ of nodes remains fixed as a system evolves, but the set $E$ of edges and the function $S$ can update. When we refer to the edge set or the state function at a certain time, we use the notation $E(t)$ and $S^t$ to explicitly indicate time-dependence, but otherwise we suppress the notation for time. When evaluating the function $S$ for a specific node, we use square brackets, so $S[i]$ indicates the state of node $i$ and $S^t[i]$ indicates the state of node $i$ at time $t$. We use $E_D$ (and $E_D(t)$ when specifying the time $t$ explicitly) to denote the set 
\begin{equation}\label{DiscordantEdgeSet}
    E_D = \{(u,v)\in E: S[u] \neq S[v]\}
\end{equation} 
of \emph{discordant edges}, which are edges between disagreeing nodes, in $E$ (and $E(t)$).
The closed neighborhood of node $i$ is 
\begin{equation}\label{NeighborSet}
    \Gamma(i,E)=\{j\in V: (i,j)\in E\}\cup \{i\}\,.
\end{equation} 
It consists of the set of nodes that are adjacent to $i$, as well as node $i$ itself. 

An important choice to make when defining the update rules for a voter model is the use of synchronous versus asynchronous updating. Synchronous updating entails updating all of the nodes in unison at each time step. By contrast, in asynchronous updating, one chooses a node (at random using a specified random process) at each step to interact with one of its neighbors, while the state of the rest of the system is constant. If the random process for choosing which node to update has a uniform distribution over the nodes, then after $N$ steps in an asynchronous voter model, one updates each node once on average in an $N$-node network. This corresponds to one time step in synchronous voter models~\cite{sood2008voter}. Therefore, when comparing asynchronous models that use a uniform random distribution to synchronous models, an asynchronous voter model evolves at a rate that is scaled by $1/N$ relative to synchronous models. The choice between synchronous and asynchronous updating is an important one, as it can have significant effects on the dynamics of voter models (beyond the time scaling), including differences in the number of absorbing states~\cite{gastner2015ising}. See~\cite{huberman1993evolutionary,roca2009evolutionary,fernandez2011update} for discussions of the effects of synchronous versus asynchronous updating in voter models and in evolutionary games. In our analysis of coevolving voter models, we find that it is simpler to define update rules in an asynchronous manner \footnote{We are not aware of any studies of coevolving voter models with synchronous updating. One challenge in defining such a model is establishing a protocol for the situation when two adjacent nodes select each other for updating and both simultaneously attempt to rewire the same edge.}. 

After choosing to do asynchronous updating, one then chooses between ``node-based'' versus ``edge-based'' updating. {In ``node-based'' voter models, in each step, one first selects a node $i$ (at random following some distribution over the set of nodes) and then chooses one of its neighbors $j$ (at random following some distribution over the set of neighbors of node $i$). If node $i$ is isolated, the system does not change in that step.} Another model choice is that ones needs to choose whether to update node $i$ or node $j$. Under ``direct node-based'' rules, one updates the state of node $i$ by copying the state of node $j$; under ``reverse node-based'' (which is also called ``invasion process'') rules, the roles are switched, so one updates the state of node $j$ by copying the state of node $i$~\cite{sood2008voter}. The two ``node-based'' rules contrasts with ``edge-based''  (i.e., ``link-based'') rules. Under ``edge-based'' rules, in each step, one first selects an edge $(i,j)$ at random following some distribution over the set of edges and then uniformly randomly selects one of the edge's incident nodes to update. That is, with equal probability, one updates node $i$ by copying the state of node $j$ or one updates node $j$ by copying the state of node $i$. The seemingly minor choice between ``direct node-based'', ``reverse node-based'', and ``edge-based'' rules can have very substantial effects on the dynamics of a voter model, including convergence time and steady-state behavior~\cite{nardini2008whos,sood2005voter,demirel2014moment,porter2016dynamical}. One can observe an immediate difference between the three rules based on how they bias the relationship between the degrees of nodes $i$ and $j$. In the ``node-based'' rules, we observe that expected degree of node $j$ is larger than the expected degree of node $i$. By contrast, in the ``edge-based'' rules, it follows from symmetry that nodes $i$ and $j$ have identical expected degrees.

In our CVMs, we use asynchronous ``edge-based'' updates, and we perform one update for each \emph{elementary time step} $t=\{1, 2, \ldots \}$. During an elementary time step, we select a discordant edge $(i,j)$ uniformly at random from $E_D$. Alternatively, one can think of choosing the edges from $E$ according to the probability mass function 
\begin{equation*}
	f_E((i,j)) = 
	\begin{cases}
		\frac{1}{|E_D|}\,, & (i,j) \in E_D\,, \\ 0\,, & (i,j) \notin E_D\,.
		\end{cases} 
\end{equation*}
The only effect on the dynamics by choosing directly from $E_D$ instead of $E$ is that we can skip steps in which nodes $i$ and $j$ already share the same state, as such steps do not affect the state of the system. This leads to a logarithmic speedup in the time to reach a steady state (compare this to the coupon-collector problem~\cite{feller1957probability}) and was called an ``efficient version'' of CVMs in~\cite{basak2015evolving}. We then select one of the two nodes (which we can take to be the one with the label $i$ without loss of generality) at random with equal probabilities to be the primary node; we take the other node ($j$) to be the secondary node. After selecting the primary node $i$, it takes a local survey of it neighbors. We measure the result of the local survey by calculating $\sigma_i := {s_i}/{k_i}$, where $s_i=|\{j: (i,j)\in E\,, S[i]=S[j]\}|$ and $k_i$ is the degree of node $i$. We also define $\bar{s_i} := k_i-s_i$, which counts the number of discordant edges that are incident to node $i$. Therefore, $\sigma_i$ is the fraction of neighbors of node $i$ that agree with $i$. For node $i$ to be selected for updating, it needs to have at least one disagreeing neighbor (i.e., at least one discordant incident edge), so $\sigma_i\in [0,1)$. Note that $\sigma_i$ is not defined for isolated nodes; however, because our CVM is edge-based, it is not possible to select an isolated node for updating.

Our nonlinear CVM has a parameter $q$ that is akin to a parameter in nonlinear $q$-voter models~\cite{castellano2009nonlinear}.  With probability $\sigma_i^q$, node $i$ performs a rewiring action, in which it deletes its edge to the chosen secondary node and then randomly forms a new edge to a node that is not currently one of its neighbors. One needs to choose a probability distribution for this random process. In the first version of our CVM, we suppose that the primary node picks a node from a network uniformly at random. (The original node $j$ from which $i$ just deleted an edge is available for selection.) This type of rewiring scheme, which yields an edge between node $i$ and any node (irrespective of its opinion state), is known as a ``rewire-to-random'' (RTR) strategy~\cite{durrett2012graph}. With complementary probability $1-\sigma_i^q$, node $i$ takes an adopting action, in which it adopts the opinion state of the chosen secondary node. We then repeat this process until there are no discordant edges in a network. If the system reaches a system state with no discordant edges at time $t^*$, then the dynamics reach a steady state, and we say that the system ``terminates'' at that system state, such that the system remains in that system state for all $t\geq t^*$. Both the rewiring and adoption actions conserve the number of edges, so $|E(t)|$ is constant in time. 
Recently, Min and San Miguel \cite{min2017fragmentation} introduced a nonlinear CVM that also incorporates such a parameter $q$. Under their ``direct node-based'' rules, once one selects a node $i$, it is with probability $(1-\sigma_i)^q$ that node $i$ performs any update at all, and then a separate parameter $p$ determines the relative probabilities of rewiring and adoption updates. This differs from our ``edge-based'' rules, in which once we select a node $i$ to update, $\sigma^q$ determines the relative probabilities of rewiring and adoption updates.

\begin{figure}
  \includegraphics[width=\linewidth]{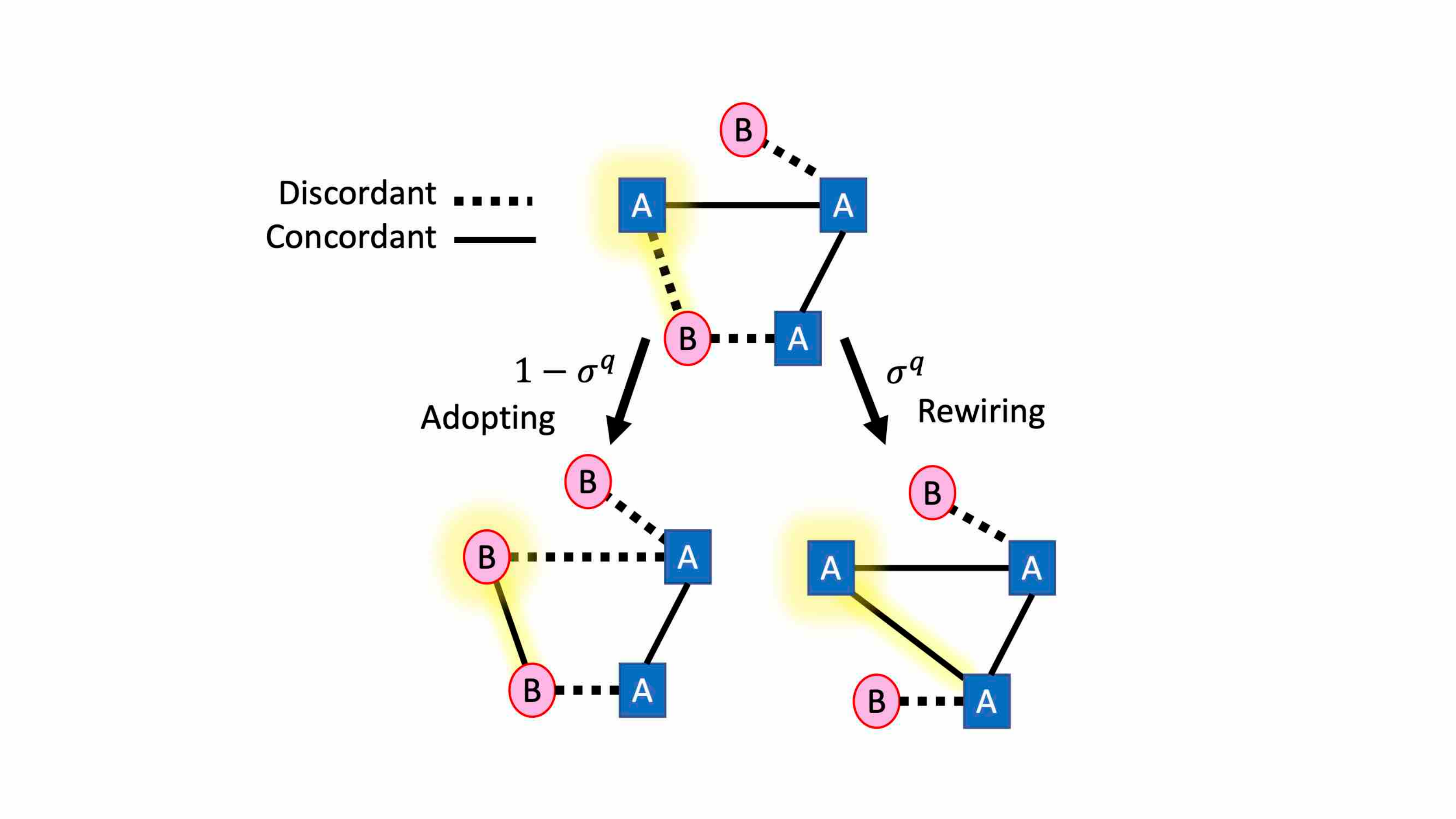}
  \caption{Schematic representation of an elementary time step in our nonlinear CVM. We highlight the selected discordant edge and the primary node. With probability $1-\sigma^q$, the primary node adopts the opinion state of its neighbor that is incident to this discordant edge. With complementary probability $\sigma^q$, the primary node performs a rewiring action. Under a rewire-to-random (RTR) scheme, there are three possible outcomes of rewiring; see \fref{fig:rewiring_schemes}. Each of the possible rewiring outcomes is equally probable. (For the rewire-to-same scheme, the depicted outcome is the only possible one, so it occurs with probability $1$ if there is a rewiring action.) The value of $\sigma$ in this example is $1/2$.
  }\label{fig:RTRschema}
\end{figure}

When $q$ is a positive integer, one can interpret our rewiring process above as the primary node randomly selecting a panel of $q$ of its neighbors (with repetition allowed). If any member of the panel is in a different state from the primary node, the latter undertakes an adopting action. Therefore, only when the panel and the primary node are all in the same state does the primary node perform a rewiring action. We summarize the rewiring process in Algorithm~\ref{alg:ftvm_rtr} and give a schematic representation of the process during one elementary time step in \fref{fig:RTRschema}. 

Because $\sigma_i\in [0,1)$ for nodes that can update, it follows that in the limit as $q\rightarrow\infty$, we recover a voter model with only adoption (and no rewiring). However, as $q\rightarrow 0^+$, we do \emph{not} recover a model with only rewiring, because for some nodes $i$, it can be the case that $\sigma_i = 0$ if all of node $i$'s incident edges are discordant; in that case, node $i$ performs an adoption action for all $q>0$. In the present paper, we also perform simulations with $q=0$; for these simulations, we take $0^0$ to be $1$ to recover a pure rewiring model.

The nonlinear CVM that we just described is an absorbing Markov chain. The absorbing system states are those in which a network has no discordant edges. Such a situation occurs when each connected component of the network is in a consensus, but it does not necessarily require all components to achieve a consensus with the same opinion state. There has been significant prior work on noisy voter models~\cite{granovsky1995noisy,carro2016noisy}, and some recent work has studied noisy CVMs with random opinion-state mutations~\cite{ji2013correlations}. In these systems, in addition to the rewiring and adoption updates, there is also a mechanism that alters the opinion states of nodes according to some random process. Incorporating such noise yields a Markov chain that no longer is absorbing, because the system can exit the systems states with component-wise consensus through random creation of discordant edges. 
{The resulting models are ergodic, so one can approximate the non-Markovian second-order (and higher-order) moment terms (see \fref{eq:momentdefs}) using Markovian terms \cite{chodrow2018local}.}

\begin{algorithm}[H]
\caption{Nonlinear Rewire-to-Random (RTR) Coevolving Voter Model}\label{alg:ftvm_rtr}
\begin{algorithmic}[1]
\Procedure{FittingInVM}{$V,E,S,q$}\Comment{{Input: Initial network and opinion states}} 
\State $E_D \gets Discordant(V,E,S);\; \; t \gets 0; \; \; Record(V,E,S,t)$ 
\While{$E_D\not=\emptyset$}\Comment{While there are disagreeing neighbors}
\State $(i,j) \gets RandomChoice(E_D)$
\State $PrimaryNode, SecondaryNode \gets RandomPermutation(i,j)$
\State $\sigma \gets LocalVote(PrimaryNode,V,E,S)$ 
\State $u \gets Uniform(0,1)$
\If{$u\leq \sigma^q$} \Comment{Rewire}
\State $E.remove(PrimaryNode,SecondaryNode)$
\State $NewNeighbor \gets RandomChoice(V\setminus \Gamma(PrimaryNode,E))$ \Comment{See Eq.~\eqref{NeighborSet}}
\State $E.add(PrimaryNode,NewNeighbor)$
\Else \Comment{Adopt}
\State $S[PrimaryNode] \gets S[SecondaryNode]$
\EndIf
\State $E_D \gets Discordant(V,E,S);\; \; t \gets t+1$;  \; $Record(V,E,S,t)$ 
\EndWhile
\EndProcedure
\end{algorithmic}
\end{algorithm}

A complete description of the system state is given by $(V,E,S)$. Even for small networks, this amount of information is difficult to study, so we seek a coarse-grained description of our CVM's dynamics~\cite{kuehn2016moment}. One type of summary is a count of ``state-specified motifs'',
which are subgraphs $H$ in which the nodes of $H$ are in specified states. Relevant counts of state-specified motifs include
\begin{enumerate}
    \item $N_X(t) = |\{i\in V: S^t[i]=X\}|$, the number of nodes $i$ at time $t$ that are in state $X\in O=\{A,B\}$\,;
        \item $N_{X,Y}(t) = |\{(i,j)\in V\times V: (i,j)\in E(t)\,, S^t[i]=X\,, S^t[j]=Y \}|$, the number of node pairs $(i,j)$ at time $t$ in which $i$ and $j$ are adjacent, node $i$ is in state $X \in O=\{A,B\}$, and node $j$ is in state $Y\in O$\,;
    \item $N_{X,Y,Z}(t) = |\{(i,j,k)\in V\times V\times V: (i,j), (j,k)\in E(t)\,, S^t[i]=X\,, S^t[j]=Y\,, S^t[k]=Z\}|$, the number of node triples $(i,j,k)$ at time $t$ in which $i$ and $j$ are adjacent, $j$ and $k$ are adjacent, node $i$ is in state $X\in O=\{A,B\}$, node $j$ is in state $Y\in O$,
    and node $k$ is in state $Z\in O$\,.
\end{enumerate}
 Note that 
\begin{align*}
 	N_A(t)+N_B(t) &= N\,, \\
	N_{AA}(t)+N_{AB}(t)+N_{BA}(t)+N_{BB}(t) &= 2|E| \,, \\
	N_{AB}(t) = N_{BA}(t) &= |E_D(t)|\,. 
\end{align*}

We refer to an instance of a triple among the $N_{XYZ}$ triples as an \emph{$XYZ$-triple}. One can also compute counts for state-specified motifs with more than three nodes, but we will not need them. Because the node set $V$ is fixed, it is convenient to examine \emph{state densities} $N_X(t)/N$, which give the fraction of nodes in a state $X$. {We are also interested in the expected values (i.e., ``moments'') of these quantities~\cite{kiss2017mathematics}. Three examples of moments are
\begin{equation}\label{eq:momentdefs}
\begin{split}
	[X] & = [X](t) = \langle N_X(t)\rangle  = \mathbb{E}[N_X(t)] \,, \\
	[XY] & = [XY](t) = \langle N_{X,Y}(t)\rangle  = \mathbb{E}[N_{X,Y}(t)]\,, \\
	[XYZ] & = [XYZ](t)  = \langle N_{X,Y,Z}(t)\rangle  =\mathbb{E}[N_{X,Y,Z}(t)] \,.
\end{split}
\end{equation}
}

We compare our rewire-to-random model (see Algorithm~\ref{alg:ftvm_rtr}) to the ``rewire-to-random'' (RTR) CVM that was described in~\cite{durrett2012graph}. Although there are many variants, we henceforth write ``the linear RTR-CVM'' when referring to this specific model. In the linear RTR-CVM, the system updates in a way that is similar to our nonlinear RTR-CVM, except that the probability for which action to take is given by a parameter $\alpha\in[0,1]$. With probability $\alpha$, a node performs a rewiring action; with probability $1-\alpha$, it adopts a neighbor's opinion. Therefore, in the linear RTR-CVM, until $E_D = \emptyset$, the count $N_A(t)$ increases by $1$ with probability $\dfrac{1-\alpha}{2}$, decreases by $1$ with probability $\dfrac{1-\alpha}{2}$, and does not change with probability $\alpha$ during each elementary time step. If we take $\alpha=0$ (entailing that the network topology never changes) and begin with a finite, connected network $(V,E)$, then $N_A(t)$ behaves equivalently to a simple, symmetric, one-dimensional (1D) random walk \footnote{In a 1D random walk, ``simple'' refers to the rule that, at each time step, the walker must move by either $+1$ or $-1$; and ``symmetric'' refers to these outcomes being equally probable. In a symmetric random walk with possible moves $+1$, $-1$, and $0$, one requires only that the first two possibilities are equiprobable to each other.} (i.e., the two possibilities both have a probability of $1/2$)~\cite{masuda2017} on the integers with boundary $\{0,N\}$. This implies that for any finite, connected network, the probability at time $t$ that opinion state $A$ eventually becomes the consensus opinion state is $\frac{N_A(t)}{N}$. {For $\alpha>0$, the quantity $N_A(t)$ behaves like a symmetric, 1D random walk with step sizes $+1$ and $-1$ that each occur with equal probability $\dfrac{1-\alpha}{2}$ and a no-move (``null'') step that occurs with probability $\alpha$.}

In the linear RTR-CVM, it is equivalent to take the view that one is choosing the type of action (rewire or adoption) \emph{before} choosing which node of the selected edge is the primary one. This makes it clearer that even when the number of nodes in state $A$ is not equal to the number of nodes in state $B$ (i.e., when $N_A(t) \neq N_B(t)$), a rewiring action causes the number $N_{AB}(t)$ of discordant edges to decrease by $1/2$ in expectation, regardless of the system state. By contrast, the effect of an adoption on the number $N_{AB}(t)$ of discordant edges does depend on the system state, and it is possible for an adoption to increase $N_{AB}(t)$ in expectation. As an extreme case, consider a star network $S_k$ with a hub node in state $A$ and $k$ leaf nodes, and suppose that one node is in state $B$ but all others are in state $A$. An updating step is guaranteed to select the network's single discordant edge; if an adoption occurs, $N_{AB}(t)$ increases by $\frac{k-3}{2}$ in expectation. However, for a system on a regular network that satisfies the conditions
\begin{equation}
    \begin{aligned}\label{eq:wellmixed}
    \sigma_i &\approx N_A(t)/N \,\text{ for all nodes }\, i\, \text{ in state }\, A\,, \\
    \sigma_j &\approx N_B(t)/N \, \text{ for all nodes } \,j\, \text{ in state }\, B\,,
\end{aligned} 
\end{equation}
an adoption causes $N_{AB}(t)$ to decreases by $1$ in expectation. We use the term \emph{locally well-mixed} for a system on any network that satisfies the conditions in \fref{eq:wellmixed}, which entails that there are no correlations between the opinion states of nodes and the network topology. In a recent paper, Lee et al.~\cite{Lee2019Homophily} defined a related quantity called \emph{social perception bias} that measures the ratio of a node's perception of the fraction of the minority to the true fraction of the minority, where a value of exactly $1$ implies perfect perception of the frequency of the minority state in a network. 
Using this terminology, one can alternatively characterize a system as locally well-mixed using the condition that all nodes have a social perception bias that is close to $1$. One can give a mathematically precise definition in the limit that the number $N$ of nodes becomes infinite. Specifically, a system is \emph{locally well-mixed} if almost all nodes have a social perception bias of $1-o(1)$ as $N\rightarrow\infty$.

In our nonlinear CVM, there is no longer a symmetry between the two nodes that are incident to the same discordant edge $(i,j)$. The local survey $\sigma_i$ of node $i$ and the local survey $\sigma_j$ of node $j$ can differ, so the probabilities for which action (rewiring or adoption) occurs depend on which node is the primary one. That is, during an elementary time step that involves node $i$ in state $A$ and node $j$ in state $B$ (i.e., after one selects the discordant edge $(i,j)$, but before selecting which node is the primary one), $N_A(t)$ either (1) increases by $1$ with probability $\dfrac{(1-\sigma_j^q)}{2}$, corresponding to node $j$ adopting node $i$'s state; (2) decreases by $1$ with probability $\dfrac{(1-\sigma_i^q)}{2}$, corresponding to node $i$ adopting node $j$'s state; or (3) remains the same with probability $\dfrac{\sigma_i^q+\sigma_j^q}{2}$, corresponding to either node $i$ rewiring or node $j$ rewiring. Therefore, although we can still view $N_A(t)$ as a 1D random walk, it is no longer symmetric, because the step probabilities can differ from each other.

The effect of this asymmetry on $N_{AB}(t)$ is more subtle. Consider a locally well-mixed system on a connected network. We also assume that one of the states, which we take to be $B$ without loss of generality, is the majority (so $N_B(t)>N_A(t)$). During an elementary time step, suppose that we select the discordant edge $(i,j)$ with node $i$ in state $A$ and node $j$ in state $B$. The local surveys then satisfy $\sigma_i<1/2<\sigma_j$. When $q=1$, this implies that if $i$ is the primary node, it is more likely to adopt than to rewire. If node $i$ adopts state $B$, then $N_{AB}(t)$ decreases by more than $1$ in expectation, because more of $i$'s neighbors are in state $B$ than in state $A$. If node $j$ is the primary node, it is more likely to rewire than to adopt. If node $j$ rewires to a node that we choose uniformly at random, then $N_{AB}(t)$ decreases by $N_B(t)/N > 1/2$ in expectation. Overall, we observe that the number $N_{AB}(t)$ of discordant edges decreases more rapidly in our nonlinear CVM than it does in the linear RTR-CVM under locally well-mixed conditions. Our nonlinear CVM has rather different dynamics when it is locally well-mixed than when it is not locally well-mixed. In \fref{sec:RTR_SBM}, we explore how to construct systems with correlations between nodes' opinion states and network topology, and we investigate how their dynamics differ from the situation in which the system is locally well-mixed. 

%%%%%%%%%%%%%%%%%%%%%%%% Simulations on ER Networks

\subsection{Simulations on \texorpdfstring{Erd\H{o}s--R\'enyi}{Erdos--Renyi} Networks} \label{sec:RTR_ER}

\begin{figure}[!tbp]
  \centering
    \subfloat[Minority State]{\includegraphics[width=0.48\linewidth]{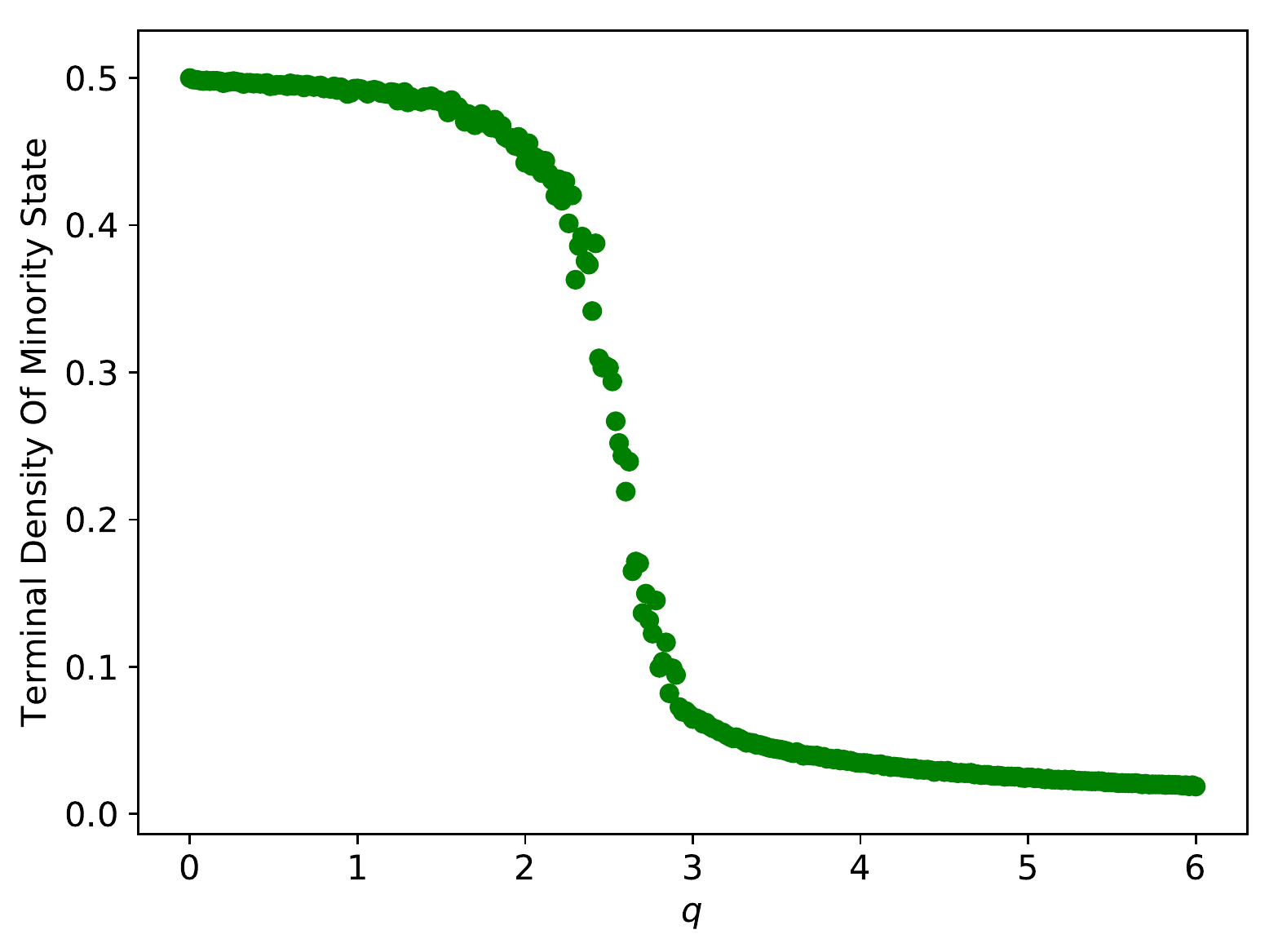}\label{fig:RTR_ER_Terminal_Minority}}
    \hfill
    \subfloat[State $A$]{\includegraphics[width=0.48\linewidth]{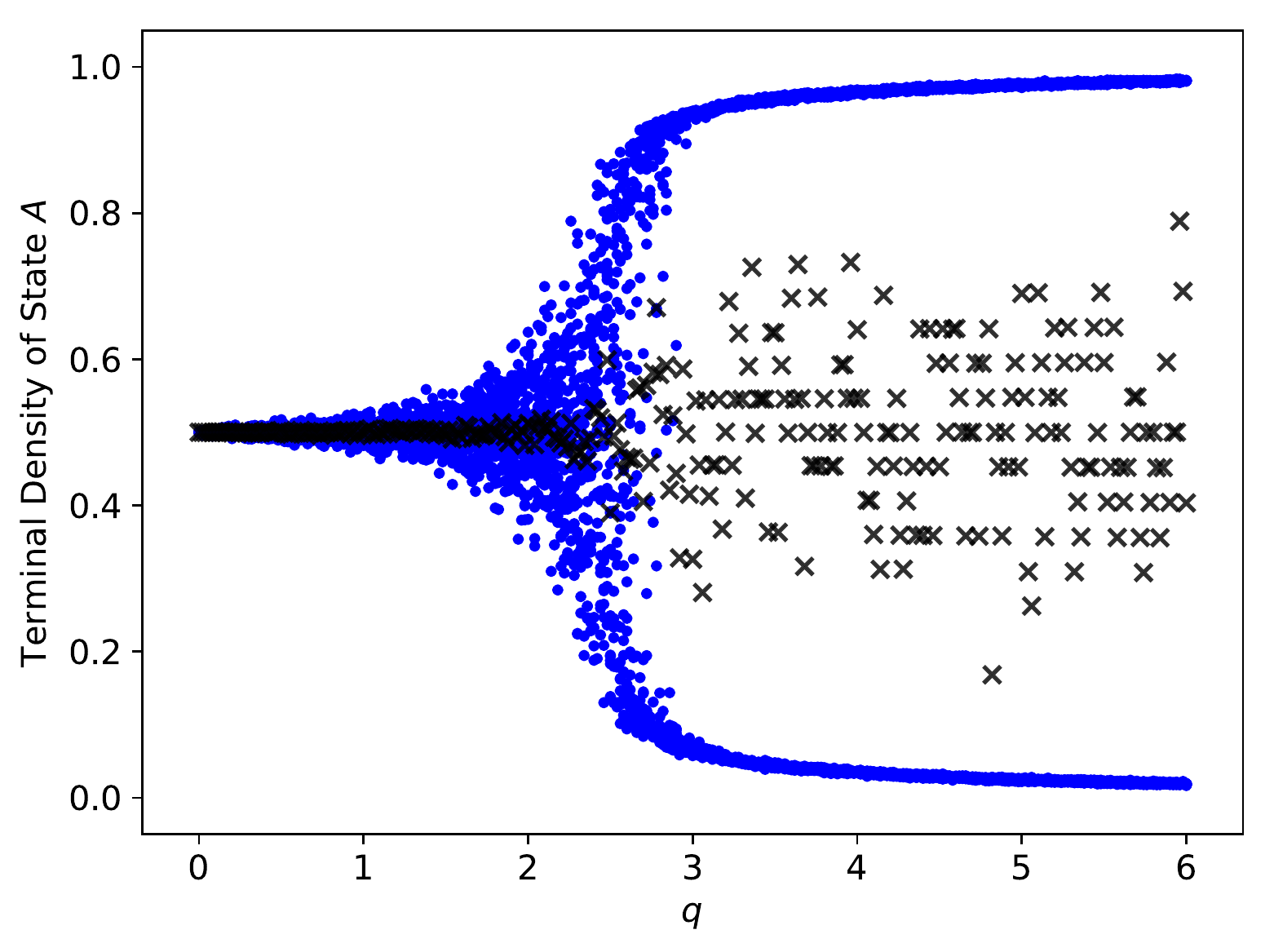}\label{fig:RTR_ER_Terminal_A}}
  \caption{Terminal density for (a) the minority state and (b) state $A$ in rewire-to-random (RTR) simulations (see Algorithm~\ref{alg:ftvm_rtr}) of our nonlinear CVM for $q\in[0,6]$ with discretization $\Delta q=0.02$. For each value of $q$, we simulate $20$ realizations. We seed each realization with a different ER network with $N=50,000$ nodes and edge probability $p=\frac{4}{N-1}$; we initialize half of the nodes in state $A$ and the other half in state $B$. In panel (a), each green point is the mean of $\min\{N_A(t)/N,N_B(t)/N\}$ at the termination of the simulation over the $20$ realizations for each value of $q$. In panel (b), each small blue dot is $N_A(t)/N$ at the termination of the simulation. Each cross ($\times$) in (b) is the mean of $N_A(t)/N$ at the termination of the simulation over the $20$ realizations for each value of $q$.}\label{fig:RTR_ER_Terminal}
\end{figure}

We begin exploring our nonlinear RTR-CVM by simulating it on Erd\H{o}s--R\'enyi (ER) $G(N,p)$ networks. We seed each realization with an ER network with $N=50,000$ nodes, half of which begin in state $A$ and the other half of which begin in state $B$. The edge probability $p$ is $\frac{4}{N-1}$, so the mean degree is $\langle k \rangle \approx 4$. Here and throughout our paper, when we simulate a voter model, we let the system evolve until it terminates in an absorbing steady state (i.e., until there are no discordant edges). This yields a ``terminal'' state. We focus on studying properties of terminal states, so we are interested in the $t\rightarrow\infty$ regime. For example, we examine the terminal state densities $N_A(t)/N$ and $N_B(t)/N$ and the terminal minority-state density $\min\{N_A(t)/N,N_B(t)/N\}$.

{In \fref[plain]{fig:RTR_ER_Terminal_Minority}, we plot the terminal minority-state density from simulations for a range of values of $q$ with discretization $\Delta q=0.02$}. When $q<1$, we observe that the minority-state density is approximately $0.5$. This implies that the network is \emph{fragmented}, in the sense that it is separated into multiple components that are disconnected from one another, such that all nodes of a component are in the same opinion state. This behavior is similar to the fragmentation observed in~\cite{durrett2012graph} for the linear RTR-CVM with sufficiently large rewiring rates. We interpret such fragmentation as individuals segregating into isolated communities, such that no pair of disagreeing individuals are neighbors of each other. This typically occurs when rewiring actions dominate the system, such that the state densities $N_X(t)/N$ do not change significantly. As we consider progressively larger values of $q$ between $1$ and $3.5$, we observe a smooth transition of an inverted `S' shape in the terminal minority-state density from approximately $0.5$ to approximately $0.03$. The small terminal minority-state densities for larger values of $q$ suggest that adoption actions dominate the system, leading to a large change in the state densities. For expository convenience, we will say that ``almost every'' node ends up in state $X$ if 90\,\% or more of the nodes are in state $X$ when a system terminates by reaching an absorbing state. Initially, $N_A(0)/N=N_B(0)/N=1/2$. However, by the end of a simulation, one of the opinion states dominates, such that $\max\{N_A(t)/N,N_B(t)/N\}\approx 1$. The other nearly vanishes, so $\min\{N_A(t)/N,N_B(t)/N\}\approx 0$, as one can see in \fref[plain]{fig:RTR_ER_Terminal_Minority}.

In \fref{fig:RTR_ER_Terminal_A}, we plot the terminal density of state $A$ from simulations for a range of values of $q$. In this plot, each of the small blue dots is the terminal value of $N_A(t)/N$ for one of the $20$ realizations that we simulate for each value of $q$. The crosses ($\times$) give the mean of the terminal quantity of $N_A(t)/N$ for each value of $q$. In the ``branching'' of the data points in this figure, we observe what appears to be a transition between a fragmentation regime (with no significant changes to state densities) and a regime with competition between the adoption and rewiring mechanisms. Because our initial state densities are $N_A(0)/N=N_B(0)/N=1/2$ in these simulations, it is equally likely for almost every node to terminate in state $A$ as it is for almost every node to terminate in state $B$. We confirm this result in the means of terminal state-$A$ densities that we plot in \fref{fig:RTR_ER_Terminal_A}.

With respect to coarse qualitative behavior, both our nonlinear RTR-CVM and the linear RTR-CVM of~\cite{durrett2012graph} have a regime --- when $q<1$ for our model, and for $1-\alpha<0.2$ for the linear RTR-CVM --- in which rewiring dominates the system, as indicated by terminal minority-state densities that are close to the starting densities of $0.5$. However, outside this regime, the two models differ significantly, as one can see by comparing \fref{fig:RTR_ER_Terminal_A} with \fref{fig:LIN_RTR_ER_Term_A}. The linear RTR-CVM appears to have a continuous (but non-smooth) phase transition from the rewiring-dominated regime to a regime in which rewiring and adoption are competitive~\cite{durrett2012graph,Silk2014Exploring,chodrow2018local}. In our nonlinear RTR-CVM, there seems to be a smooth transition between a regime in which rewiring dominates and a regime in which adoption dominates.

%%%%%%%%%%%%%%%%%%%%%%%% Approximations

\subsection{Approximations}\label{sec:approx_rtr}

Although mean-field approximations have been unable thus far to produce precise quantitative results in previous work on CVMs, they have been useful for exploring some of their qualitative behavior~\cite{Kimura2008Coevolutionary, pugliese2009heterogeneous, bohme2011analytical, gleeson2012accuracy, nardini2008whos}. Writing a mean-field approximation of our model will require developing a mean-field analog of $\sigma_i$. We consider a state-heterogeneous mean, so we separately average the local surveys of nodes in state $A$ and average
the local surveys of nodes in state $B$. First, we consider the {unweighted mean}
\begin{equation}\label{eq:arithemticmeansigma}
    \overline{\sigma_A} = \frac{\sum_{i:S[i]=A,\, k_i \neq 0} \sigma_i}{ \sum_{i:S[i]=A,\, k_i \neq 0} 1}\,.
\end{equation}
The {unweighted mean} $\overline{\sigma_A}$ is useful, because it has a simple interpretation and it is well-approximated \footnote{This approximation is exact for non-empty, degree-regular networks with at least one node in state $A$.} by $\frac{N_{AA}(t)}{N_{AA}(t)+N_{AB}(t)}$. We will use the {unweighted mean} $\overline{\sigma_A}$ in \fref{sec:Illusion}, but it is not well-suited as a mean-field approximation of our nonlinear RTR-CVM. In edge-based models such as ours and the linear RTR-CVM of~\cite{durrett2012graph}, nodes are not equally likely to be selected for an update. Instead, one is more likely to select nodes that have several neighbors of the opposing state, as they contribute more edges to the set $E_D$ of discordant edges (from which we sample uniformly). Therefore, a suitable mean-field analog of $\sigma$ should weight nodes based on their probability of being selected. The probability that we select a particular node $i$ to be the primary node in an update is 
\begin{equation}
    \frac{1}{2}\frac{k_i-s_i}{N_{AB}(t)}=\frac{1}{2}\frac{\bar{s_i}}{N_{AB}(t)}\,,
\end{equation}
where $\bar{s_i}$ is the number of discordant edges that are incident to node $i$ and the $1/2$ accounts for the random choice between primary and secondary node. Therefore, our mean-field analogs of $\sigma$, which we denote by $\sigma_X$ for $X \in \{A,B\}$, are
\begin{align}\label{eq:sigma_equation_1}
	    \sigma_A &= \sum_{\{i:S[i]=A\,,\, k_i \neq 0\}} \sigma_i\frac{1}{2} \frac{k_i-s_i}{N_{AB}(t)} \\ 
	    &= \frac{1}{2N_{AB}(t)}\left( N_{AA}(t)- \sum_{\{i:S[i]=A\,,\, k_i \neq 0\}} \frac{s_i^2}{k_i}\right)\,, \\
	\sigma_B &= \frac{1}{2N_{AB}(t)}\left( N_{BB}(t)- \sum_{\{i:S[i]=B\,,\, k_i \neq 0\}} \frac{s_i^2}{k_i}\right)\,.
\end{align}

The equation for the first moment, which we express in terms of state $A$, is
\begin{equation}\label{eq:moment_equation}
    \frac{d [A]}{dt} = ([AB]+[BA])(\sigma_A^q -\sigma_B^q)\,.
\end{equation}
Equation \eqref{eq:moment_equation} arises from taking the difference of the ``incoming rate''  (i.e., nodes that change their state \emph{to} $A$) minus the ``outgoing rate'' (i.e., nodes that change their state \emph{from} $A$) to determine the net rate of change of nodes in state $A$. Nodes in state $B$ that are adjacent to a node in state $A$ adopt state $A$ at rate $1-\sigma_B^q$. Nodes in state $A$ that are adjacent to a node in state $B$ adopt state $B$ at rate $1-\sigma_A^q$. Summing these rates over all nodes gives $([AB]+[BA])(1-\sigma_A^q)$. Similarly, the rate at which nodes in state $B$ adopt state $A$ is $([AB]+[BA])(1-\sigma_B^q)$.

\Fref[plain]{eq:moment_equation} indicates that the local surveys, which we capture in our mean-field approximation by $\sigma_A$ and $\sigma_B$, have a global effect on the drift of opinion states, as they control the sign of $\frac{d [A]}{dt}$. (By contrast, $\frac{d [A]}{dt} = 0$ in the linear RTR-CVM of~\cite{durrett2012graph}.) This suggests that network structure plays a more prominent role in how $[A]$ and $[B]$ evolve in our nonlinear RTR-CVM than in the linear RTR-CVM. For example, consider a network with two communities, where one community is densely connected and consists of $C_A$ nodes in state $A$ and the other community is sparsely connected and consists of $C_B$ nodes in state $B$. We suppose that the second community is larger than the first (i.e., $C_B>C_A$). We also suppose that the two communities are linked to each other (in a way that we will make more precise in \fref{sec:RTR_SBM}). When we select a discordant edge $(i,j)$ with node $i$ in state $A$ and node $j$ in state $B$, the local surveys satisfy $\sigma_i > \sigma_j$. On average, at least initially, we expect that more nodes in state $B$ convert to state $A$ than the reverse. However, the values of $\sigma_A$ and $\sigma_B$ can change rapidly in non-obvious ways as the system evolves, potentially reversing the sign of $\frac{d [A]}{dt}$. Therefore, it is not guaranteed that such a two-community network will terminate in a state with a large fraction of nodes in state $A$. In fact, as we will see in \fref{sec:RTR_SBM}, whether this occurs depends on the nonlinearity parameter $q$. 

%%%%%%%%%%%%%%%%%%%%%%%% Simulations on Stochastic Block Models

\subsection{Simulations on Stochastic Block Models} \label{sec:RTR_SBM}

To explore how community structure impacts the dynamics of our nonlinear RTR-CVM, we simulate it on a network that we seed with communities using a stochastic block model (SBM)~\cite{fortunato2016community,newman2018book}. We assign the $N$ nodes into two communities, which we call community $a$ and community $b$. Community $a$ consists of $cN$ nodes and community $b$ consists of $(1-c)N$ nodes, with $c<1/2$. That is, community $b$ has more nodes than community $a$. In our discussion, we often refer to an SBM with two seeded communities as a ``two-community structure''. We seed all nodes in community $a$ with state $A$ and all nodes in community $b$ with state $B$. 

%%%%

\subsubsection{Two-Community SBM} \label{sec:RTR_SBM_two_communities}

\begin{figure}[ht]
  \centering
    \subfloat[Two-Community Structure]{\includegraphics[width=0.48\linewidth]{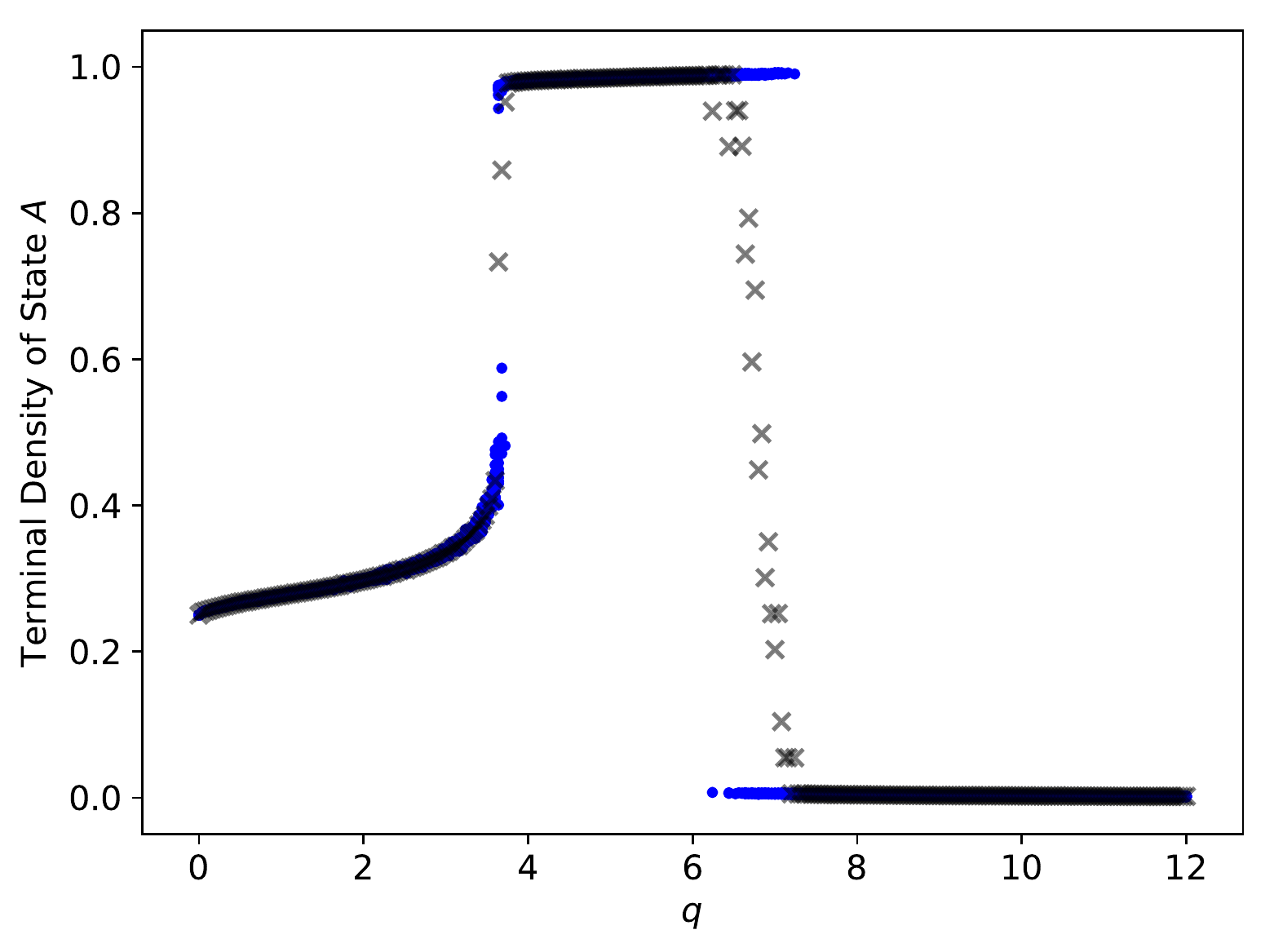}\label{fig:RTR_SBM_TWO_COM_TERMINAL_A}}
    \hfill
    \subfloat[Core--Periphery Structure]{\includegraphics[width=0.48\linewidth]{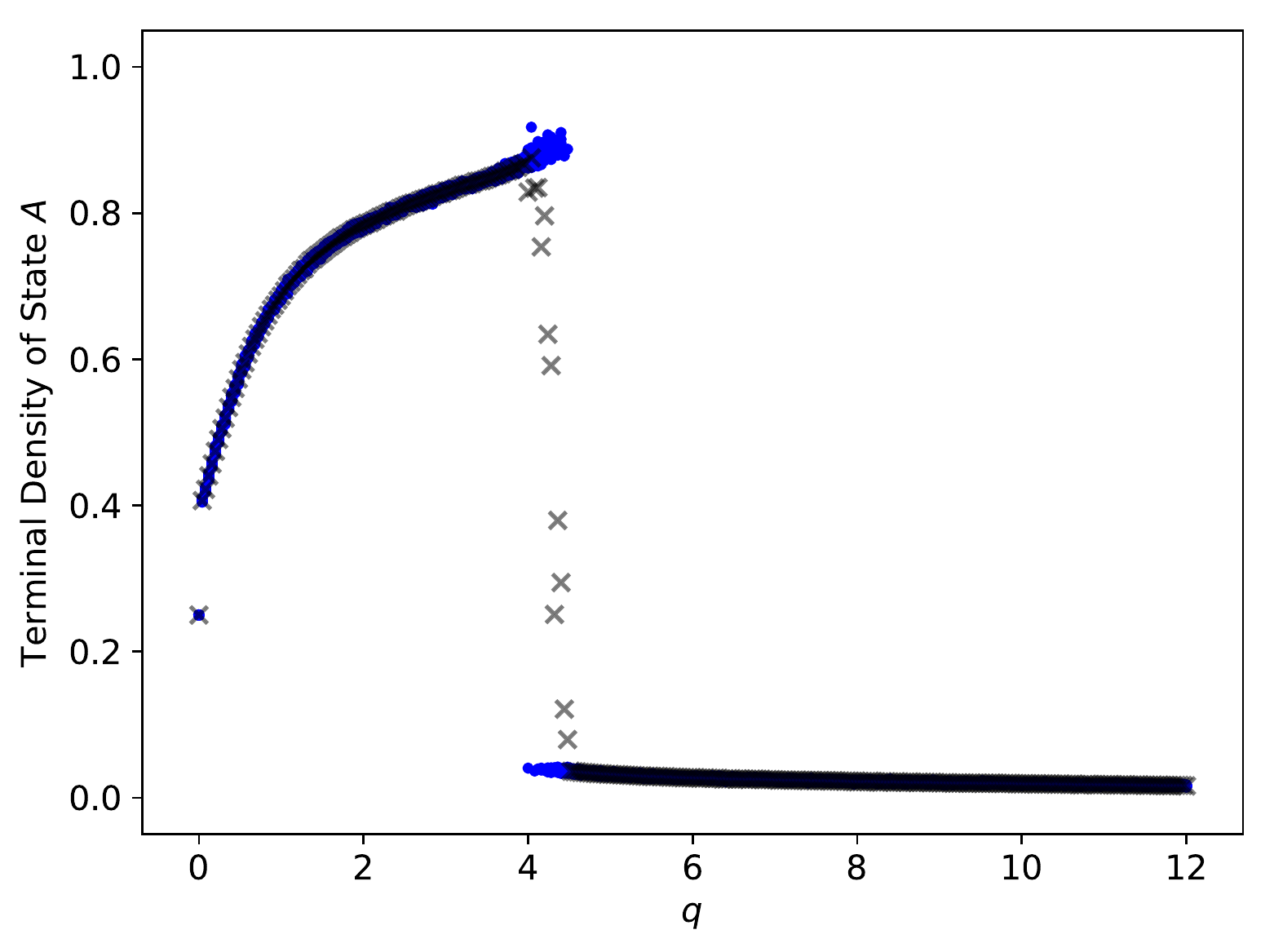}\label{fig:RTR_SBM_Terminal_A_CP}}
  \caption{Terminal density of state $A$ from rewire-to-random (RTR) simulations (see Algorithm~\ref{alg:ftvm_rtr}) for our nonlinear CVM with $q\in[0,12]$ and discretization $\Delta q=0.04$. For each value of $q$, we simulate $20$ realizations. In (a), we seed each realization with two-community SBM networks, as described in \Fref{sec:RTR_SBM_two_communities}. In (b), we seed each realization with an SBM network with core--periphery structure, as described in \Fref{sec:RTR_SBM_CP}. We plot individual realizations with blue dots and means with $\times$ symbols.}\label{fig:RTR_SBM_final_A}
\end{figure}

To create a two-community network in which the smaller community (i.e., community $a$) is denser than than the the larger community (i.e., community $b$), the edge-probability matrix 
\begin{equation}\label{eq:SBM_edge_prob_matrix}
	P = \begin{pmatrix} P_{aa} & P_{ab}\\ P_{ba} & P_{bb} \end{pmatrix}
\end{equation}
has probabilities that satisfy $P_{aa}>P_{bb}>P_{ab}=P_{ba}$. We initialize our simulations with networks with $c=1/4$, and we set the SBM parameters to be $P_{aa}=\frac{12}{cN-1}$, $P_{bb}=\frac{4}{(1-c)N-1}$ (so that $N_{AA}(0)\approx N_{BB}(0)$), and $P_{ab}={1}/{N}$. In our simulations, we check that the expectations of $\sigma_A$ and $\sigma_B$ satisfy the inequality $\mathbb{E}[\sigma_A(0)]  > \mathbb{E}[\sigma_B(0)]$. See \fref[plain]{appsec:MFA_Sigma} for details. Accordingly, we expect that, at least initially, the density $N_A(t)/N$ of state $A$ increases as the system evolves. In \fref[plain]{fig:RTR_SBM_Steps_Vs_A}, we plot $N_A(t)/N$ during the first $10,000$ elementary time steps of simulations for nine different values of $q$. In this plot, we show the initial (and transient) dynamics, rather than the full temporal evolution of our simulation to termination. The plot confirms the initial increase in $N_A(t)/N$. However, in \fref{fig:RTR_SBM_TWO_COM_TERMINAL_A}, we observe that, despite this initial increase, the terminal value of $N_A(t)/N$ depends on the nonlinearity parameter $q$.

\begin{figure}
  \includegraphics[width=.8\linewidth]{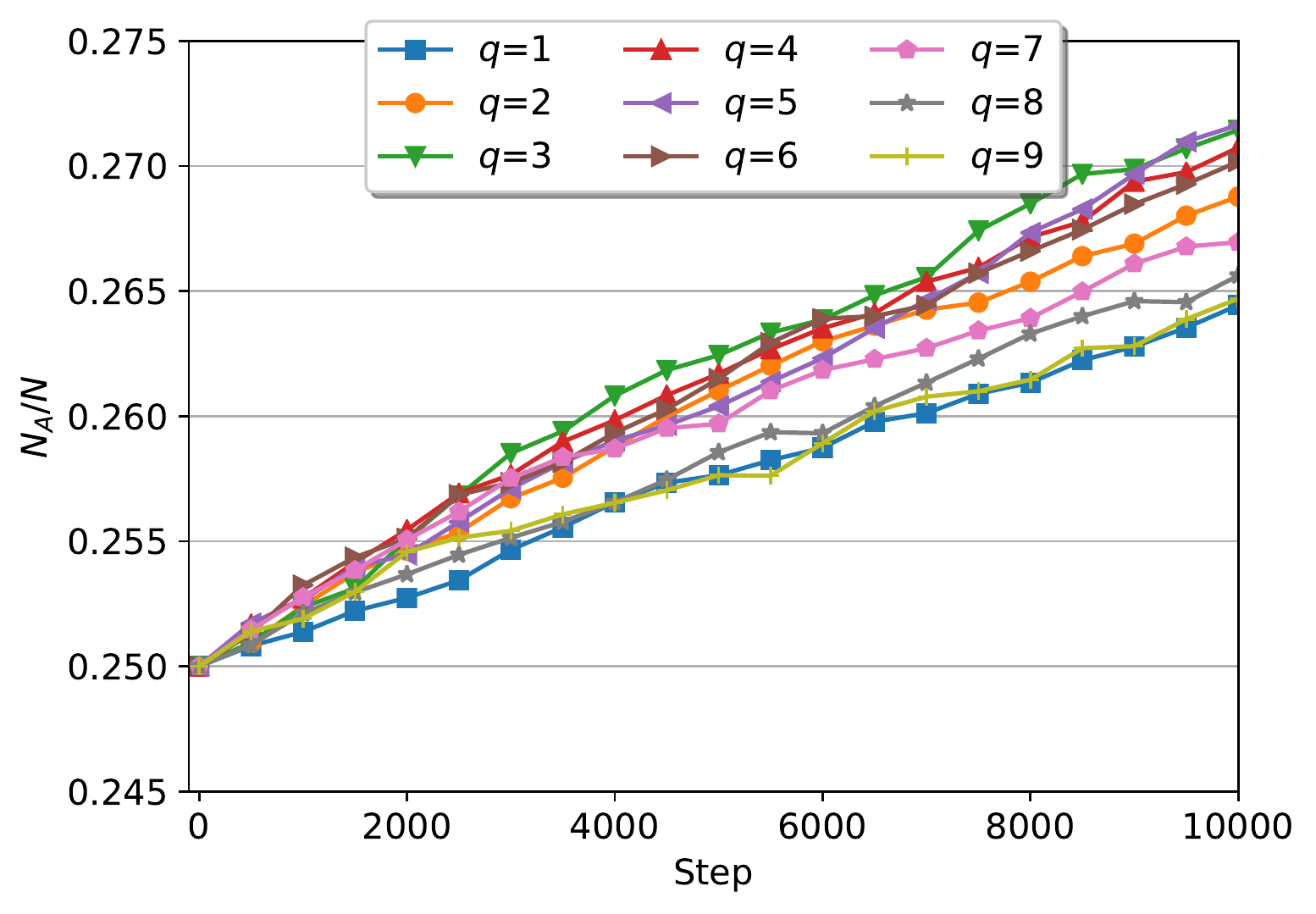}
    \caption{Density $N_A(t)/N$ of state $A$ for the first $10,000$ elementary time steps for nine values of $q$ in rewire-to-random (RTR) simulations (see Algorithm~\ref{alg:ftvm_rtr}) of our nonlinear CVM. We seed each of the nine realizations with two-community SBM networks, as described in \fref{sec:RTR_SBM_two_communities}. We show only initial and transient dynamics. For the terminal behavior of the simulations, see \fref{fig:RTR_SBM_TWO_COM_TERMINAL_A}.
    }\label{fig:RTR_SBM_Steps_Vs_A}
\end{figure}

In \fref[plain]{fig:RTR_SBM_TWO_COM_TERMINAL_A}, we plot the terminal state-$A$ density from $20$ simulations for each $q$.
For $q \in [0,3.8)$, between approximately 25\,\% and 60\,\% of the nodes terminate in state $A$, implying that the network fragments into multiple components. We observe what appears to be a hybrid phase transition~\cite{vazquez2008generic,lee2016hybrid} at $q^c \approx 3.8$, where there is a higher-order transition (i.e., at least second order) for $q\leq  q^c$ and a first-order transition $q> q^c$ because the terminal density is nearly constant (wiyh $N_A/N \approx 1$) for $q \in (q^c,6.2)$. This implies that almost every node terminates in state $A$ in all of our realizations, suggesting that rewiring dynamics are not fragmenting the system into multiple components.  For $q \in (6.2,7)$, either almost every node terminates in state $A$ or almost every node terminates in state $B$, and the latter occurs in progressively more realizations as we increase $q$. Although we do not show this in \fref[plain]{fig:RTR_SBM_TWO_COM_TERMINAL_A}, the system has an additional regime for sufficiently large $q$. In this regime, the adoption action dominates and the system behaves like a voter model that does not coevolve with network structure (see \fref{appsec:RTR_big_q_appendix}). In this situation, 
state $A$ becomes the consensus opinion with probability $N_A(0)/N=c=1/4$.

In \fref{fig:LIN_RTR_SBM_TWO_COM_Term_A}, we repeat our experiment using the linear RTR-CVM seeded with a two-community SBM network with the same parameter values. We observe that \fref[plain]{fig:LIN_RTR_SBM_TWO_COM_Term_A} resembles the outcome of initializing the linear RTR-CVM with an ER network (see \fref{fig:LIN_RTR_ER_Term_A_1-3}). This suggests that, with respect to terminal state densities, the linear RTR-CVM may be less sensitive than our nonlinear RTR-CVM to initial community structure in a network.

As we show in \fref{appsec:MFA_Sigma}, we are able to numerically approximate the quantity $\mathbb{E}[\sigma_A(0)^q-\sigma_B(0)^q]$ for two-community SBM networks. We find that it depends on the parameters $q$, $c$, $P_{aa}$, $P_{ab}$, and $P_{ba}$. As the system evolves, however, it becomes more challenging to track $\mathbb{E}[\sigma_A(t)^q-\sigma_B(t)^q]$ over time $t$. Nevertheless, from \fref{fig:RTR_SBM_TWO_COM_TERMINAL_A}, we know that the temporal evolution is affected by the value of the nonlinearity parameter $q$.

%%%%%%

\subsubsection{Core--Periphery Structure} \label{sec:RTR_SBM_CP}

We now investigate the dynamics of our nonlinear RTR-CVM with $\sigma_A(0) \gg \sigma_B(0)$ and $N_A(0)\ll N_B(0)$ using a core--periphery configuration of the SBM~\cite{csermely2013structure,rombach2017core}. Instead of having two communities, our initial network has a small, densely-connected core of nodes in state $A$ and a large periphery of sparsely-connected nodes in state $B$. We label the core as block $a$ and the periphery as block $b$.  In this core--periphery network, the SBM probabilities in \fref{eq:SBM_edge_prob_matrix} satisfy $P_{aa}>P_{ab}>P_{bb}$. In this scenario, a node $i$'s local survey $\sigma_i$ differs from the true global densities. For core nodes, $\sigma_i>1/2$, so such nodes believe that their state (namely, state $A$) is the majority state, even though it is not. Conversely, for the peripheral nodes, $\sigma_j<1/2$, so such nodes believe that their state (i.e., state $B$) is the minority state, even though it is not. When $q=1$, if we select a core node as the primary node in an update, it is more likely to rewire than to adopt, at least initially. However, as most nodes are in state $B$ initially, the core node is likely to rewire to another node in state $B$. If we select a peripheral node as the primary node for an update, it is more likely to adopt than to rewire. At least initially, such adoptions convert peripheral nodes from the majority state $B$ to the minority state $A$. 

In our simulations of our nonlinear RTR-CVM on SBM networks with core--periphery structure, we take $3 N_A(0)=N_B(0)$, and we set the probabilities to be $P_{aa}=\frac{20}{N_A(0)-1}$, $P_{bb}=\frac{1}{N_B(0)-1}$, and $P_{ab}=\frac{5}{N}$. In \fref{fig:RTR_SBM_Terminal_A_CP}, we plot the terminal density of state $A$ for various values of $q$ from simulations on networks with $N=50,000$ nodes. We observe a transition in the qualitative dynamics in the range $q \in [4,4.5]$. As we increase $q$ from $0$ to $4$, there are progressively more nodes that terminate in state $A$ before the network fragments, with approximately $90$\,\% the nodes terminating in state $A$ when $q=4$. For $q\in (4,4.5)$, the system appears to exhibit a phase transition that is similar to that of the ostensible hybrid phase transition of \fref{fig:RTR_SBM_Terminal_A_CP}.

In this case, however, the hybrid transition is discontinuous. In \fref{appsec:RTR_big_q_appendix}, we explore $q\in[0,100]$. For $q\in(4.5,39)$, we find that almost every node terminates in state $B$ (see \fref{fig:RTR_SBM_Terminal_A_CP_big_q}). 

%%%%%%%%%%%%%%%%%%%%%%%% Majority and Minority Illusion %%%%%%%

\subsubsection{Majority and Minority Illusions} \label{sec:Illusion}

Recent work by Lerman et al.~\cite{lerman2016majority} on the ``majority illusion'' in social networks examined the phenomenon of distorted local observations when a state that is globally rare in a network may be dramatically overrepresented in the local neighborhoods of many individuals. Using a model of threshold opinion dynamics, Lerman et al.\ illustrated that majority illusions can accelerate the spread of states that are initially rare. For our work with binary opinion states, we find it useful to distinguish between two different types of ``illusions''. By a \emph{majority illusion}, we mean the phenomenon of nodes in a minority state perceiving their state to be in the majority. Analogously, by a \emph{minority illusion}, we mean the phenomenon of nodes in the majority state perceiving their state to be in a minority. For binary opinion states, the minority illusion implies that nodes in the majority state incorrectly perceive that the minority state is held by the majority of nodes.

In our nonlinear CVM, a node's local survey $\sigma$ is based on a sample of the global population. In a locally well-mixed system (see \fref{eq:wellmixed}), the sample leads to good estimates of the global densities $N_A(t)/N$ and $N_B(t)/N$. However, when seeding the system as in \fref{sec:RTR_SBM_two_communities} and \fref{sec:RTR_SBM_CP}, the samples are biased initially. {In \fref[plain]{fig:RTR_SBM_Terminal_Steps_Sigma_Proportions_TWO_COM} and \fref[plain]{fig:RTR_SBM_Terminal_Steps_Sigma_Proportions_CP}, we plot the mean of the local surveys $\overline{\sigma_A}$ (solid blue curves) and $\overline{\sigma_B}$ (solid red curves) versus elementary time steps and compare them to the true global densities, $N_A(t)/N$ (dashed blue curves) and $N_B(t)/N$ (dashed red curves), for simulations on systems that we seed with two-community structure and core--periphery structure, respectively.} We calculate the unweighted means $\overline{\sigma_A}$ and $\overline{\sigma_B}$ from \fref{eq:arithemticmeansigma}, so we are treating all local surveys equally. In mathematical language, assuming that state $A$ is in the minority (i.e., $N_A(t)/N < 1/2$), the majority illusion occurs when $\overline{\sigma_A}>1/2$. Analogously, assuming that state $B$ is in the majority (i.e., $N_B(t)/N > 1/2$), the minority illusion occurs when $\overline{\sigma_B}<1/2$. In Table~\ref{Tab:Illusion}, we summarize how we seed networks with different types of ``illusions'' using an SBM network with $N_A(0)=cN$ and $c<{1}/{2}$.

\begin{center}
\begin{table}[ht]

 \begin{tabular}{||p{5cm} | p{5cm}||} 
 \hline
 Illusion  & Edge Probabilities \\ 
 \hline\hline
 No Illusion  & $P_{aa}=P_{ab}=P_{bb}$ \\ 
 \hline
 Majority Illusion for $A$  &  $\frac{c}{1-c}P_{aa}\gg  P_{ab}$ \\ 
 \hline
 Minority Illusion for $B$ & $\frac{1-c}{c} P_{bb}\ll  P_{ab}$ \\ 
 \hline
 Both Illusions & $\frac{1-c}{c} P_{bb}\ll  P_{ab} \ll \frac{c}{1-c}P_{aa}$  \\ 
 \hline
\end{tabular}
  \caption{Summary of SBM parameters that we use to seed a network with a majority illusion, a minority illusion, both types of illusions, or neither illusion using an SBM network with $N_A(0)=cN$ nodes in state $A$ and $c<{1}/{2}$. }
    \label{Tab:Illusion}
\end{table}
\end{center}

In \fref[plain]{fig:RTR_SBM_Terminal_Steps_Sigma_Proportions_TWO_COM}, we seed a network using the SBM two-community structure that we described in \fref{sec:RTR_SBM_two_communities}. Initially, the larger community (which has $3/4$ of the nodes) consists of nodes in state $B$, and the smaller community (which has the remaining $1/4$ of the nodes) consists of nodes in state $A$. However, the local surveys of the nodes in the smaller community lead them to perceive state $A$ as the majority state and thus state $B$ as the minority. Similarly, the local surveys of the nodes in the larger community lead them to perceive state $B$ as the majority state and thus state $A$ as the minority. In other words, the larger community of nodes (which are in the majority state) correctly believe that their state is in the majority. However, the smaller community of nodes (which are in the minority state) experience a majority illusion, as they incorrectly believe that their state is in the majority.

\begin{figure}[ht]
  \includegraphics[width=.7\linewidth]{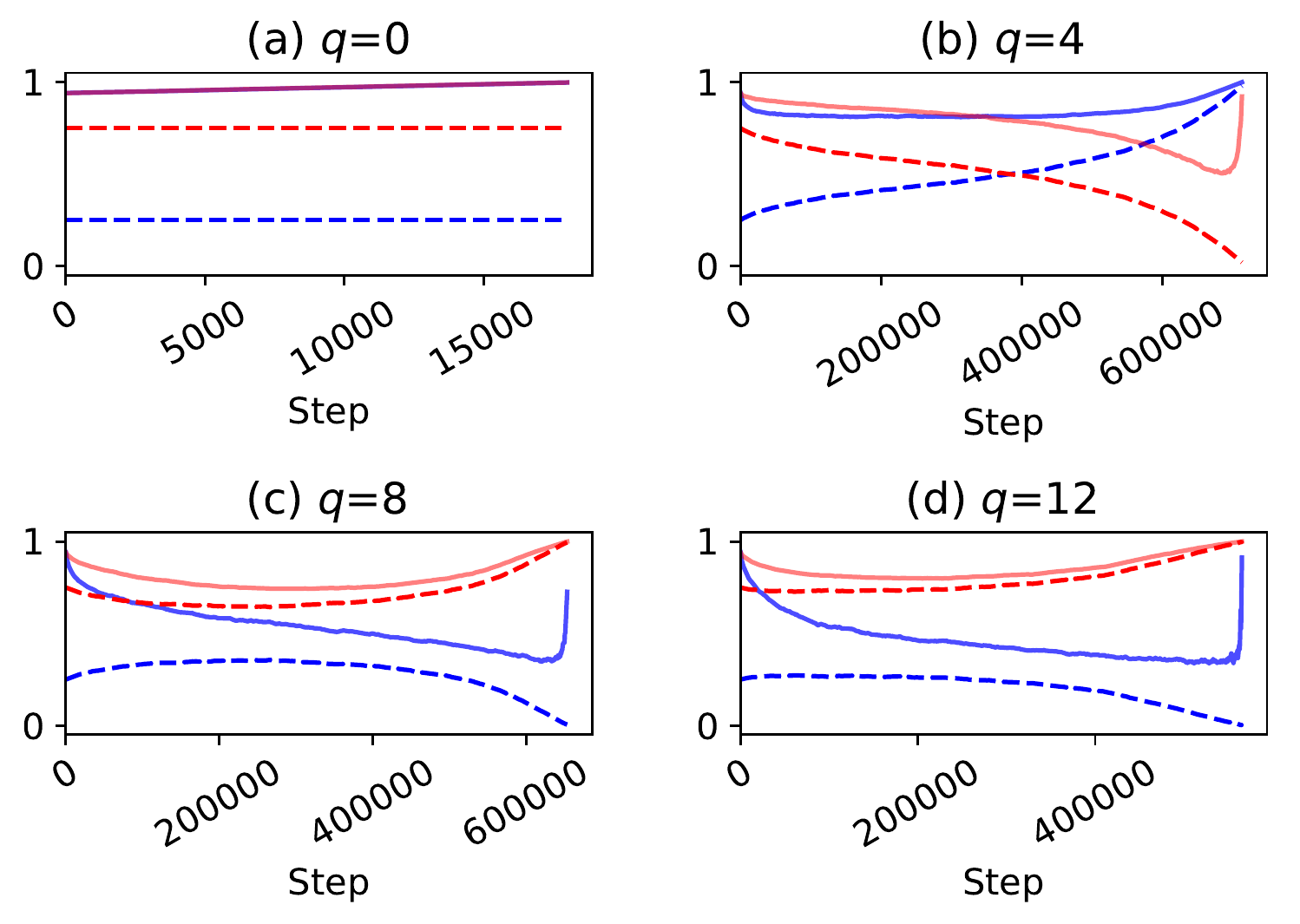}
    \caption{{Unweighted means} [$\overline{\sigma_A}$ (solid blue curve) and $\overline{\sigma_B}$ (solid red curve)] of local surveys for nodes in states $A$ and $B$ (see \fref{eq:arithemticmeansigma} for details) and global state densities [$N_A/N$ (dashed blue curve) and $N_B/N$ (dashed red curve)] of states $A$ and $B$ versus elementary time steps in four rewire-to-random (RTR) simulations (see Algorithm~\ref{alg:ftvm_rtr}) of our nonlinear CVM. We seed each realization using the SBM two-community structure that we described in \fref{sec:RTR_SBM_two_communities}. 
     }\label{fig:RTR_SBM_Terminal_Steps_Sigma_Proportions_TWO_COM}
\end{figure}

As we noted in \fref{fig:RTR_SBM_TWO_COM_TERMINAL_A}, the effect on the terminal densities of initializing the system with a majority illusion depends on the value of the nonlinearity parameter $q$. For $q=0$, in which only rewiring occurs, the state densities do not change, but the system fragments, such that each node only has neighbors that share its state. Therefore, $\overline{\sigma_A}$ and $\overline{\sigma_B}$ increase towards $1$, and the majority illusion of state $A$ increases in severity. For $q=4$, the illusion becomes a reality, in the sense that $N_A/N$ increases to match $\overline{\sigma_A}$. The network ultimately reaches an absorbing state with most nodes in state $A$, but there are still small clusters of nodes in state $B$, so $\overline{\sigma_B}$ increases to $1$ towards the end of the simulation (because we take the mean over only these nodes). For $q=8$ and $q=12$, the nodes ``wise up'' in the sense that $\overline{\sigma_A}$ decreases to match $N_A/N$. The network ultimately reaches an absorbing state with most nodes in state $B$, but there are still small clusters of nodes in state $A$ that cause $\overline{\sigma_A}$ to increase to $1$ towards the end of our simulations. 

In \fref[plain]{fig:RTR_SBM_Terminal_Steps_Sigma_Proportions_CP}, we seed the system using the SBM core--periphery structure that we described in \fref{sec:RTR_SBM_CP}. Initially, all peripheral nodes (which constitute $3/4$ of the nodes) are in state $B$, and the core nodes (which constitute the remaining $1/4$ of the nodes) are in state $A$. The local surveys of the peripheral nodes lead them to perceive state $B$ as the minority state and thus state $A$ as the majority. Similarly, the local surveys of the core nodes lead them to perceive state $A$ as the majority state and thus state $B$ as the minority. In other words, the core nodes (which are in the minority state) experience a majority illusion, incorrectly believing that their state is in the majority. Conversely, the peripheral nodes (which are in the majority state) experience a minority illusion, incorrectly believing that their state is in the minority.

\begin{figure}[ht]
  \includegraphics[width=.7\linewidth]{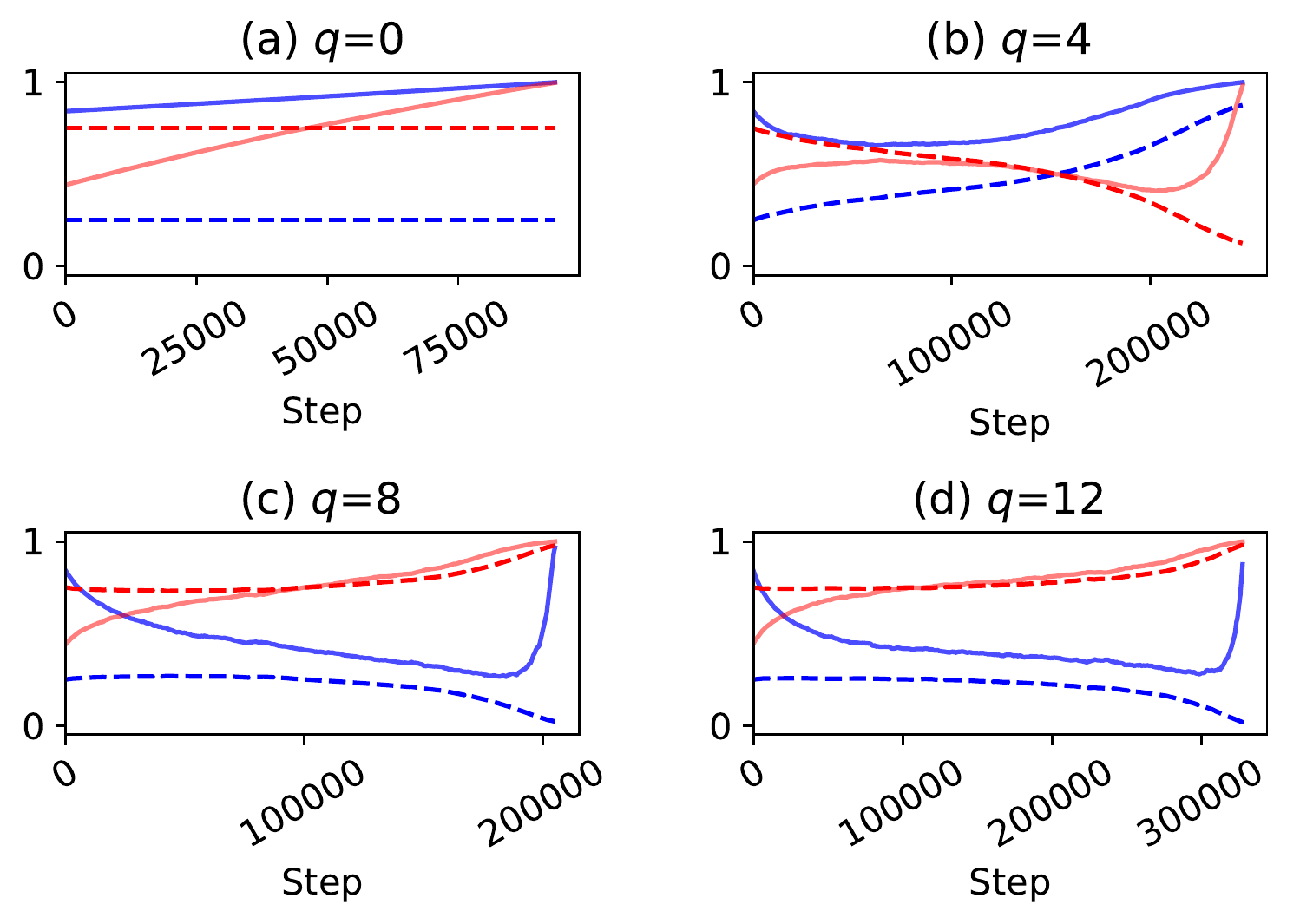}
    \caption{{Unweighted means} [$\overline{\sigma_A}$ (solid blue curve) and $\overline{\sigma_B}$ (solid red curve)] of local surveys for nodes in states $A$ and $B$ (see \fref{eq:arithemticmeansigma} for details) and global state densities [$N_A/N$ (dashed blue curve) and $N_B/N$ (dashed red curve)] for states $A$ and $B$ versus elementary time steps in four rewire-to-random (RTR) simulations (see Algorithm~\ref{alg:ftvm_rtr}) of our nonlinear CVM. We seed each realization using the SBM core--periphery structure that we described in \fref{sec:RTR_SBM_CP}.
     }\label{fig:RTR_SBM_Terminal_Steps_Sigma_Proportions_CP}
\end{figure}

As we saw for two-community structure, the effect of the majority and minority illusions depends on the value of the nonlinearity parameter $q$ in our networks with core--periphery structure. For $q=0$, in which only rewiring occurs, state densities do not change, but the network fragments, such that each node only has neighbors that share its state. Therefore, $\overline{\sigma_A}$ and $\overline{\sigma_B}$ increase towards $1$, and the majority illusion for state $A$ increases in severity, but the minority illusion for state $B$ dissipates. For $q=4$, the majority illusion for state $A$ becomes a reality, in the sense that $N_A/N$ increases to match $\overline{\sigma_A}$. The minority illusion for state $B$ also becomes a reality, in the sense that $N_B/N$ decreases to match $\overline{\sigma_B}$ initially. However, towards the end of a simulation, $\overline{\sigma_B}$ increases because there are still small clusters of nodes in state $B$ and $\overline{\sigma_B}$ is a mean over only these nodes. For $q=8$ and $q=12$, the nodes wise up to both illusions, as $\overline{\sigma_A}$ decreases to match $N_A/N$ and $\overline{\sigma_B}$ increases to match $N_B/N$. The system ultimately reaches an absorbing state with most nodes in state $B$, but there are still small clusters of nodes in state $A$ that cause $\overline{\sigma_A}$ to increase to $1$ towards the end of our simulations. 

These examples demonstrate that, under certain conditions, seeding our nonlinear RTR-CVM with illusions can lead to the spreading of initially rare states. For instance, when we seed the system with an SBM two-community network such that there is a majority illusion but not a minority illusion and take the value of the nonlinearity parameter to be $q=4$, almost every node adopts the initially rare state in all realizations of our simulations. However, under other conditions, seeding the system with illusions can also stifle the spread of initially rare states. For example, when we seed the system with core--periphery structure such that there is both a majority illusion and a minority illusion and take the value of the nonlinearity parameter to be $q=8$, the rare state vanishes almost entirely in all realizations of our simulations. This behavior contrasts strongly with what we observe in the linear RTR-CVM, in which the probability that an initially rare state spreads, conditioned on the event that a state does indeed spread, is equal to the initial fraction of nodes in the rare state. In other words, the adoption rate parameter $\alpha$ in the linear RTR-CVM affects whether some opinion state spreads to almost every node, but the initial state densities determine the probability of which state it will be. By contrast, the nonlinearity parameter $q$ in our nonlinear RTR-CVM affects not only whether some opinion state spreads to almost every node but also the probabilities for which state it will be.  See \fref{appsec:Linear_CVM_Simulations} for details.

%%%%%%%%%%%%%%%%%%%%%%%%
%%% Rewire-to-Same
%%%%%%%%%%%%%%%%%%%%%%%%

\section{Rewire-to-Same}
\label{sec:RTS}

%%%%%%%%%%%%%%%%%%%%%%%% Rewire-to-Same Model Introduction %%%%%

\subsection{Model}

In this section, we explore our nonlinear CVM with a ``rewire-to-same" (RTS) scheme. We give a formal description of it in Algorithm~\ref{alg:ftvm_rts} in \fref{appsec:RTSAlgo}. The key difference from the RTR scheme is that when rewiring occurs in the RTS scheme, the primary node deletes its discordant edge to the secondary node and then forms an edge with a node that we choose uniformly at random from the set of nodes in the same state as the primary node. In \fref{appsec:Linear_CVM_Simulations_RTS}, we compare our nonlinear RTS-CVM of Algorithm~\ref{alg:ftvm_rts} to the linear RTS-CVM of~\cite{durrett2012graph}. Based on previous work~\cite{basu2017evolving,Silk2014Exploring}, it seems that RTS schemes have been more difficult to analyze quantitatively than RTR schemes in linear CVMs.

To fully specify the RTS scheme, we start by examining a peculiarity of the RTS scheme: What happens when a rewiring action cannot take place, because the primary node is already adjacent to all nodes in its state (including the trivial case in which there are no other nodes in its state)? This situation is likely to arise if a system approaches consensus or if a network is densely connected (specifically, if the mean degree satisfies $\langle k \rangle \geq {N}/{2}$)~\cite{basu2017evolving,basak2015evolving}. There are several possible rules to employ, and the choice of rule may affect both the outcome and the analysis. Possible specifications include the following: 
\begin{enumerate}
    \item[(i)] stipulate that there is no replacement edge;
    \item[(ii)] stipulate that we instead perform an RTR operation;
    \item[(iii)] stipulate that the recently deleted discordant edge reforms;
    \item[(iv)] stipulate that the recently deleted discordant edge reforms and that the primary node instead performs an adoption action; and
    \item[(v)] stipulate that we allow multi-edges, self-edges, or both.
\end{enumerate}

Each of these choices either introduces a new mechanism, such as an edge deletion, or it changes the class of allowed networks. (Previously, we demanded that networks have neither self-edges nor multi-edges.) We choose to use option (i) of letting no replacement edge form, such that $|E(t)|$ is no longer a conserved quantity. By contrast, recent work on CVMs on dense random graphs allowed the formation of multi-edges~\cite{basu2017evolving}.

%%%%%%%%%%%%%%%%%%%%%%%% Rewire-to-Same Simulations %%%%%

\subsection{Simulations}

\begin{figure}
  \includegraphics[width=.7\linewidth]{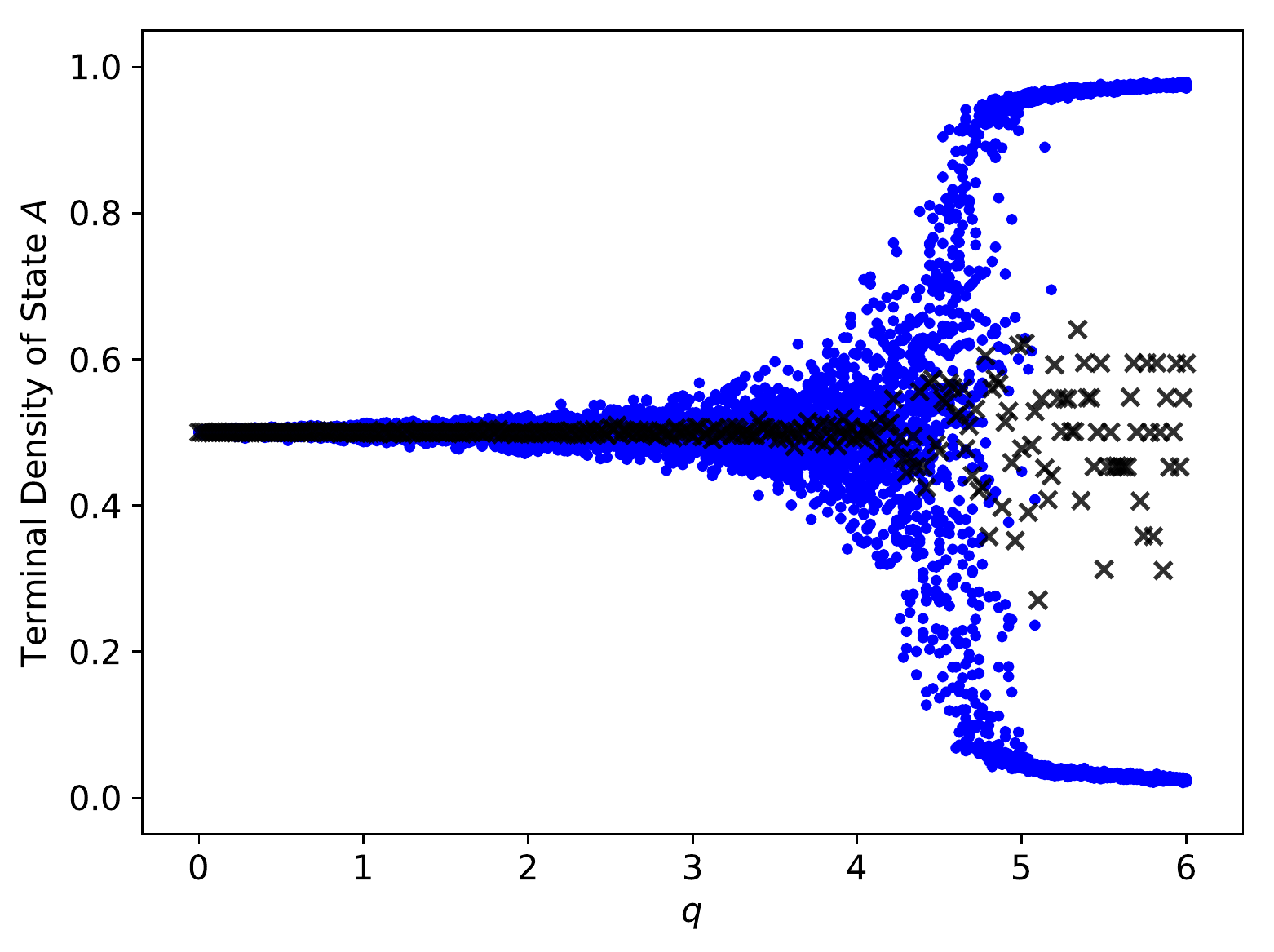}
    \caption{Terminal density of state $A$ in rewire-to-same (RTS) simulations (see Algorithm~\ref{alg:ftvm_rts}) for our nonlinear CVM for $q\in[0,6]$ with an increment of $\Delta q= 0.02$. For each value of $q$, we simulate $20$ realizations. We seed each realization with a different ER network with $N=50,000$ nodes and an edge probability of $p=\frac{4}{N-1}$, and we initialize half of the nodes in state $A$ and the other half in state $B$. We plot individual realizations with blue dots and means with $\times$ symbols.}\label{fig:RTS_ER_Final_A}
\end{figure}

We first simulate our nonlinear RTS-CVM on ER networks. In \fref{fig:RTS_ER_Final_A}, we observe similar qualitative behavior as we did in our nonlinear RTR-CVM (see \fref{fig:RTR_ER_Terminal_A}). Specifically, there seems to be a continuous transition between fragmentation and consensus regimes. One difference is that the transition occurs at about $q\approx4.5$ in the RTS version of the model. The similarity of these two models contrasts starkly with results for the linear CVM, whose behavior differs significantly under the two rewiring schemes (see \fref{fig:LIN_RTR_ER_Term_A} and \fref{fig:lin_rts_terminal_A_ER}). Prior research on linear CVMs suggests that the rewiring scheme affects linear CVMs significantly because of strong correlations between network structure and node states that arise from RTS actions, but not from RTR actions~\cite{demirel2014moment}. By contrast, the similarity of results between the RTR and RTS variants for our nonlinear CVM suggests that such correlations play a less prominent role in our model than they do in previously studied linear CVMs.

\begin{figure}
  \includegraphics[width=\linewidth]{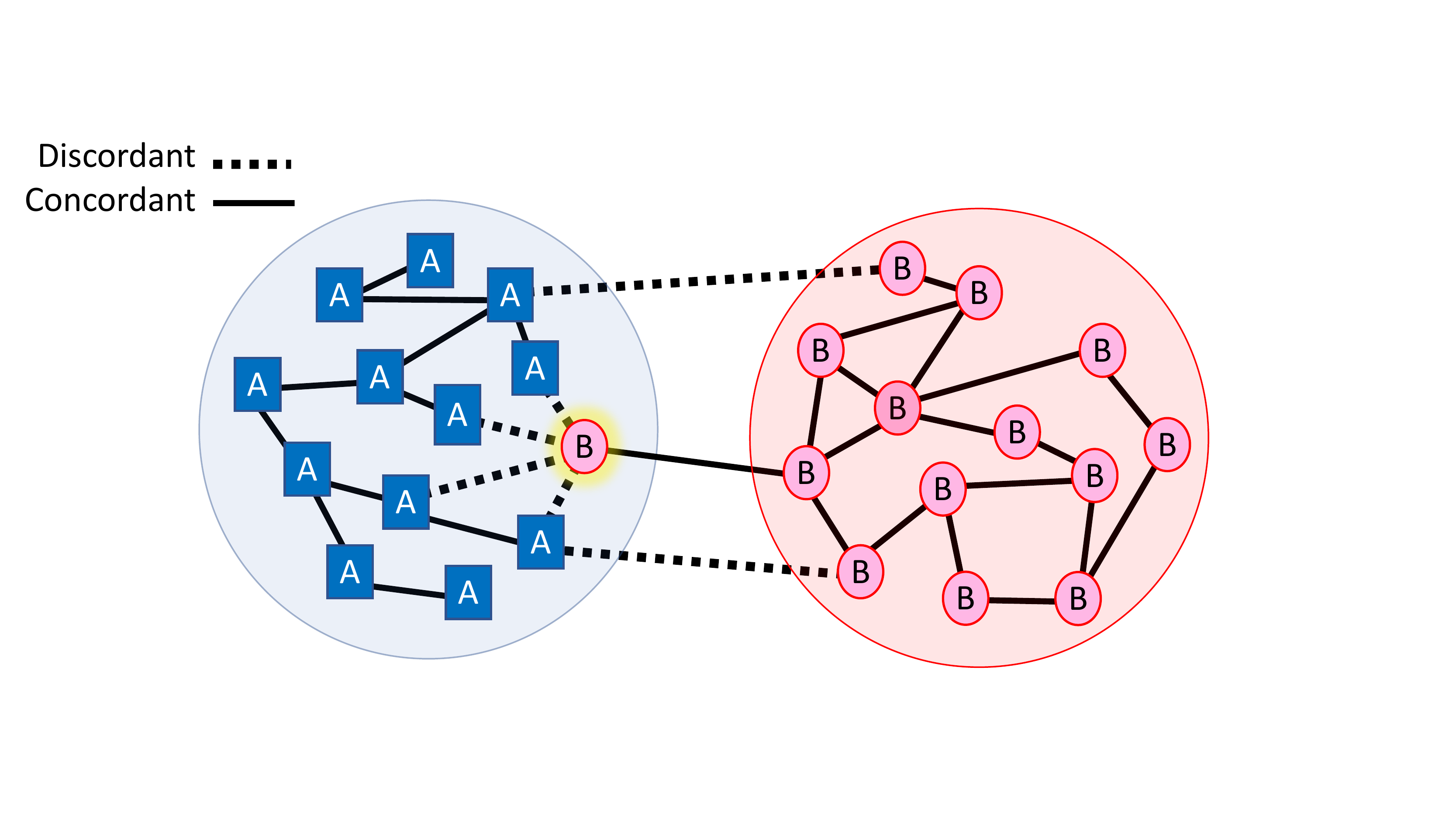}
    \caption{``Typical'' configuration near fragmentation for a linear RTS-CVM. Several discordant edges are associated with very few nodes.
    In this example, the highlighted node changes from state $A$ to state $B$, and all of its previously concordant edges become discordant edges, which induces $ABA$-triple correlations. [This illustration is our version of Figure 5 of~\cite{demirel2014moment}.]
    }\label{fig:Dem_near_end}
\end{figure}

The sketch in \fref{fig:Dem_near_end} represents what Demirel et al.\ ~\cite{demirel2014moment} reported as a ``typical'' configuration that is near fragmentation for a linear RTS-CVM. As the system evolves, nodes group into communities, which are connected to each other by only a few edges. Occasionally, there is an adoption that creates many discordant edges that are concentrated at one node. This leads to a disproportionately large number of $ABA$-triples that are concentrated on one node and constitutes a strong three-node correlation. However, in our nonlinear RTS-CVM, such situations occur much less frequently than they do in the linear RTS-CVM of~\cite{durrett2012graph}. In fact, in our nonlinear CVM, the more $ABA$-triples that a node in state $A$ creates by adopting state $B$, the less likely it is to adopt state $B$. To illustrate this observation, suppose that node $i$ is in state $A$ and has at least two neighbors in state $A$. If we select node $i$ for an update, it adopts state $B$ with probability $1-\sigma_i^q$, where $\sigma_i$ is the fraction of nodes that are adjacent to node $i$ and are also in state $A$. Consequently, the probability of node $i$ creating $ABA$-triples by adopting state $B$ decreases as the number of concordant edges that are incident to node $i$ (i.e. the edges that would form part of the $ABA$-triples) increases.

\begin{figure}
  \includegraphics[width=.7\linewidth]{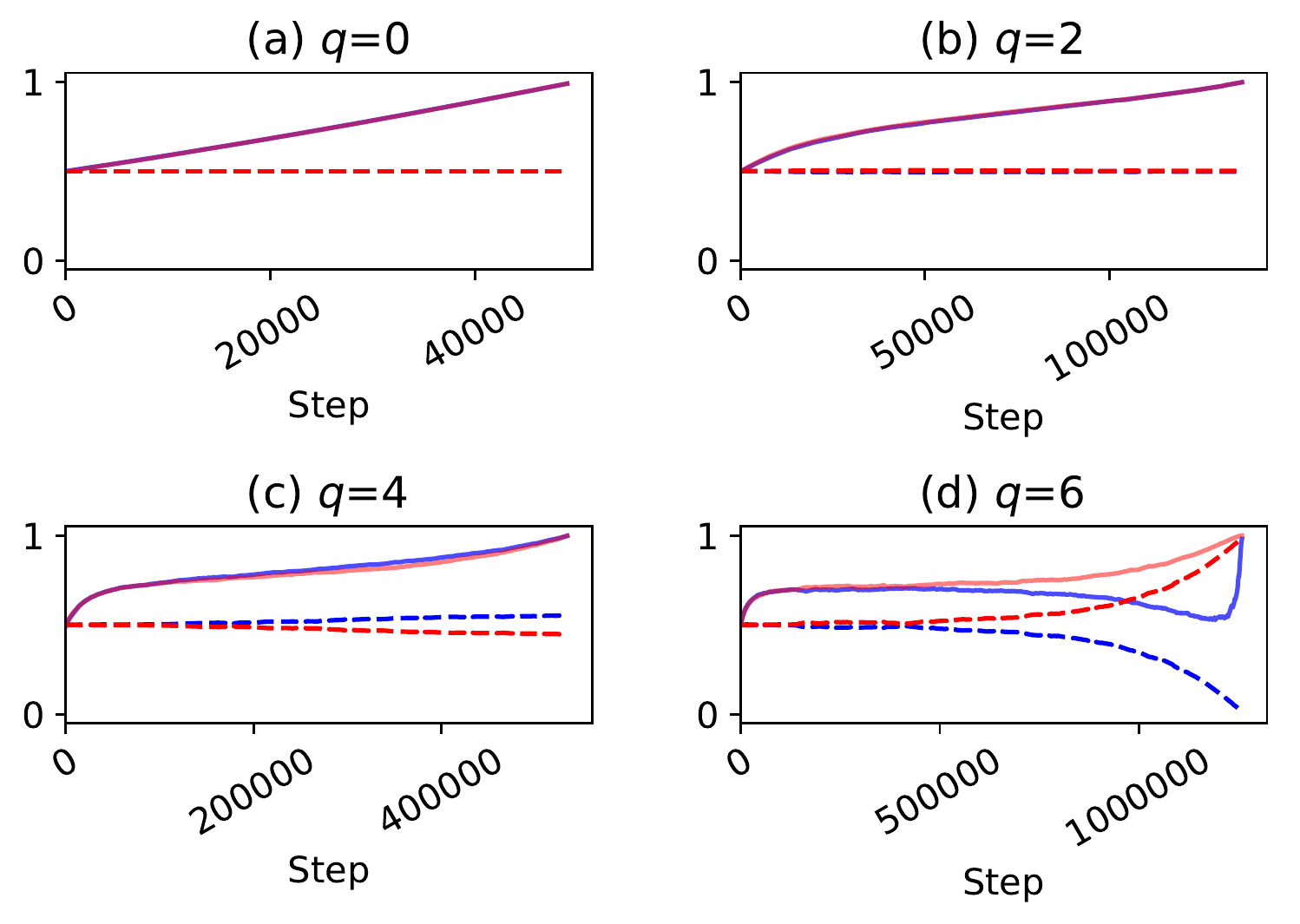}
    \caption{{Unweighted means} [$\overline{\sigma_A}$ (solid blue curve) and $\overline{\sigma_B}$ (solid red curve)] of local surveys for nodes in states $A$ and $B$ (see \fref{eq:arithemticmeansigma} for details) and global state densities [$N_A/N$ (dashed blue curve) and $N_B/N$ (dashed red curve) for states $A$ and $B$, respectively] versus elementary time steps in four RTS simulations (see Algorithm~\ref{alg:ftvm_rts}) in our nonlinear CVM. We seed each realization with an ER network with $N=50,000$ nodes and edge probability $p=\frac{4}{N-1}$, and we initialize half of the nodes in state $A$ and the other half in state $B$. 
    }\label{fig:RTS_ER_Sigma}
\end{figure}

In \fref{fig:RTS_ER_Sigma}, we plot the evolution of {unweighted means}, $\overline{\sigma_A}$ and $\overline{\sigma_B}$, of local surveys and compare them to global state densities (as in \fref{sec:Illusion}). We observe fragmentation in our simulations with $q=0$, $q=2$, and $q=4$. Fragmentation causes the local surveys of nodes to become distorted, because nodes are in clusters and have no neighbors in a different state. This leads to a weak form of a majority illusion for nodes in each state: although roughly half of the nodes are in state $A$ and roughly half are in state $B$, almost every node perceives its own state to be in the majority. In our simulations with $q=6$, we again initially observe a weak majority illusion for nodes in both states. As the system evolves, the density of state $A$ decreases and $\overline{\sigma_A}$ decreases commensurately, but nodes in state $A$ still exhibit a majority illusion. The system ultimately reaches an absorbing state with most nodes in state $B$, but there are still small clusters of nodes in state $A$ that cause $\overline{\sigma_A}$ to increase to $1$ towards the end of our simulations.

\begin{figure}[ht]
  \centering
    \subfloat[Two-Community Structure]{\includegraphics[width=0.48\linewidth]{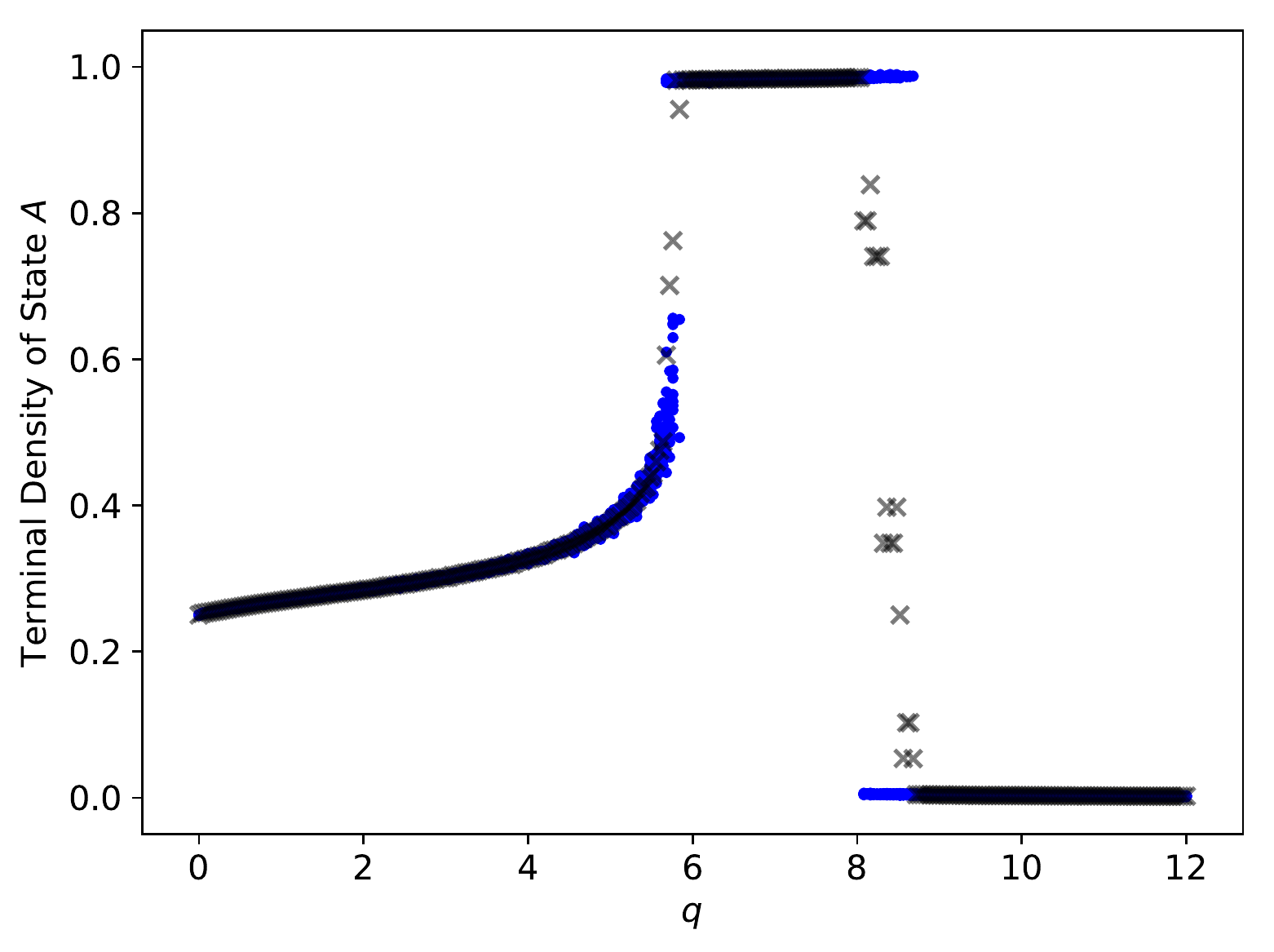}\label{fig:RTS_SBM_Final_A_Two_Com}}
    \hfill
    \subfloat[Core--Periphery Structure]{\includegraphics[width=0.48\linewidth]{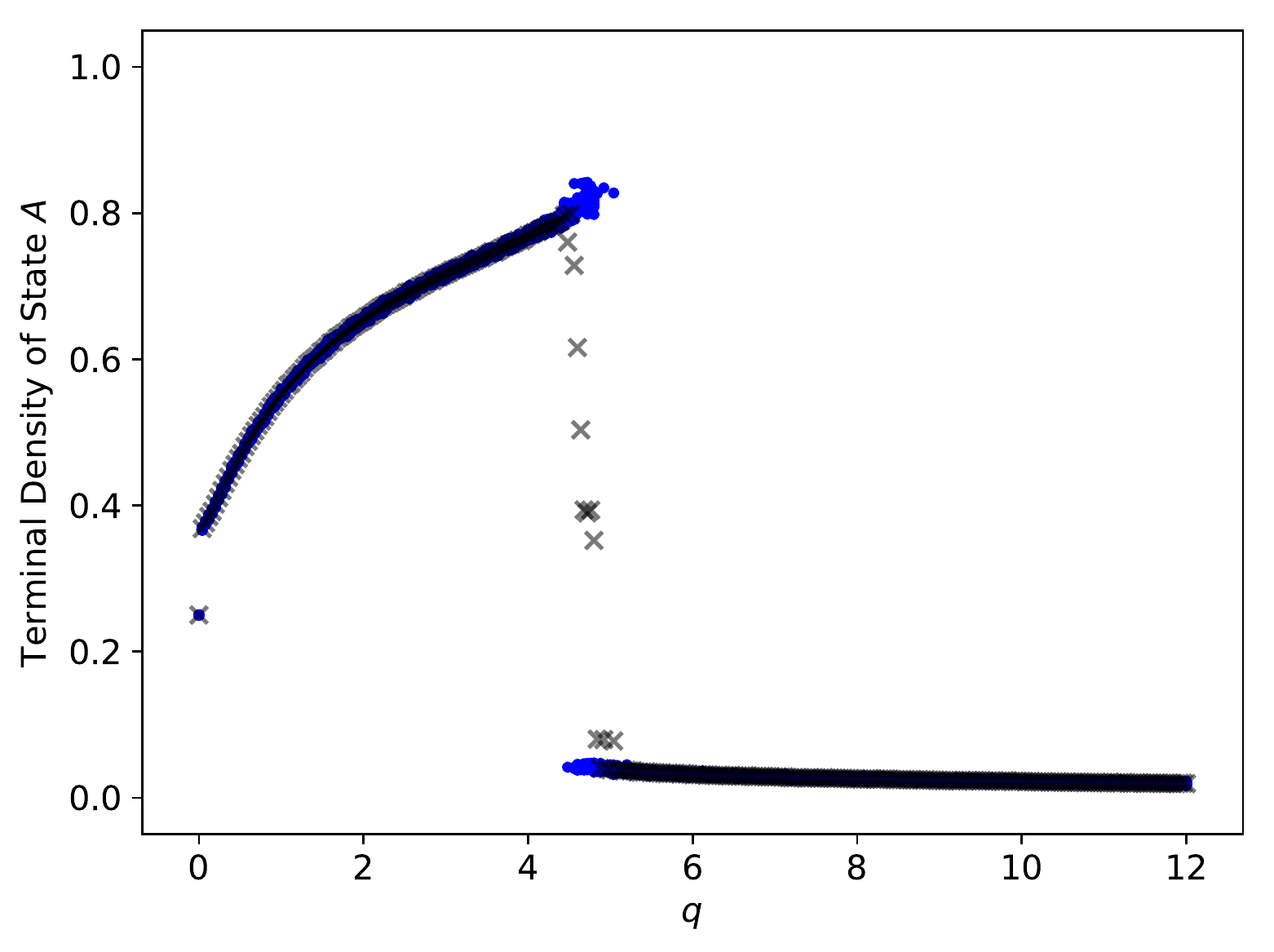}\label{fig:RTS_SBM_Final_A_CP}}
  \caption{Terminal density of state $A$ for RTS simulations (see Algorithm~\ref{alg:ftvm_rts}) of our nonlinear CVM for nonlinearity parameters $q\in[0,12]$ with an increment of $\Delta q = 0.04$. For each value of $q$, we simulate $20$ realizations. In (a), we seed our simulations with the two-community SBM networks that we described in \fref{sec:RTR_SBM_two_communities}. In (b), we seed our simulations with the SBM core--periphery structure that we described in \fref{sec:RTR_SBM_CP}. We plot individual realizations with blue dots and means with $\times$ symbols.}\label{fig:RTS_SBM_final_A}
\end{figure}

We conclude this section on our nonlinear RTS-CVM by conducting simulations that we seed with (1) two-community structure and (2) core--periphery structure using SBM networks (see \fref{sec:RTR_SBM}). In \fref{fig:RTS_SBM_Final_A_Two_Com}, we observe that our nonlinear RTS-CVM initialized with two-community structure exhibits qualitatively similar long-time behavior as our nonlinear RTR-CVM initialized with the same two-community structure  (see \fref{fig:RTR_SBM_TWO_COM_TERMINAL_A}). Nevertheless, simulations using these two rewiring schemes do exhibit quantitative differences, such as in the locations of the transitions between regimes with qualitatively different terminal statistics. In \fref{fig:RTS_SBM_Final_A_CP}, we also observe qualitatively similar results for simulations seeded with core--periphery structure for both the RTS and RTR versions of our nonlinear CVM (see \fref{fig:RTR_SBM_Terminal_A_CP}).

%%%%%%%%%%%%%%%%%%%%%%%%
%%% Rewire-to-None
%%%%%%%%%%%%%%%%%%%%%%%%

\section{Rewire-to-None} \label{sec:RTN}

%%%%%%%%%%%%%%%%%%%%%%%% Rewire-to-None Model Introduction %%%%%%

\subsection{Model}

We now examine our nonlinear CVM with edge deletion, which we call ``rewire-to-none'' (RTN) to parallel the rewire-to-random and rewire-to-same terminology. In this RTN-CVM, adoption occurs with probability $1-\sigma_i^q$, and edge deletion occurs with probability $\sigma_i^q$; there are no replacement edges. We give a precise description of this model in \fref{appsec:RTNAlgo}. 

To the best of our knowledge, a RTN scheme has not been studied previously using linear CVMs, although a bounded-confidence opinion model with edge deletion (to model unfollowing on social media) was examined very recently in~\cite{sasahara2019inevitability}. In \fref{appsec:LINRTNCVM}, we investigate a linear CVM with our RTN scheme both analytically and computationally. There are many reasons to study an RTN mechanism in opinion models. In some sense, the edge-deletion mechanism is simpler than mechanisms that require additional parameters and specification of a rewiring rule \footnote{However, edge deletion can pose a mathematical challenge, because the number of edges and the mean degree are no longer constant~\cite{kiss2017mathematics}.}. Moreover, edge deletion may also more accurately model internet social dynamics than rewiring, because individuals perform actions such as ``unfriending'' or ``unfollowing'' without necessarily ``friending'' or ``following'' another account~\cite{john2015don, geschke2019triple,skoric2018predicts}. Edge deletion is also an important network mechanism in the structural evolution of social networks~\cite{farajtabar2015coevolve,saavedra2008asymmetric}.

%%%%%%%%%%%%%%%%%%%%%%%% Rewire-to-None Simulations

\subsection{Simulations}

\begin{figure}
  \includegraphics[width=.7\linewidth]{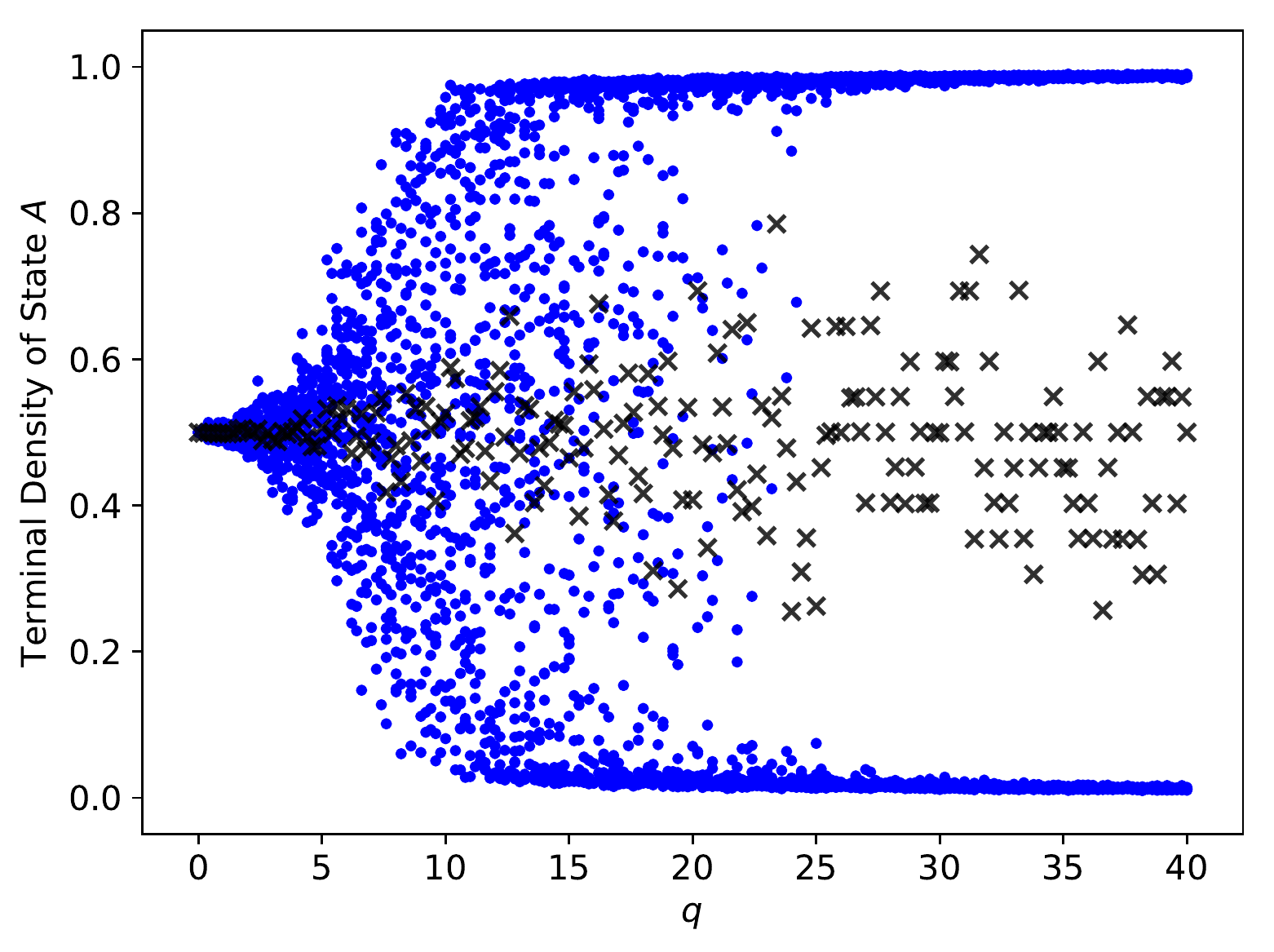}
    \caption{Terminal density of state $A$ in rewire-to-none (RTN) simulations (see Algorithm~\ref{alg:ftvm_rtn}) of our nonlinear CVM for $q\in[0,40]$ with an increment of $\Delta q = 0.2$. For each value of $q$, we simulate $20$ realizations. We seed each realization with a different ER network with $N=25,000$ nodes and an edge probability of $p=\frac{4}{N-1}$, and we initialize half of the nodes in state $A$ and the other half in state $B$. We plot individual realizations with blue dots and means with $\times$ symbols.}\label{fig:RTN_ER_Terminal_A}
\end{figure}

We seed our nonlinear RTN-CVM with ER $G(N,p)$ networks with $N=25,000$ nodes and an edge probability of $p=\frac{4}{N-1}$, and we initialize half of the nodes in state $A$ and the other half of the nodes in state $B$. We plot the terminal state density of $A$ in \fref{fig:RTN_ER_Terminal_A}. When $q=0$, no adoption occurs, so state densities do not change before the network fragments. We find that with edge deletion (i.e., the RTN mechanism), fragmentation of a network into disconnected components can occur for a wide range of $q$ values up to at least $q=20$. {This is a larger range than what we observed for the RTR and RTS schemes for our nonlinear CVM. For the RTR scheme, we did not observation fragmentation for $q\gtrapprox3$; for the RTS scheme, we did not observe fragmentation for $q\gtrapprox5.5$.}

\begin{figure}[!tbp]
  \centering
    \subfloat[Two-Community Structure]{\includegraphics[width=0.48\linewidth]{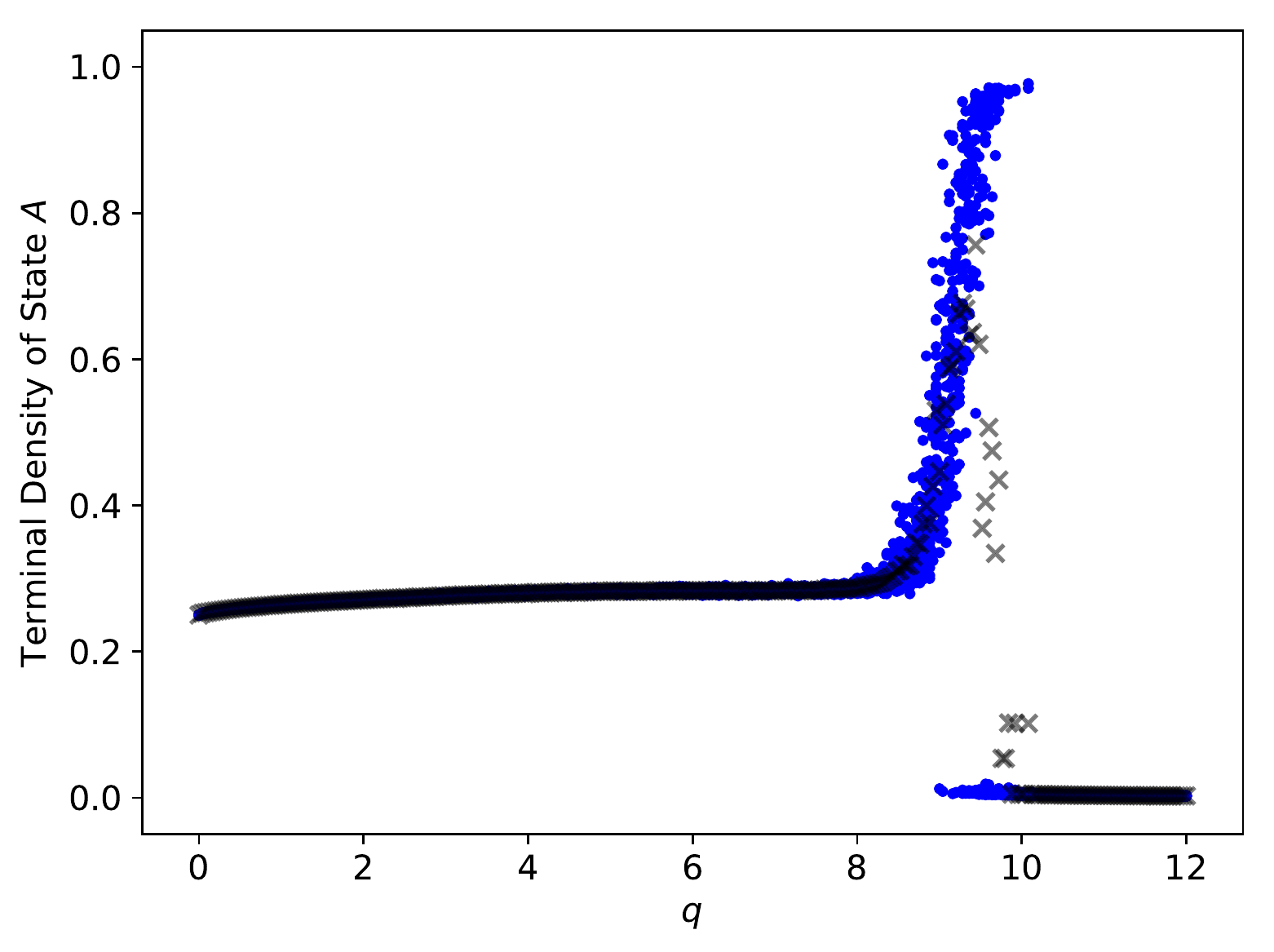}\label{fig:RTN_Two_Com_A}}
    \hfill
    \subfloat[Core--Periphery Structure]{\includegraphics[width=0.48\linewidth]{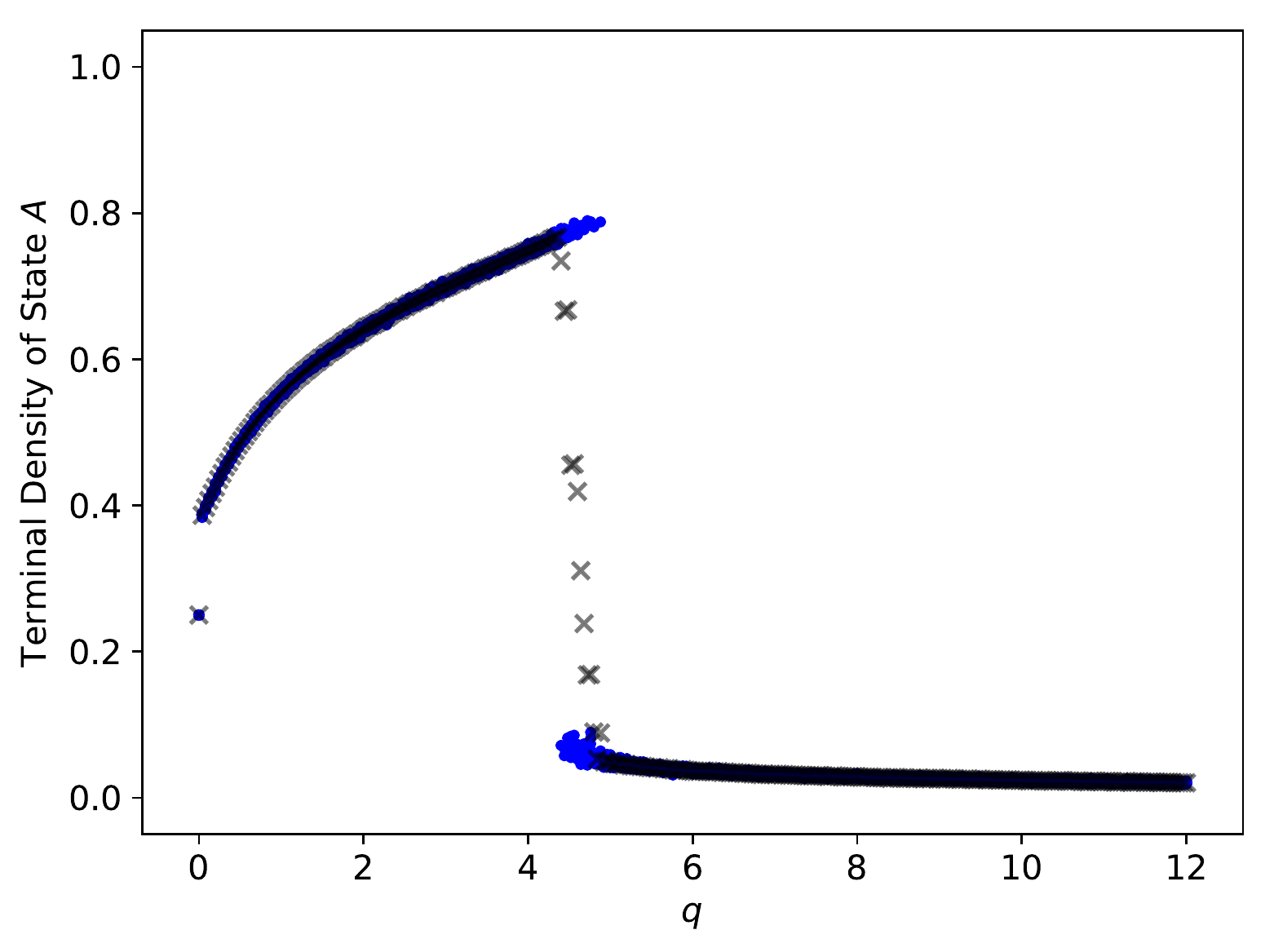}\label{fig:RTN_SBM_cp_A}}
  \caption{Terminal density of state $A$ in simulations of our nonlinear RTN-CVM for $q\in[0,12]$ with an increment of $\Delta q = 0.04$. For each value of $q$, we simulate $20$ realizations. In (a), we seed the system with the two-community SBM networks that we described in \fref{sec:RTR_SBM_two_communities}. In (b), we seed the system with the SBM core--periphery networks that we described in \Fref{sec:RTR_SBM_CP}. We plot individual realizations with blue dots and means with $\times$ symbols.}\label{fig:RTN_SBM_A}
\end{figure}

We conduct simulations using two-community SBM networks (see \fref{sec:RTR_SBM_two_communities}) to seed the system. We plot the terminal density of state $A$ in \fref{fig:RTN_Two_Com_A}; for $q \in (0,8.5)$, we observe that the two communities separate from each other and that there are no signficant changes to the densities of opinion states. In contrast to the RTR scheme that we illustrated in \fref{fig:RTR_SBM_TWO_COM_TERMINAL_A} and the RTS scheme in \fref{fig:RTS_SBM_Final_A_Two_Com}, for the nonlinear RTN-CVM, we do not observe any values of $q$ in which almost every node terminates in state $A$ for every realization of a simulation. For $q\approx 9.52$, state $A$ spreads to most nodes in most realizations, with a mean terminal density of approximately $0.75$ over the $20$ realizations. However, in some realizations for $q\approx 9.52$, almost every node terminates in state $B$. 

Finally, we simulate our nonlinear RTN-CVM model using SBM core--periphery networks (see \fref{sec:RTR_SBM_CP}) to seed the system. In \fref{fig:RTN_SBM_cp_A}, we observe that the RTN scheme produces terminal behavior that is qualitatively very similar to what we observed with the RTR (see \fref{fig:RTR_SBM_Terminal_A_CP}) and RTS (see \fref{fig:RTS_SBM_Final_A_CP}) schemes.

%%%%%%%%%%%%%%%%%%%%%%%%
%%% Conclusion
%%%%%%%%%%%%%%%%%%%%%%%%

\section{Conclusions and Discussion} \label{sec:Conclusion}

We explored a novel nonlinear coevolving voter model in which nodes take local information into consideration for their update actions, and we examined variants of our model with three different rewiring schemes: rewire-to-same, rewire-to-random, and rewire-to-none (i.e., ``unfriending''). In our nonlinear CVM, updates are edge-based and occur asynchronously. An updating node surveys its neighbor and records the fraction $\sigma$ that share its state. With probability $\sigma^q$, for a nonlinearity parameter $q$, the node rewires its discordant connection; otherwise, with complementary probability $1 - \sigma^q$, the node adopts a new state. 

By conducting extensive numerical simulations, we observed that our nonlinear CVM exhibits qualitatively similar characteristics as the linear CVM of~\cite{durrett2012graph} with respect to terminal state densities when both models are initialized on ER networks with equal state densities $N_A(0)/N=N_B(0)/N=1/2$. For example, both types of models possess a regime with rapid fragmentation into communities of different opinion states and a regime in which the system reaches a consensus. However, when we seed our nonlinear CVM with more complicated network structures, such as ones with community structure or core--periphery structure, we observed striking differences between our nonlinear CVM and the aforementioned linear CVM. In these scenarios, when the nodes have distorted views of local densities --- such that they believe that they are in the majority or minority when the opposite is true --- the value of the nonlinearity parameter $q$ has a major effect on terminal state densities. For certain values of $q$ and initial network topologies, the initially minority state consistently became the consensus in our simulations; for other values of $q$, the initially majority state consistently became the consensus. Although further analysis is necessary (especially of finite-size effects), our work suggests that on certain networks, our nonlinear CVM exhibits a rich assortment of phase transitions. The impact of initial network topology on terminal state densities distinguishes our nonlinear CVM from the linear CVM of~\cite{durrett2012graph}. We also demonstrated that unlike the linear CVM of~\cite{durrett2012graph}, which is very sensitive to the choice of rewiring mechanism, our nonlinear CVM yields qualitatively similar behavior with both the RTR and RTS mechanisms.

Our nonlinear CVM also exhibits fascinating manifestations of both majority and minority illusions. For example, we observed that majority illusions can arise as a system evolves, and we also found that such illusions can resolve in different ways (e.g., by becoming true or by nodes wising up) for different values of the nonlinearity parameter $q$.

Our investigation of our nonlinear CVM raises several interesting questions. For example, we noted in \fref{sec:RTS} and \fref{sec:RTN} that the differences in model behavior from rewiring schemes in our nonlinear CVM are far less pronounced than they are in the examined linear CVM, and it is desirable to develop a mechanistic understanding of this qualitative difference between these families of models. It will also be interesting to develop precise conditions that determine when majority and minority illusions arise in our nonlinear CVM. Such illusions can either accelerate or stifle the spread of rare states, so it is important to develop an understanding of the mechanisms that lead to these effects. To examine these ideas further, it will be interesting to explore the dynamics of our models on a larger variety of networks, such as those that were developed recently by Stewart et al.\  \cite{stewart2019information}. It is also desirable to extend tools for approximating the dynamics of linear CVMs (such as approximate master equations~\cite{gleeson2013binary} and pair approximations~\cite{jedrzejewski2017pair}) to nonlinear CVMs.

There are also many fascinating ways to extend our nonlinear CVM. We anticipate that it will be particularly interesting to incorporate ideas from recent efforts that have examined the effects of noise (e.g., random state mutations)~\cite{ji2013correlations}, hipsters (in the form of nodes that try to be in a minority)~\cite{juul2019hipsters}, and zealots (in the form of nodes that do not change states)~\cite{klamser2017zealotry}. Another worthwhile direction is to study adaptive opinion models with continuous opinions (e.g., using a bounded-confidence mechanism)~\cite{kozma2008consensus, sasahara2019inevitability,brede2019}. 

Opinions and social networks are coupled to each other intimately in a complex way. Developing and refining models for coevolving opinions and social networks can help improve understanding of not only their relationships with each other but also their impact on political and social polarization, echo chambers, and other social phenomena.

%%%%%%%%%%%%%%%%%%%%%%%%
%%% Acknowledgements
%%%%%%%%%%%%%%%%%%%%%%%%

\section*{Acknowledgements}

We thank Andrea Baronchelli, Heather Zinn Brooks, Michelle Feng, James Gleeson, Thilo Gross, Istvan Kiss, Yi Ming Lai, Michael Lindstrom, Joel Miller, William Oakley, Alice Schwarze, Samuel Scarpino, and Bill Shi for helpful discussions and comments. We also thank two anonymous referees for helpful comments. YHK was supported by MURI N00014-17-S-F006. YHK and MAP acknowledge support from the National Science Foundation (grant number 1922952) through the Algorithms for Threat Detection (ATD) program.

%%%%%%%%%%%%%%%%%%%%%%%%
%%% Appendices
%%%%%%%%%%%%%%%%%%%%%%%%

\appendix
\renewcommand{\thesubsection}{\arabic{subsection}}

%%%%%%%%%%%%%%%%%%%%%%%%
%%% Mean-Field Calculations Appendix
%%%%%%%%%%%%%%%%%%%%%%%%

\section{ Mean-Field Approximation for Local Surveys}
\label{appsec:MFA_Sigma}

\counterwithin{figure}{section}

In \fref{sec:approx_rtr}, we explored a mean-field approximation of our nonlinear CVM with an RTR scheme. 
This necessitated finding a suitable mean-field analog of the local surveys $\sigma_i$. We considered a state-heterogeneous approximation, in which we separately average over nodes in state $A$ and state $B$. Denoting these approximations as $\sigma_A$ and $\sigma_B$, we found in \fref{eq:sigma_equation_1} that
\begin{equation}
    \sigma_A = \frac{1}{2N_{AB}}\left( N_{AA}- \sum_{i:S[i]=A,\, k_i \neq 0} \frac{s_i^2}{k_i}\right)
\end{equation}
and
\begin{equation}
  \sigma_B=\frac{1}{2N_{AB}}\left( N_{BB}- \sum_{i:S[i]=B,\, k_i \neq 0} \frac{s_i^2}{k_i}\right)\,.
\end{equation}
Equivalently, we can write
\begin{equation}
    \sigma_A=\sum_{i:S[i]=A, \,\bar{s_i} \neq 0} \dfrac{s_i\bar{s_i}}{(s_i+\bar{s_i})\sum_{j:S[j]=A} \bar{s_j}}
\end{equation}
and 
\begin{equation}
    \sigma_B=\sum_{i:S[i]=B,\, \bar{s_i} \neq 0} \dfrac{s_i\bar{s_i}}{(s_i+\bar{s_i})\sum_{j:S[j]=B} \bar{s_j}}\,.
\end{equation}

We seek to compute $\mathbb{E}[\sigma_A(0)]$ for systems that we seed with two-community SBM networks. (If we take all edge probabilities in the SBM to be equal, we obtain ER networks.) We calculate
\begin{equation} \label{Eq:ExpectedValueSigma}
	\begin{split}
\mathbb{E}[\sigma_A(0)] & = \mathbb{E}\left[\sum_{i:S[i]=A,\, \bar{s_i} \neq 0} \dfrac{s_i\bar{s_i}}{(s_i+\bar{s_i})\sum_{j:S[j]=A} \bar{s_j}}\right] \\
& = \sum_{i:S[i]=A,\, \bar{s_i} \neq 0} \mathbb{E}\left[ \dfrac{s_i\bar{s_i}}{(s_i+\bar{s_i})\sum_{j:S[j]=A} \bar{s_j}}\right] \\
& \approx \left(N_A(0) (1-\exp(P_{ab}N_B(0)))\right) \\
	&\qquad\times\mathbb{E}\left.\left[ \dfrac{s_i\bar{s_i}}{(s_i+\bar{s_i})\sum_{j:S[j]=A} \bar{s_j}}\right\vert \bar{s_i} > 0\right] \\
& = \left(N_A(0) (1-\exp(P_{ab}N_B(0)))\right) \\
	&\qquad\times\mathbb{E}\left.\left[ \dfrac{s_i\bar{s_i}}{(s_i+\bar{s_i})\left(\bar{s_i}+\sum_{j:S[j]=A,\, j\neq i} \bar{s_j}\right)}\right\vert \bar{s_i} > 0\right]\,.
	\end{split}
\end{equation}
The quantity $s_i$ is the random number of neighbors of node $i$ that have the same opinion as $i$;
it is distributed binomially with parameters $n=N_A(0)-1$ and $p=P_{aa}$. The quantity $\bar{s_i}$ is the random number of neighbors of node $i$ that have a different opinion from $i$; it is distributed binomially with parameters $n=N_B(0)$ and $p=P_{ab}$. 
There are $N_A(0)$ nodes in state $A$; with probability $(1-(1-P_{ab})^{N_B(0)})$, each such node has at least one discordant edge. In the limit $N \rightarrow \infty$, there are $\left(N_A(0) [1-\exp(P_{ab}N_B(0)]\right)$ nodes in state $A$ with at least one discordant neighbor. We use this expression as a large-$N$ approximation in the third line of Eq.~\ref{Eq:ExpectedValueSigma}; in this line, we replace summing over nodes that are in state $A$ and are incident to at least one discordant edge by multiplying by the expected number of such nodes.

Edges in our SBM networks are independent of each other, so $s_i$, $\bar{s_i}$, and $\bar{s_j}$ are all independent random variables. For convenience, we define the notation $z := \sum_{j:S[j]=A, \,j\neq i} \bar{s_j}$. The sum of independent and identically distributed
binomial random variables is another binomial random variable, so $z$ is distributed binomially with parameter values $n=(N_A(0)-1)N_B(0)$ and $p=P_{ab}$. Using the law of the unconscious statistician, we obtain
\begin{equation}\label{eq:expected_sigma_2}
\begin{split}
	&E\left[\left. \dfrac{s_i\bar{s_i}}{(s_i+\bar{s_i})\left(\bar{s_i}+z\right)}\right\vert \bar{s_i} > 0\right]  \\
	&=\frac{\sum_{t_i,\,\bar{t_i}\geq 1,\, t} \left(\dfrac{t_i\bar{t_i}}{(t_i+\bar{t_i})(\bar{t_i}+z)}P(s_i=t_i)P(\bar{s_i}=\bar{t_i})P(z=t)\right)}{1-P(\bar{s_i}=0)}\,.
\end{split}
\end{equation}
With \fref{eq:expected_sigma_2}, we can numerically approximate $\mathbb{E}[\sigma_A(0)]$ for both ER and two-community SBM networks.

%%%%%%%%%%%%%%%%%%%%%%%%
%%% Nonlinear CVM Algorithms Appendix
%%%%%%%%%%%%%%%%%%%%%%%%

\section{Algorithms for Nonlinear Coevolving Voter Models} \label{appsec:RTS_RTN_Algos} 
\counterwithin{figure}{section}

%%%%%%%%%%%%%%%%%%%%%%%% Rewire-To-Same Model Algorithm %%%%

\subsection{Rewire-To-Same (RTS) Model} \label{appsec:RTSAlgo} 

In Algorithm~\ref{alg:ftvm_rts}, we present the precise rules for our nonlinear CVM with a rewire-to-same scheme (i.e., for the nonlinear RTS-CVM). Specifically, we use an RTS scheme that stipulates that there is no replacement edge when a primary node attempts to rewire, but is unable to do so (see \fref{sec:RTS}). We use the notation $V_X$ to denote the set of nodes that are in state $X$.

\begin{algorithm}[H]
\caption{Nonlinear Rewire-to-Same (RTS) Coevolving Voter Model}\label{alg:ftvm_rts}
\begin{algorithmic}[1]
\Procedure{FittingInVM}{$V,E,S,q$}\Comment{{Input: Initial network and opinion states}} 
\State $E_D \gets Discordant(V,E,S);\; \; t \gets 0; \; \; Record(V,E,S,t)$ 
\State $V_{A}\gets GetNodesByState(V,S,A);\; \; V_{B}\gets GetNodesByState(V,S,B)$
\While{$E_D\not=\emptyset$}\Comment{While there are discordant neighbors}
\State $(i,j) \gets RandomChoice(E_D)$
\State $PrimaryNode, SecondaryNode \gets RandomPermutation(i,j)$
\State $\sigma \gets LocalVote(PrimaryNode,V,E,S)$ 
\State $u \gets Uniform(0,1)$
\State $PotentialNewNeighbors \gets V_{S[PrimaryNode]}\setminus \Gamma(PrimaryNode,E)$
\If{$u\leq \sigma^q$} \Comment{Rewire}
\State $E.remove(PrimaryNode,SecondaryNode)$
\If{$PotentialNewNeighbors \neq \emptyset$} 
\State $NewNeighbor \gets RandomChoice(PotentialNewNeighbors)$
\State $E.add(PrimaryNode,NewNeighbor)$
\EndIf
\Else \Comment{Adopt}
\State $S[PrimaryNode] \gets S[SecondaryNode]$
\State $V_{A}\gets GetNodesByState(V,S,A);\; \; V_{B}\gets GetNodesByState(V,S,B)$
\EndIf
\State $E_D \gets Discordant(V,E,S);\; \; t \gets t+1$;  \; $Record(V,E,S,t)$
\EndWhile
\EndProcedure
\end{algorithmic}
\end{algorithm}

%%%%%%%%%%%%%%%%%%%%%%%% Rewire-To-None Model Algorithm %%%%

\subsection{Rewire-To-None (RTN) Model} \label{appsec:RTNAlgo} 

In Algorithm~\ref{alg:ftvm_rtn}, we present the precise rules for our nonlinear CVM with a rewire-to-none (i.e., edge-deletion) scheme. We use the acronym RTN-CVM for this model.

\begin{algorithm}[H]
\caption{Nonlinear Rewire-to-None (RTN) Coevolving Voter Model}\label{alg:ftvm_rtn}
\begin{algorithmic}[1]
\Procedure{FittingInVM}{$V,E,S,q$}\Comment{{Input: Initial network and opinion states}} 
\State $E_D \gets Discordant(V,E,S);\; \; t \gets 0; \; \; Record(V,E,S,t)$ 
\While{$E_D\not=\emptyset$}\Comment{While there are discordant neighbors}
\State $(i,j) \gets RandomChoice(E_D)$
\State $PrimaryNode, SecondaryNode \gets RandomPermutation(i,j)$
\State $\sigma \gets LocalVote(PrimaryNode,V,E,S)$ 
\State $u \gets Uniform(0,1)$
\If{$u\leq \sigma^q$} \Comment{``Rewire"}
\State $E.remove(PrimaryNode,SecondaryNode)$
\Else \Comment{Adopt}
\State $S[PrimaryNode] \gets S[SecondaryNode]$
\EndIf
\State $E_D \gets Discordant(V,E,S);\; \; t \gets t+1$;  \; $Record(V,E,S,t)$
\EndWhile
\EndProcedure
\end{algorithmic}
\end{algorithm}

%%%%%%%%%%%%%%%%%%%%%%%%
%%% Linear CVMs Appendix
%%%%%%%%%%%%%%%%%%%%%%%%

\section{Linear Coevolving Voter Models} \label{appsec:Linear_CVM_Simulations}
\counterwithin{figure}{section}

%%%%%%%%%%%%%%%%%%%%%%%% Simulations of linear RTR-CVM %%%%%%

\subsection{Linear Rewire-to-Random (RTR) CVM} \label{appsec:Linear_CVM_Simulations_RTR} 

\begin{figure}[!tbp]
  \centering
    \subfloat[$N_A(0)/N=1/2$]{\includegraphics[width=0.48\linewidth]{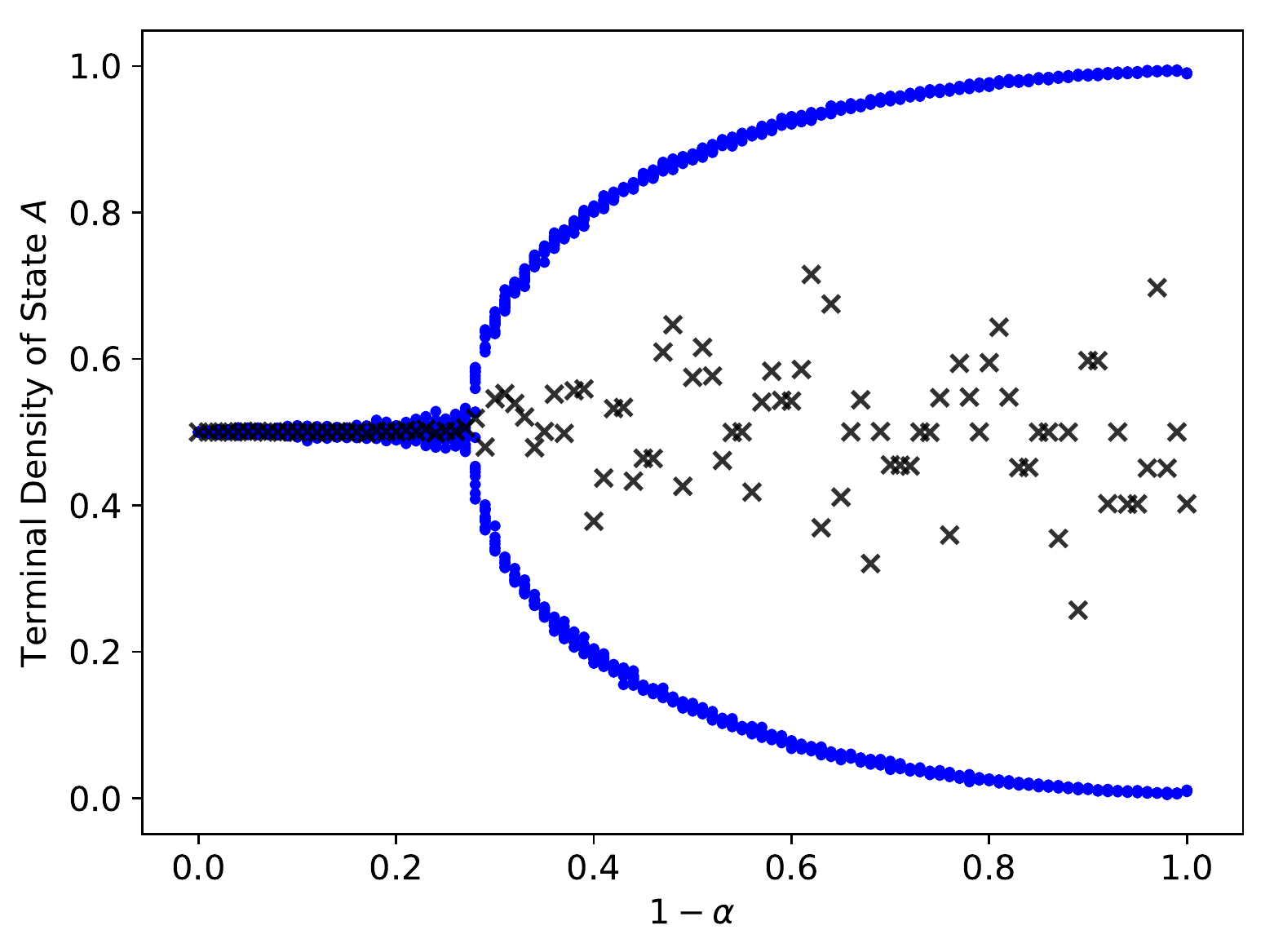}\label{fig:LIN_RTR_ER_Term_A}}
    \hfill
    \subfloat[$N_A(0)/N=1/4$]{\includegraphics[width=0.48\linewidth]{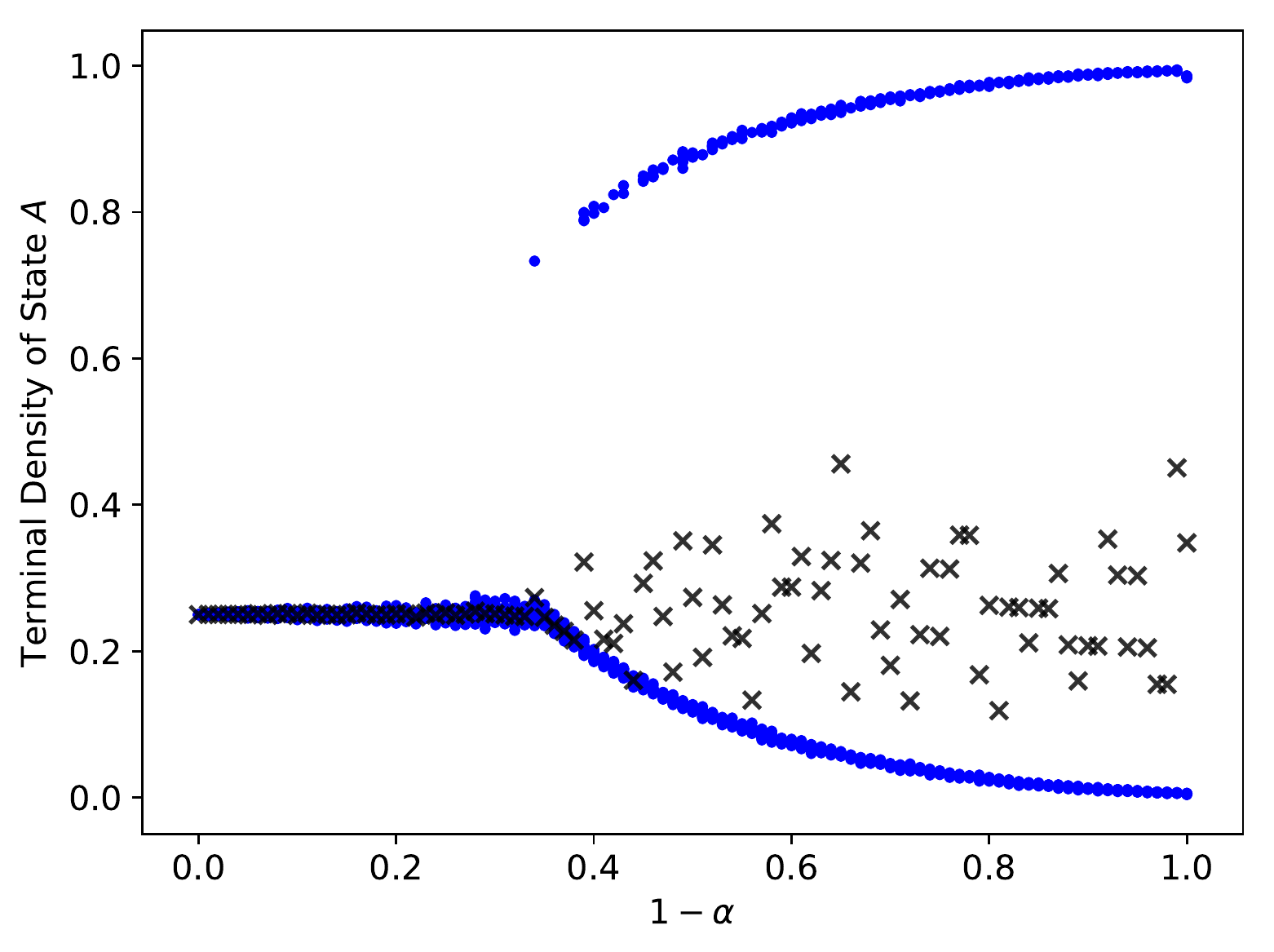}\label{fig:LIN_RTR_ER_Term_A_1-3}}
  \caption{Terminal density of state $A$ in simulations for the linear RTR-CVM from~\cite{durrett2012graph} for $\alpha \in [0,1]$ with a step size of $\Delta \alpha = 0.01$. For each value of $\alpha$, we simulate $20$ realizations. We seed each realization with a different ER network with $N=20,000$ nodes and an edge probability of $p=\frac{4}{N-1}$. In (a), we initialize half of the nodes in state $A$ and half of the nodes in state $B$. In (b), we initialize $1/4$ of the nodes in state $A$ and $3/4$ of the nodes in state $B$. Each blue dot indicates a value of $N_A(t)/N$ at the termination of a simulation.  For each $\alpha$, each $\times$ symbol is the terminal mean of $N_A(t)/N$ over the $20$ realizations.
   Note that the horizontal axis is $(1-\alpha)$.}\label{fig:LIN_RTR_ER_Term_A_both}
\end{figure}

We compare the simulation results for our nonlinear CVM with a rewire-to-random scheme to the linear rewire-to-random CVM that was studied in~\cite{durrett2012graph}. We seed the system using ER $G(N,p)$ networks with $N=20,000$ nodes and an edge probability of $p=\frac{4}{N-1}$. In \fref{fig:LIN_RTR_ER_Term_A}, we initialize half of the nodes in state $A$ and the other half in state $B$. In \fref{fig:LIN_RTR_ER_Term_A_1-3}, we initialize $1/4$ of the nodes in state $A$ and the other $3/4$ of nodes in state $B$. In \fref{fig:LIN_RTR_ER_Term_A_both}, we plot the terminal density of state $A$ for $20$ realizations of the simulations for each value of $q$. In this figure, we also show the means of terminal densities for each $q$.

In \fref{fig:LIN_RTR_ER_Term_A}, when $1-\alpha \lessapprox  0.25$, rewiring actions dominate; the fraction of nodes that terminate in state $A$ is approximately constant, with a value of $0.5$. For progressively larger values of $1-\alpha$ (i.e., as we consider a progressively smaller rewiring rate $\alpha$ and hence a progressively larger adoption rate), rewiring and adoption actions begin competing and the plot appears to branch, with one branch decreasing to $0$ and the other increasing to $1$ as $1-\alpha \rightarrow 1$. This illustrates that, by the time the system terminates, there are larger changes to the state densities of the system for progressively larger $1-\alpha$. Because the system begins with $N_A(0)/N=N_B(0)/N=1/2$, terminating along either branch (i.e., whether there is a positive or negative change for $N_A(t)/N$ or $N_B(t)/N$) is equally probable, as indicated by the values of the means of the terminal state-$A$ densities.

When $1-\alpha = 1$, no rewiring occurs, so isolated nodes do not change their opinion state. In our simulations in \fref{fig:LIN_RTR_ER_Term_A}, the seed ER networks have an expected mean degree of $4$. In our realizations, the seed networks have a largest connected component (LCC) that consists of approximately $98.2\%$ of the nodes on average, and most of the remaining approximately $1.8\%$ of the nodes have degree $0$. When the system terminates and the nodes in the LCC settle in either state $A$ or state $B$, approximately $0.9$\,\% of nodes are in the other opinion state, because we initialize half of the nodes in each state and those nodes never update their states.

In \fref{fig:LIN_RTR_ER_Term_A_1-3}, we observe behavior that is qualitatively similar to that in \fref{fig:LIN_RTR_ER_Term_A}. For $1-\alpha \lessapprox  0.35$, rewiring actions dominate; the fraction of nodes that terminate in state $A$ is approximately constant, with a value of $0.25$. For progressively larger values of $1-\alpha$, rewiring and adoption actions begin competing and the plot appears to branch, with one branch decreasing to $0$ (indicating a negative change to $N_A(t)/N$) and the other increasing to $1$ (indicating a positive change to $N_A(t)/N$) as $1-\alpha \rightarrow 1$. Unlike in \fref{fig:LIN_RTR_ER_Term_A}, we begin with $N_A(0)/N=1/4$, so we expect only $1/4$ of the realizations to terminate along the upper branch (i.e., with a positive change to $N_A(t)/N$), which is what we observe by examining the means of the terminal state-$A$ densities. 

\begin{figure}[!tbp]
  \centering
    \subfloat[Two-Community Structure]{\includegraphics[width=0.48\linewidth]{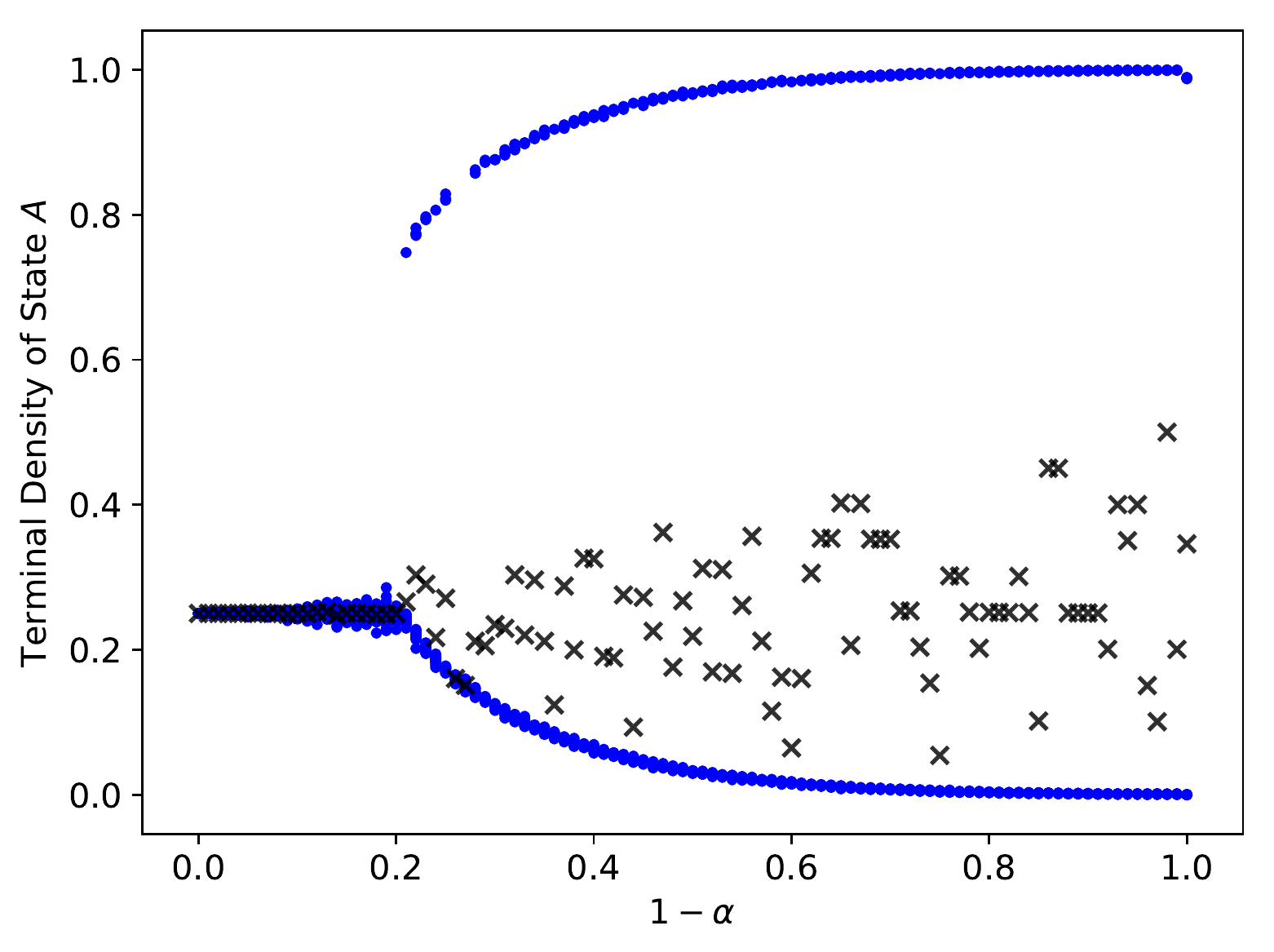}\label{fig:LIN_RTR_SBM_TWO_COM_Term_A}}
    \hfill
    \subfloat[Core--Periphery Structure]{\includegraphics[width=0.48\linewidth]{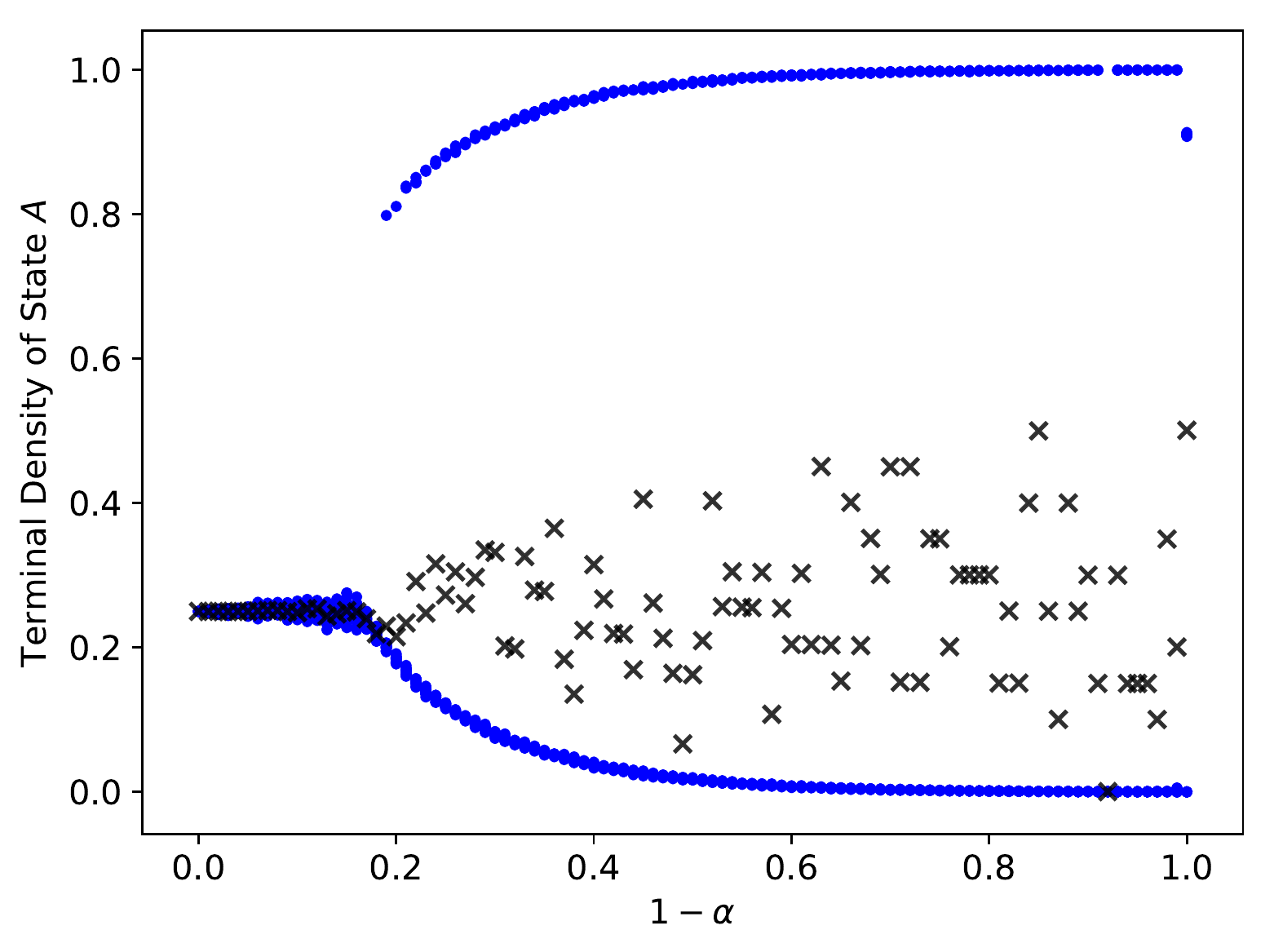}\label{fig:LIN_RTR_CP_Term_A}}
  \caption{Terminal density of state $A$ in simulations of the linear RTR-CVM from~\cite{durrett2012graph} for $\alpha \in [0,1]$ with a step size of $\Delta \alpha = 0.01$.  For each value of $\alpha$, we simulate $20$ realizations. In (a), we seed each realization with a different two-community SBM network (as described in \fref{sec:RTR_SBM_two_communities}). In (b), we seed each realization with a different core--periphery structure using SBM networks (as described in \fref{sec:RTR_SBM_two_communities}). We plot individual realizations with blue dots and means with $\times$ symbols.}\label{fig:LIN_RTR_SBM_Term_A}
\end{figure}

We also consider a linear RTR-CVM in which we seed the system with two-community structure using SBMs. We use the same parameter values as in \fref{sec:RTR_SBM_two_communities}, so $N_A(0)/N=1/4$ of the nodes are in state $A$. In \fref{fig:LIN_RTR_SBM_TWO_COM_Term_A}, we plot the terminal density of state $A$ from our simulations. The plot has roughly the same shape as when we seeded the linear CVM with an ER network (see \fref{fig:LIN_RTR_ER_Term_A_1-3}). When $1-\alpha \lessapprox 0.2$, rewiring actions dominate; the terminal minority-state densities are constant, with a value of $0.25$. Adoption actions compete with rewiring actions when $1-\alpha\gtrapprox0.2$, and the plot appears to branch, with one branch decreasing to $0$ and the other increasing to $1$ as $1-\alpha \rightarrow 1$. This illustrates that, by the time the system terminates, there are progressively larger changes to the state densities of the system for progressively larger values of $1-\alpha$. Because the system begins with $N_A(0)/N=1/4$, terminating along the upper branch (i.e., there is a positive change for $N_A(t)/N$) occurs in approximately $1/4$ of the realizations, as indicated by the values of the means of the terminal state-$A$ densities.

Finally, in \fref{fig:LIN_RTR_CP_Term_A}, we plot simulations of the linear RTR-CVM in which we seed the system with core--periphery structure.  We use the same parameter values as in \fref{sec:RTR_SBM_CP}. The plot's similarity to \fref{fig:LIN_RTR_SBM_TWO_COM_Term_A} illustrates an insensitivity of this linear RTR-CVM to some types of initial network structure.  With the parameter values of our seed core--periphery networks, approximately $10$\,\% of the nodes begin in state $B$ and are isolated. When $1-\alpha=1$, there is no rewiring, so these nodes remain isolated and thus do not change state. This explains why in the realizations in which state $A$ spreads to almost every node, it only reaches approximately $90$\.\% of nodes in the system.

There are values of $\alpha$ in \fref{fig:LIN_RTR_ER_Term_A_1-3} and \fref{fig:LIN_RTR_SBM_Term_A} for which none of our realizations terminate with almost all nodes in state $A$. The probability that a realization terminates along the top branch (signifying a net positive change in the density of state $A$) is $N_A(0)/N=1/4$, so we expect on rare occasions (specifically, with probability $(3/4)^{20} \approx0.0032$) that all $20$ realizations for a particular value of $\alpha$ terminate along the bottom branch (i.e., with a net negative change in the density of state $A$).

%%%%%%%%%%%%%%%%%%%%%%%% Simulations of linear RTS-CVM %%%%%

\subsection{Linear Rewire-to-Same CVM} \label{appsec:Linear_CVM_Simulations_RTS}

\begin{figure}
  \includegraphics[width=.7\linewidth]{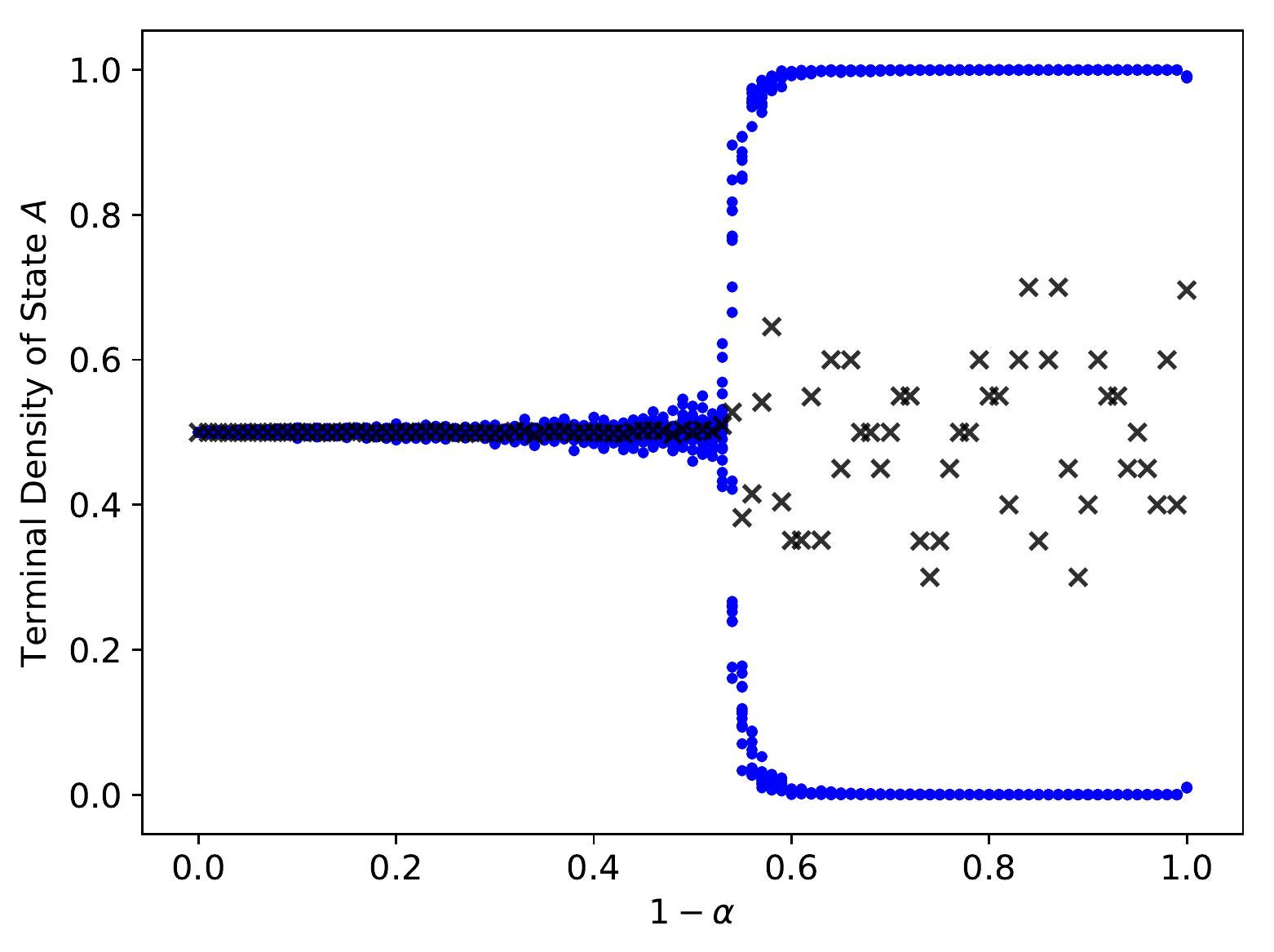}
    \caption{Terminal density of state $A$ in simulations of the linear RTS-CVM from~\cite{durrett2012graph} for $\alpha \in [0,1]$ with a step size of $\Delta \alpha = 0.01$. For each value of $\alpha$, we simulate $20$ realizations. We seed each realization with a different ER network with $N=20,000$ nodes, an edge probability of $p=\frac{4}{N-1}$, half of the nodes in state $A$, and half of the nodes in state $B$. We plot individual realizations with blue dots and means with $\times$ symbols.}\label{fig:lin_rts_terminal_A_ER}
\end{figure}

We compare the results of simulations of our nonlinear RTS-CVM to simulations of the linear RTS-CVM from~\cite{durrett2012graph}. In \fref{fig:lin_rts_terminal_A_ER}, we observe what appears to be a discontinuous phase transition for a critical value of $\alpha$. For $1-\alpha \lessapprox 0.57$, rewiring dominates and state densities do not change significantly. However, when $1-\alpha \gtrapprox 0.57$, almost every node terminates in the same state. The system begins with $N_A(0)/N=N_B(0)/N=1/2$, so it is equally probable for almost every node to terminate in state $A$ or almost every node to terminate in state $B$. As we noted in \fref{appsec:Linear_CVM_Simulations_RTR}, when $1-\alpha = 1$, there are no rewiring actions, so isolated nodes do not change their opinion state.

\begin{figure}[!tbp]
  \centering
    \subfloat[Two-Community Structure]{\includegraphics[width=0.48\linewidth]{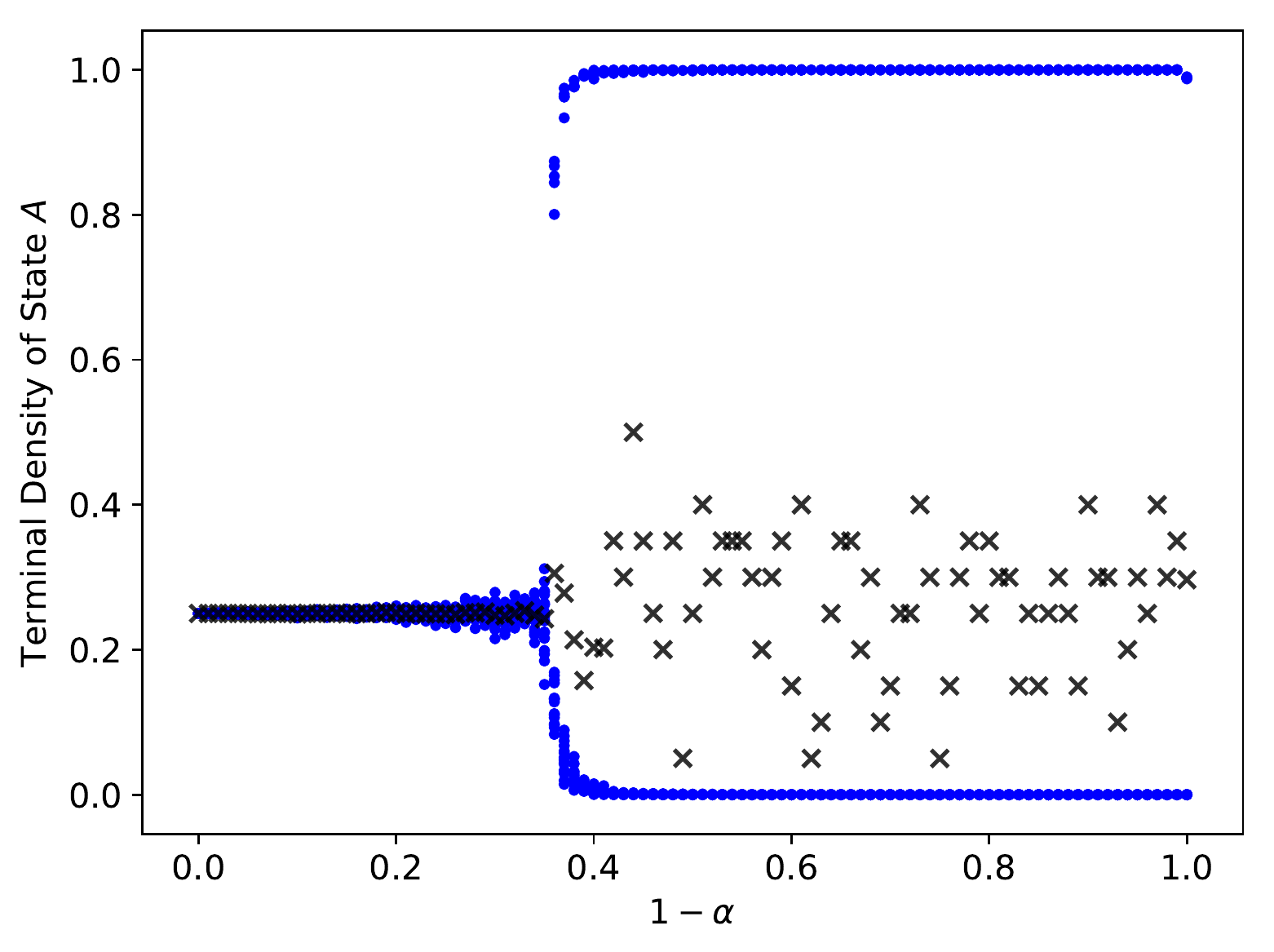}\label{fig:LIN_RTS_SBM_TWO_COM_Term_A}}
    \hfill
    \subfloat[Core--Periphery Structure]{\includegraphics[width=0.48\linewidth]{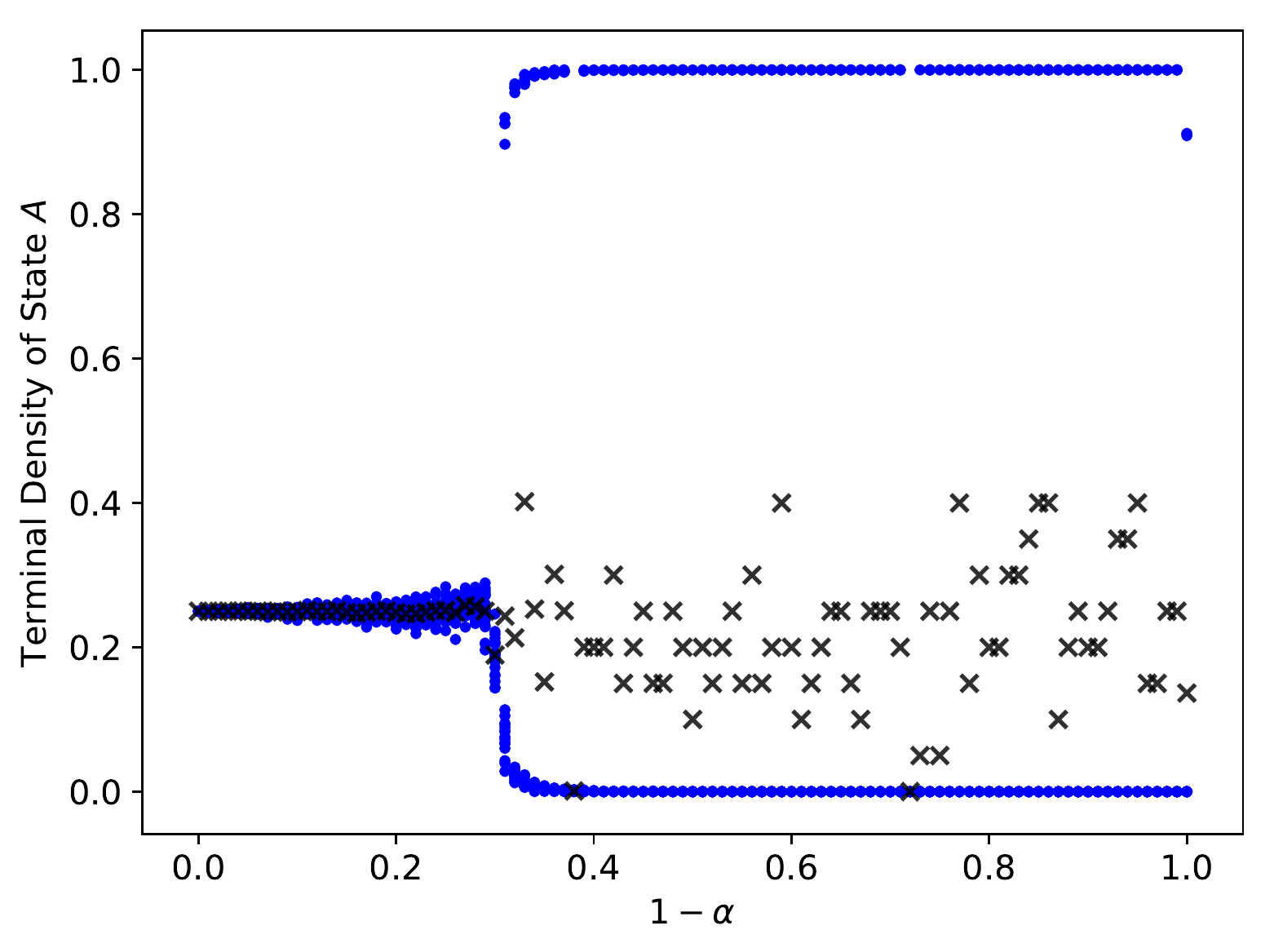}\label{fig:LIN_RTS_CP_Term_A}}
  \caption{Terminal density of state $A$ in simulations of the linear RTS-CVM from~\cite{durrett2012graph} for $\alpha \in [0,1]$ with a step size of $\Delta \alpha = 0.01$. For each value of $\alpha$, we simulate $20$ realizations. In (a), we seed each realization with a different two-community SBM network (as described in \fref{sec:RTR_SBM_two_communities}). In (b), we seed each realization with a different core--periphery structure using SBM networks (as described in \fref{sec:RTR_SBM_two_communities}). We plot individual realizations with blue dots and means with $\times$ symbols.}\label{fig:LIN_RTS_SBM_Term_A}
\end{figure}

In \fref{fig:LIN_RTS_SBM_Term_A}, we plot the terminal state-$A$ densities from simulations of the linear RTS-CVM from~\cite{durrett2012graph} on networks that we seed with two-community structure (see \fref{sec:RTR_SBM_two_communities}) and on networks that we seed with core--periphery structure (see \fref{sec:RTR_SBM_two_communities}). In both cases, we observe qualitatively similar results as in \fref{fig:lin_rts_terminal_A_ER}, which again suggests that this linear CVM is less sensitive than our nonlinear CVM to some initial network structures. It also suggests that this linear CVM is more sensitive than our nonlinear CVM to the choice of rewiring mechanism.

%%%%%%%%%%%%%%%%%%%%%%%% Linear Rewire-to-None Coevolving Voter Model and Simulations %%%%%%

\subsection{Linear Rewire-to-None CVM}
\label{appsec:LINRTNCVM} 

We briefly discuss some results from our simulations of linear CVMs with edge deletion (i.e., a rewire-to-none scheme). As in the linear RTR-CVM (see Appendix~\ref{appsec:Linear_CVM_Simulations_RTR}), this rule involves picking a discordant edge uniformly at random from the set of discordant edges, choosing one of the nodes to be the primary node uniformly at random, and then either deleting the discordant edge with probability $\alpha$ or having the primary node change states with probability $1-\alpha$. As in the RTS scheme (see \fref{sec:RTS}), $|E(t)|$ is not conserved. However, in the linear RTN-CVM, edge deletions occur at a fixed rate $\alpha$, so the expected number of edges is $\mathbb{E}[|E(t)|]=|E(0)|-\alpha t$, which is valid until the system terminates, after which the number of edges is constant.

\begin{figure}
  \includegraphics[width=.7\linewidth]{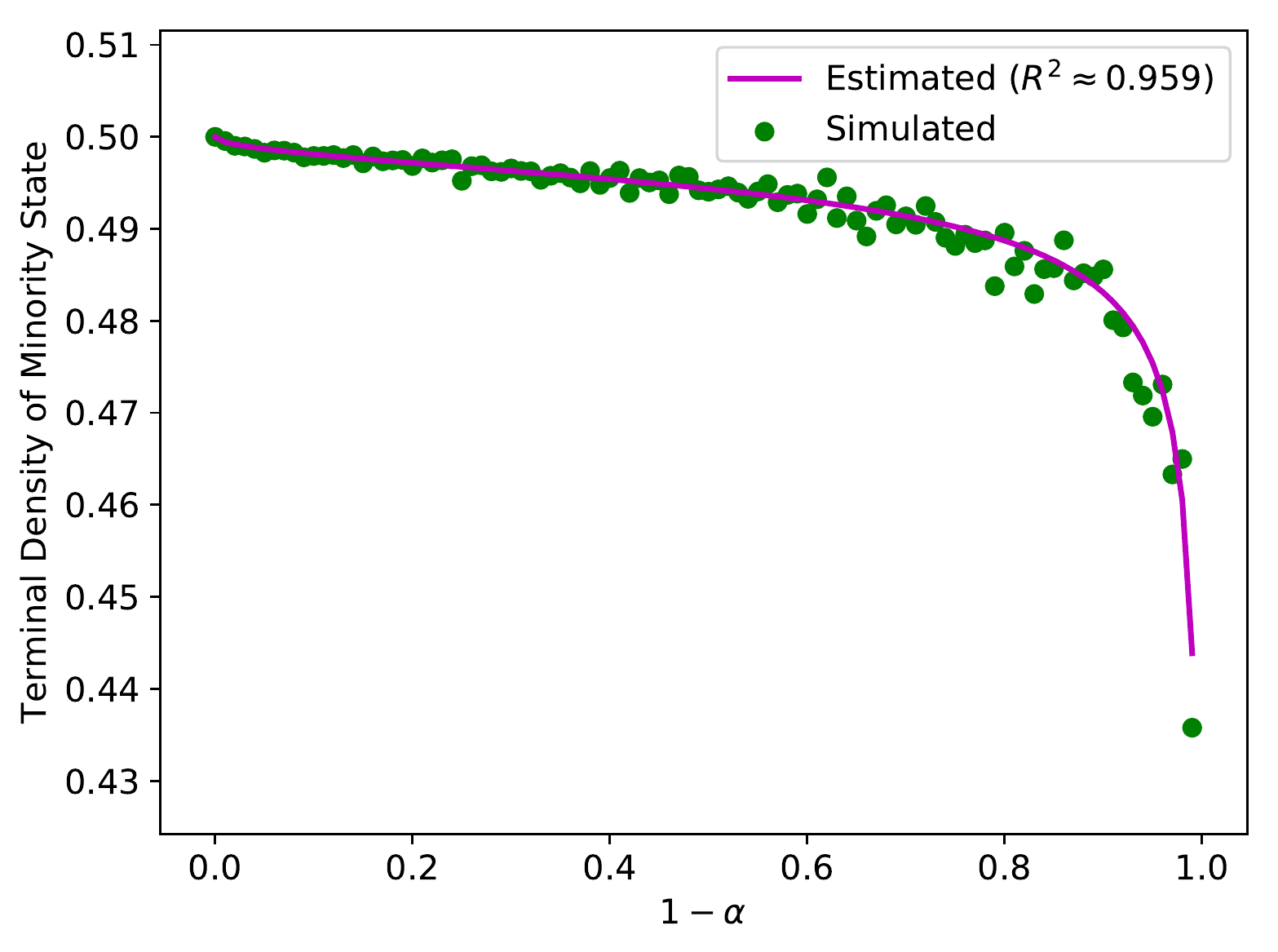}
    \caption{Terminal minority-state density in simulations of a linear CVM with edge deletion (i.e., using a rewire-to-none scheme) for $\alpha \in [0,1]$ with a step size of $\Delta \alpha = 0.01$. For each value of $\alpha$, we simulate $20$ realizations. We seed each realization with a different ER network with $N=20,000$ nodes and an edge probability of $p=\frac{4}{N-1}$. Half of the nodes start in state $A$, and the other half start in state $B$. Each green point is the mean over $20$ realizations for a given value of $\alpha$. The magenta curve is the estimated terminal minority-state density from \fref{eq:RTN_ER_estimator}.
     }\label{fig:lin_rtn_terminal_minority_ER}
\end{figure}

In \fref{fig:lin_rtn_terminal_minority_ER}, we plot the terminal minority-state density of our simulations. We seed the system with an ER $G(N,p)$ network with $N=20,000$ nodes and an edge probability of $p=\frac{4}{N-1}$.
Initially, there are approximately $20,000$ discordant edges. We compute an estimate for the terminal minority-state density in terms of $\alpha$ by assuming that the adoption mechanism does not significantly increase or decrease the number $N_{AB}(t)$ of discordant edges. Because we delete discordant edges at a constant rate $\alpha$, we expect the system to terminate in approximately $\frac{E_D(0)}{\alpha}$ elementary time steps. During this time, the expected number of adoption actions is $\frac{(1-\alpha)E_D(0)}{\alpha}$. Each adoption action increases the number $N_A(t)$ of nodes in state $A$ by $1$ with probability $1/2$ and decreases it by $1$ with probability $1/2$. We can thus think of $N_A(t)$ as a simple, symmetric random walk during the steps when adoption occurs. For a simple, symmetric random walk that starts at the origin, the expected distance of the walker to the origin after $n$ steps is $\sqrt{\dfrac{2n}{\pi}}$~\cite{feller1957probability}. Therefore, after $\frac{(1-\alpha)E_D(0)}{\alpha}$ steps, we expect that $N_A(t)$ has either increased or decreased by $\frac{1}{N}\sqrt{\frac{2(1-\alpha)E_D(0)}{\pi \alpha}}$. 
 
In \fref{fig:lin_rtn_terminal_minority_ER}, we include a plot of our estimate 
\begin{equation}\label{eq:RTN_ER_estimator}
    \frac{1}{2}-\frac{1}{N}\sqrt{\frac{2(1-\alpha)E_D(0)}{\pi \alpha}}
\end{equation} 
for the terminal minority-state density, where the $1/2$ term comes from the initial densities of $1/2$ and we subtract from this value because we are %studying
calculating the minority-state density. The plot illustrates that our estimate captures the behavior of the linear RTN-CVM accurately, with a coefficient of determination of $R^2 \approx 0.959$.

%%%%%%%%%%%%%%%%%%%%%%%%
%%% Large-$q$ Simulations
%%%%%%%%%%%%%%%%%%%%%%%%

\section{Simulations of our Nonlinear Coevolving Voter Model for Large Values of \texorpdfstring{$q$}{q}}
\label{appsec:RTR_big_q_appendix}

\begin{figure}[ht]
  \includegraphics[width=.7\linewidth]{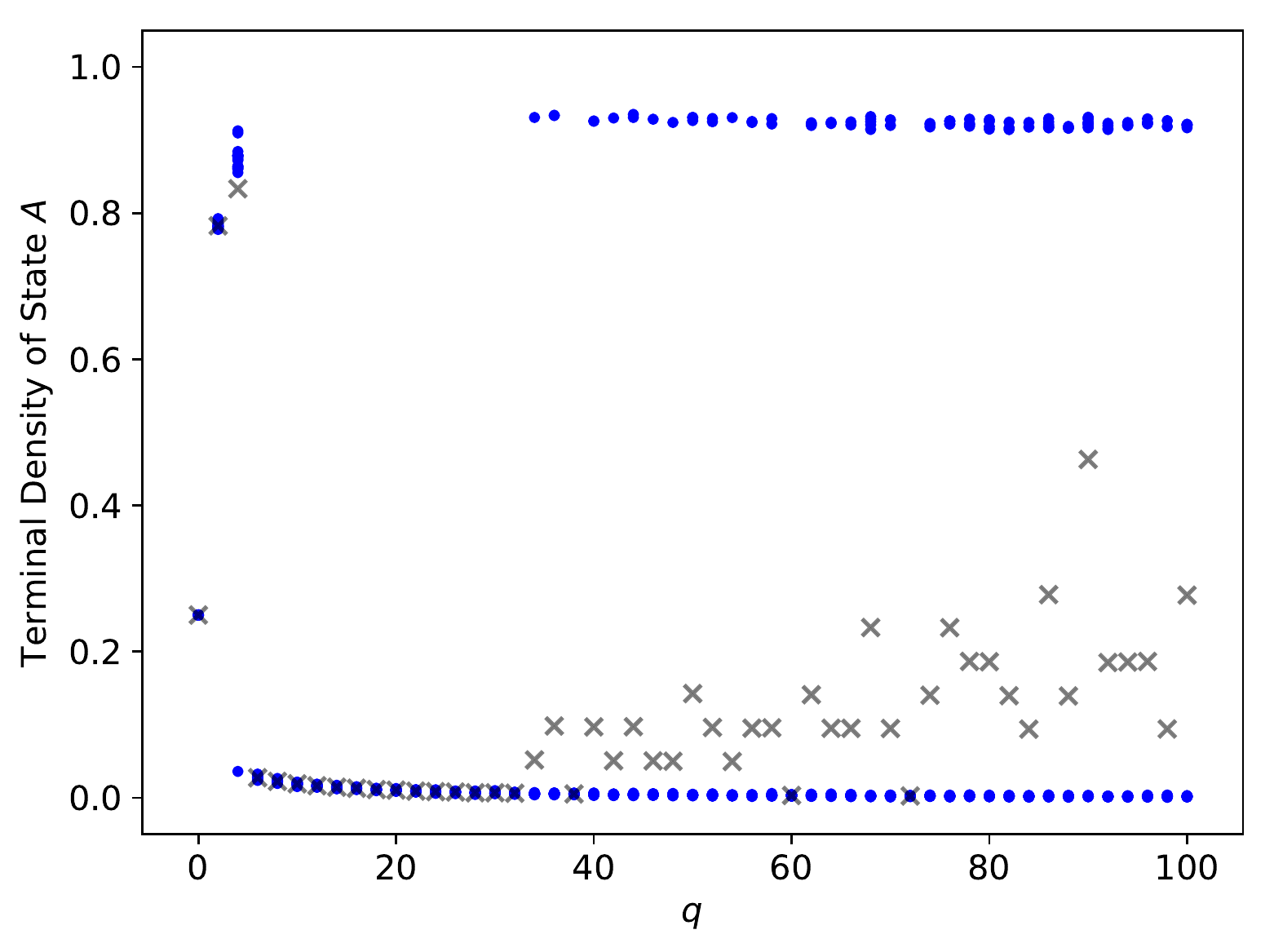}
    \caption{Terminal density of state $A$ in simulations of our nonlinear RTR-CVM (see Algorithm~\ref{alg:ftvm_rtr}) for $q\in[0,100]$ with a step size of $\Delta q= 2$. For each value of $q$, we simulate $20$ realizations. We seed each realization with a different SBM network with core--periphery structure (see \Fref{sec:RTR_SBM_CP}), but now there are $N=20,000$ nodes. We plot individual realizations with blue dots and means with $\times$ symbols.
        }\label{fig:RTR_SBM_Terminal_A_CP_big_q}
\end{figure}

In our simulations of nonlinear CVMs on SBM networks with two-community structure and core--periphery structure, we observed regimes of $q$ values in which almost every node terminates in state $B$, which initially has density $3/4$, for all $20$ realizations. See \fref{fig:RTR_SBM_TWO_COM_TERMINAL_A} and \fref{fig:RTR_SBM_Terminal_A_CP} for two examples. This regime extends past $q=12$, but we know that as $q\rightarrow \infty$, we must recover a voter model that does not coevolve with network structure. In this appendix, we repeat one of our simulations for large values of $q$ to improve our understanding of this behavior.

We repeat our experiment from \fref{sec:RTR_SBM_CP} of our nonlinear RTR-CVM seeded with core--periphery structure, but now we consider $q\in [0,100]$ and take our network to have $N=20,000$ nodes. In \fref{fig:RTR_SBM_Terminal_A_CP_big_q}, we plot the terminal density of state $A$. In this case, we see that when $q\gtrapprox38$, the adoption mechanism begins to dominate and state $A$ becomes competitive with state $B$. In some trials, state $A$ spreads to almost every node. For sufficiently large values of $q$, adoption actions should completely dominate and the system should behave like a voter model that does not coevolve with network structure; in such a scenario, almost every node terminates in the same state. Specifically, we expect that almost every node terminates in state $A$ in $N_A(0)/N=1/4$ of the realizations. In our computations, we observe this scenario for $q\gtrapprox 80$.

As we noted for \fref{fig:LIN_RTR_CP_Term_A}, due to our initial networks, approximately $10$\,\% of the nodes start in state $B$ and are isolated. Therefore, in the regimes that are dominated by adoption actions, the system terminates with these nodes still in state $B$, even when state $A$ has spread to almost every node in the LCC.

%%%%%%%%%%%%%%%%%%%%%%%%
%%% References
%%%%%%%%%%%%%%%%%%%%%%%%

%\bibliographystyle{apsrev4-2}
%\bibliography{VM_biblio.bib}

%%%%%%%%

%apsrev4-2.bst 2019-01-14 (MD) hand-edited version of apsrev4-1.bst
%Control: key (0)
%Control: author (72) initials jnrlst
%Control: editor formatted (1) identically to author
%Control: production of article title (-1) disabled
%Control: page (0) single
%Control: year (1) truncated
%Control: production of eprint (0) enabled
%

%%%%%%%

%%%%%%

\end{document}